\documentclass[fleqn,usenatbib]{mnras}
\usepackage[T1]{fontenc}
\usepackage{ae,aecompl}
\usepackage{graphicx}
\usepackage{amsmath}
\usepackage{amssymb}
\usepackage{txfonts}

\title[Expected IMBHs: Early-type galaxies]{Expected intermediate-mass black
  holes in the Virgo cluster. I. Early-type galaxies}

\author[Graham et al.]{
Alister W.\ Graham$^{1}$\thanks{E-mail: AGraham@swin.edu.au} 
and 
Roberto Soria$^{2,3,4}$ 
\\
$^1$Centre for Astrophysics and Supercomputing, Swinburne University of
Technology, Hawthorn, VIC 3122, Australia.\\
$^2$National Astronomical Observatories, Chinese Academy of Sciences,
Beijing 100012, China.\\
$^3$International Centre for Radio Astronomy Research, Curtin University, GPO
Box U1987, Perth, WA 6845, Australia.\\
$^4$Sydney Institute for Astronomy, School of Physics A28, The University of
Sydney, Sydney, NSW 2006, Australia.
}

\date{Accepted 2018 December 7. Received 2018 November 28; in original form
  2018 August 8}

\pubyear{2018}

\begin{document}
\label{firstpage}
\pagerange{\pageref{firstpage}--\pageref{lastpage}}
\maketitle

\begin{abstract}

We expand upon the AMUSE-Virgo survey which imaged 100 early-type Virgo
cluster galaxies
with the {\it Chandra X-ray Observatory}, and we place an
emphasis on potential intermediate-mass black holes (IMBHs).
Virgo early-type galaxies with absolute magnitudes
$\mathfrak{M}_B\gtrsim-20.5$ mag
have $B$-band luminosities that scale with the square of the stellar velocity
dispersion: $L_B\propto\sigma^2$.
We show that the non-linear `super-quadratic' relation $M_{\rm bh}\propto
L_B^2$--$L_B^{2.5}$ from
Graham \& Scott yields black hole masses, $M_{\rm bh}$,
that agree with the $M_{\rm bh}$-$\sigma$ relation
down to at least $M_{\rm bh}=10^4\,M_{\odot}$.
We predict that 30 of the 100 galaxies have $M_{\rm bh}\le10^5\,M_{\odot}$,
with IC~3602 having $M_{\rm bh}=10^4\,M_{\odot}$ and
IC~3633 having $M_{\rm bh}=$(6--8)$\times10^3\,M_{\odot}$.
We additionally revise the black hole Eddington ratios, and
their scaling with black hole mass,
and we report a point-like {\it Chandra} source at the nucleus of five
additional galaxies (NGC~4382, NGC~4387, NGC~4417, NGC~4467, and NGC~4472).
Moreover, three of the galaxies predicted here to host an IMBH
have a point-like {\it Chandra} source near their nucleus:
IC~3442 ($M_{\rm bh}=2\times10^5\,M_{\odot}$);
IC~3492 ($M_{\rm bh}=5\times10^4\,M_{\odot}$); and
IC~3292 ($M_{\rm bh}=6\times10^4\,M_{\odot}$). 
Furthermore, IC~3442 and IC~3292 host a nuclear star cluster that is
expected to house an IMBH.  Finally, we
present the ($B-K$)-$\mathfrak{M}_K$ colour-magnitude diagram and
discuss the implications for the $M_{\rm bh}$-$L_K$ and
$M_{\rm bh}$-$M_{\rm *,galaxy}$ relations, revealing why
stripped galaxies, especially rare compact elliptical
galaxies, should be excluded from $M_{\rm bh}$--$L$ scaling relations.

\end{abstract}


\section{Introduction}

Black holes with masses that are intermediate between those created by
supernova today ($\lesssim$100--300 solar masses: Belczynski et al.\ 2010;
Crowther et al.\ 2010) and the `supermassive black holes' (SMBHs: $M_{\rm
  bh} > 10^5\, M_{\odot}$) regularly detected at the centres of galaxies (e.g.\ Jiang et
al.\ 2011; Reines et al.\ 2013; Graham \& Scott 2015, and references therein), 
remain tantalisingly elusive.  To date, there are few good `intermediate-mass 
black hole' (IMBH) candidates at the centres of galaxies (Valencia-S.\ et al.\ 2012; Baldassare et
al.\ 2015; Graham et al.\ 2016; Nguyen et al.\ 2017, 2018), but there has been an
increasing number of claims for off-centre candidates (e.g.\ Colbert \&
Mushotzky 1999; Ptak \& Griffiths 1999; Farrell et al.\ 2009, 2014; Soria et
al.\ 2010; Webb et al.\ 2010, 2014, 2017; Liu et al.\ 2012; Secrest et
al.\ 2012; Sutton et al.\ 2012; Kaaret \& Feng 2013; Miller et al.\ 2013; Cseh
et al.\ 2015; Mezcua et al.\ 2015, 2018b; Oka et al.\ 2015; Pasham et al.\ 2015).
While this {\it might} be a clue to the formation location, and in turn
formation mechanism, of IMBHs, recent analysis suggests that many of these
non-central X-ray sources are super-Eddington stellar-mass black holes rather
than sub-Eddington IMBHs (Feng \& Soria 2011; Kaaret et al.\ 2017). 

Ongoing advances in our knowledge of the (central black hole mass)-(host galaxy
luminosity, $L$) scaling relation have revealed the need to re-examine 
past studies which estimated the masses of black holes, at the centres of
low-luminosity galaxies, using a nearly linear $M_{\rm bh}$--$L$ scaling
relation.  
Those near-linear relations were defined using galaxies with black hole masses
primarily greater than $10^7 - 10^8 \, M_{\odot}$ (e.g.\ Dressler 1989; Yee
1992; Kormendy \& Richstone 1995; Magorrian et al.\ 1998; McLure \& Dunlop
2002; Marconi \& Hunt 2003; Ferrarese \& Ford 2005; Graham 2007).  However, 
Graham \& Scott (2013) observed that, for black hole
masses less than $10^7 - 10^8 \, M_{\odot}$, 
the $B$-band ($\log M_{\rm bh}$)--($\log L_{\rm
  B,spheroid}$) relation has a slope equal to 2 to 2.5 rather than $\approx$1, which they
referred to as a `super-quadratic' $M_{\rm bh}$--$L$ relation as it was slightly steeper than
quadratic.  Given that $L_{\rm B}\propto\sigma^2$ --- where $\sigma$ is the
stellar velocity dispersion --- for low-mass
early-type galaxies (ETGs: Davies et al.\ 1983), the steeper than linear
$M_{\rm  bh}$--$L_{\rm B}$ relation is unavoidable if it is to be consistent with the 
$M_{\rm bh} \propto \sigma^{\beta}$ relation with its reported exponent
$\beta$ equal to 4--5 
(Ferrarese \& Merritt 2000; Gebhardt et al.\ 2000; 
Merritt \& Ferrarese 2001).  Moreover, and of relevance for this
investigation, this steeper $M_{\rm bh}$--$L_{\rm B}$
scaling relation predicts smaller black holes in low-mass galaxies than
the linear $M_{\rm bh}$--$L_{\rm B}$ relation, and suggests that 
some past estimates of black hole masses have been too high in low-mass ETGs,
perhaps missing the IMBH population. 

IMBHs would not only provide a bridging mass to the current bimodal
distribution of known black hole masses, but their existence, or lack thereof, may
hold clues to the seeding and early co-evolution of today's SMBHs and their host galaxies.
Furthermore, IMBHs provide interesting laboratories for the production of both
gravitational waves (Hils \& Bender 1995; Amaro-Seoane et al.\ 2007; Mapelli
et al.\ 2012) and flares from stellar tidal disruption events (Hills 1975;
Frank \& Rees 1976; Peterson \& Ferland 1986; Rees 1988; Komossa \& Bade 1999;
Komossa 2015; Lin et al.\ 2018; Fragione et al.\ 2018).

As a part of the early effort to search for IMBHs, and to additionally measure
the distribution of black hole accretion rates plus their duty cycle,
substantial observing time (453.6 ks) was awarded during {\it Chandra} Cycle~8 to a
Large Project titled `The Duty Cycle of Supermassive Black Holes: X-raying
Virgo' (Proposal ID 08900784, P.I.\ Treu).  The project imaged 84 early-type
galaxies in the Virgo cluster, and combined this with archival X-ray data for
the remaining 14 galaxies that comprised the sample of 100 ETGs which had
recently been imaged at optical wavelengths with the {\it Hubble Space
  Telescope} for the `Advanced Camera for Surveys Virgo Cluster Survey'
(ACSVCS; C\^ot\'e et al.\ 2004; Ferrarese et al.\ 2006).  Having identified
which galaxies have nuclear X-ray activity, and consequently likely host an
AGN, the Project needed to estimate the masses of these AGN's black holes.
Gallo et al.\ (2008) provided two sets of predictions for this, using the
galaxies' (i) velocity dispersion and (ii) $B$-band stellar luminosity.

At low masses, the Project's black hole mass estimates from the near-linear
$M_{\rm bh}$--$L_{\rm B,galaxy}$ relation were found to be systematically
larger than those derived using the $M_{\rm bh}$--$\sigma$ relation, see also
Wandel (1999) and Ryan et al.\ (2007) who had previously reported on this
pattern, as did Coziol et al.\ (2011).  
The difference in mass was as large as three orders of magnitude 
(Gallo et al.\ 2008, their figure~4) and left the Project unable to provide
clarity as to the black hole masses.  None of the black hole mass estimates
obtained from their adopted $M_{\rm bh}$--$L_{\rm B,galaxy}$ relation were
less than $10^6\, M_{\odot}$.  This not only impacted impressions as to the
commonality of IMBHs at the centres of galaxies, but it additionally had
consequences for their measurement of Eddington ratios (Gallo et al.\ 2010).
 
In this paper, we provide updated predictions for the masses of the central
black holes in the 100 early-type galaxies belonging to the ACSVCS survey.
The structure of this paper is such that the galaxy sample is briefly
introduced in the following section, along with the galaxies' associated magnitudes
and velocity dispersions. In subsection~\ref{Sec_X}, we present new X-ray
detections, since Gallo et al.\ (2010), for six of the ETGs, including two
dwarf galaxies.  In subsections~(\ref{Sec_sig}) and (\ref{Sec_mag}), we detail
how the black hole masses are predicted here using the galaxies' velocity
dispersion and $B$-band absolute magnitude, respectively.
Section~\ref{Sec_IMBH} presents our predicted
black hole masses, and highlights select galaxies considered to have IMBHs.
In Section~(\ref{Sec_Edd}), we combine the updated black hole masses and X-ray
fluxes to compute new Eddington ratios.  We subsequently report on how the
average Eddington-scaled X-ray luminosity varies with black hole mass.
Section~(\ref{Sec_CM}) presents the $B-K$ colours for the galaxy sample,
enabling a prediction of what the $K$-band $M_{\rm bh}$--$L_{\rm K,galaxy}$
relation should look like, and also how the 
$M_{\rm bh}$--$M_{*,galaxy}$ relation involving the galaxy stellar mass should
look. 
Finally, we provide a discussion, with an emphasis on IMBHs, in
Section~(\ref{Sec_Disc}).

\section{Data}

\subsection{Magnitudes and velocity dispersions}\label{Sec_mags}

The AGN Multiwavelength Survey of Early-Type Galaxies in the Virgo Cluster
(AMUSE-Virgo: Gallo et al.\ 2008, 2010; Leipski et al.\ 2012) used (surface
brightness fluctuation)-based distances from Mei et al.\ (2007), central
stellar velocity dispersions\footnote{As desired, the overwhelming majority of
  the stars contributing to these stellar velocity dispersion measurements are
  (hopefully) largely unaffected by the black hole.}  collated by MacArthur et
al.\ (2008), and derived new (Galactic extinction)-corrected apparent $B$-band
galaxy magnitudes, for the 100 ACSVCS galaxies (C\^ot\'e et al.\ 2004).  
The $B$-band magnitudes were
based on the {\it HST/ACS} $g$ and $z$ band galaxy magnitudes published by
Ferrarese et al.\ (2006), who had also reported on the presence of any (light
profile)-flattened cores seen in these passbands.  Following Bonfini et
al.\ (2018), who report on dust-dimmed cores as opposed to those with a
central stellar deficit, we reassign NGC~4552's (VCC~1632's) designation from
a core-S\'ersic galaxy (Graham et al.\ 2013) 
to a S\'ersic galaxy given that its `core' is apparently due to dust.
Of the 8 Virgo cluster galaxies reported to have a flattened core by Ferrarese
et al.\ (2006), this galaxy had the faintest apparent $B$-band magnitude. 
 
\setcounter{figure}{0}
 
\begin{figure}
 \includegraphics[trim=1.6cm 2cm 7.3cm 4.5cm, width=.83\columnwidth]{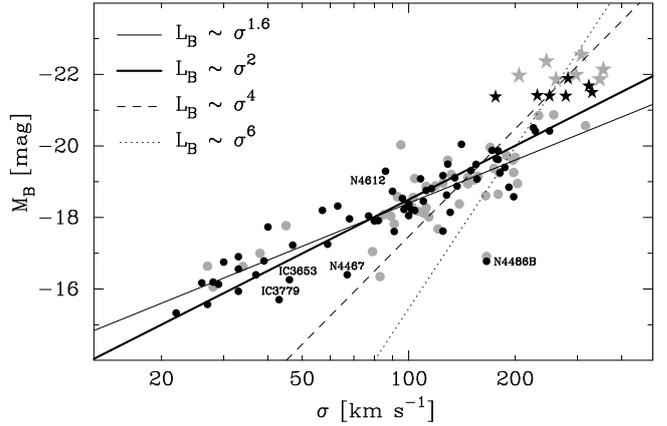}
    \caption{Galaxy $B$-band magnitude versus stellar velocity dispersion (54
      grey points from Gallo et al.\ 2008; 67 black points from Hyperleda 2018
      and the RC3).  Stars denote galaxies with partially depleted cores.
}
\label{Fig1} 
\end{figure}

The AMUSE-Virgo project used the velocity dispersions and $B$-band galaxy
magnitudes to provide two predictions for the masses of the centrally-located
black holes in these galaxies, which they subsequently combined with their
{\it Chandra} X-ray data.  The latter consisted of 5.4 ks (90 minute)
exposures with the Advanced CCD Imaging Spectrometer (ACIS) detector.  In
Figure~(\ref{Fig1}), we have plotted the absolute magnitudes against the
stellar velocity dispersions for the 54 galaxies considered to have `secure'
velocity dispersions by Gallo et al.\ (2008).  We additionally plot the
galaxies using the slightly updated velocity dispersions available in
Hyperleda\footnote{http://leda.univ-lyon1.fr} (Paturel et al.\ 2003) as of
January 2018, including new values for 13 more galaxies, and using the
$B$-band magnitudes (Vega) obtained from the {\it Third Reference Catalogue of
  Bright Galaxies} (RC3: de Vaucouleurs et al.\ 1991) as tabulated in the
NASA/IPAC Extragalactic Database
(NED)\footnote{http://nedwww.ipac.caltech.edu}.  
We corrected the RC3 magnitudes 
for Galactic extinction using the values from Schlafly \&
Finkbeiner (2011), as tabulated in NED, and converted into absolute magnitudes
using the distances from Mei et al.\ (2007).  For ease of reference, these
values are collated in the Appendix Tables.

It is apparent from Figure~(\ref{Fig1}) that the galaxy $B$-band
luminosity-(velocity dispersion) relation for ETGs is not described by a
single power-law (see also figure~3 in Chilingarian \& Mamon 2008).  While
Minkowski (1962) discovered a correlation between galaxy luminosity and
velocity dispersion, he refrained from fitting a slope until more data became
available at fainter magnitudes.  Nowadays, one can appreciate how one's
galaxy magnitude selection criteria can influence the slope observed in one's
data set.  Faber \& Jackson (1976) were the first to fit a relation to
Minkowski's correlation, famously reporting that $L_{\rm galaxy}\propto
\sigma^4$ for ETGs.  Studying luminous ETGs, Schechter (1980) and Malumuth \&
Kirshner (1981) reported that $L_{\rm galaxy}\propto \sigma^5$, with a more
recent study reporting that $L_{\rm galaxy}\propto \sigma^{6.5\pm1.3}$ (Lauer et
al.\ 2007).  Tonry (1981) found that the inclusion of fainter galaxies
resulted in a power-law such that $L_{\rm galaxy}\propto \sigma^3$, while
studies which excluded the luminous pressure-supported ETGs found that $L_{\rm
  galaxy}\propto \sigma^2$ (Davies et al.\ 1983; Held et al.\ 1992; de Rijcke
et al.\ 2005; Matkovi\'c \& Guzm\'an 2005; Kourkchi et al.\ 2012).  A fuller
review of the $L-\sigma$ relation 
can be found in Graham (2013), and the extension to dwarf spheroidal
galaxies with velocity dispersions from 3 to 12 km s$^{-1}$ can be seen in
Toloba et al.\ (2014, their figure~17).

The data in Figure~(\ref{Fig1}) reveals that the ETGs whose cores have not
been partially depleted of stars --- or perhaps it is those galaxies with
large-scale rotating disks --- roughly follow the relation $L_{\rm
  B,galaxy}\propto \sigma^2$.  One may wonder if $L_{\rm B,galaxy} \propto
\sigma^{1.6}$ provides a better description, however, the appearance of such a
shallow slope is likely due to sample selection effects such that we do not
have velocity dispersion measurements in galaxies fainter than $\approx -15.5$
mag ($B$-band).  Indeed, studies which have included fainter galaxies than us
do not observe such a shallow slope (e.g.\ Toloba et al.\ 2014).  The
existence of a continuous log-linear relation seen in Figure~(\ref{Fig1}) for
the S\'ersic galaxies, i.e.\ those without partially depleted cores, is seen
in almost all other diagrams involving the physical parameters of ETGs
(e.g.\ metallicity, colour, kinematics, globular clusters, etc.).  The only
exception is when one uses the effective `half-light' parameters, which
results in a continuous {\it curved} relation due to the systematically
changing S\'ersic index (see Graham 2013 for an explanation).  While there are
not sufficient numbers of galaxies in our dataset with partially depleted
cores, one can appreciate from Figure~(\ref{Fig1}) how a (galaxy
sample)-dependent exponent of 4, 5, 6 or greater was obtained from past
studies that included galaxies at the luminous-end of the $B$-band
luminosity-(velocity dispersion) diagram.

\begin{figure*}
$
\begin{array}{ccc}
 \includegraphics[angle=90, trim=0cm 3.0cm 0.cm 3.0cm, height=5cm]{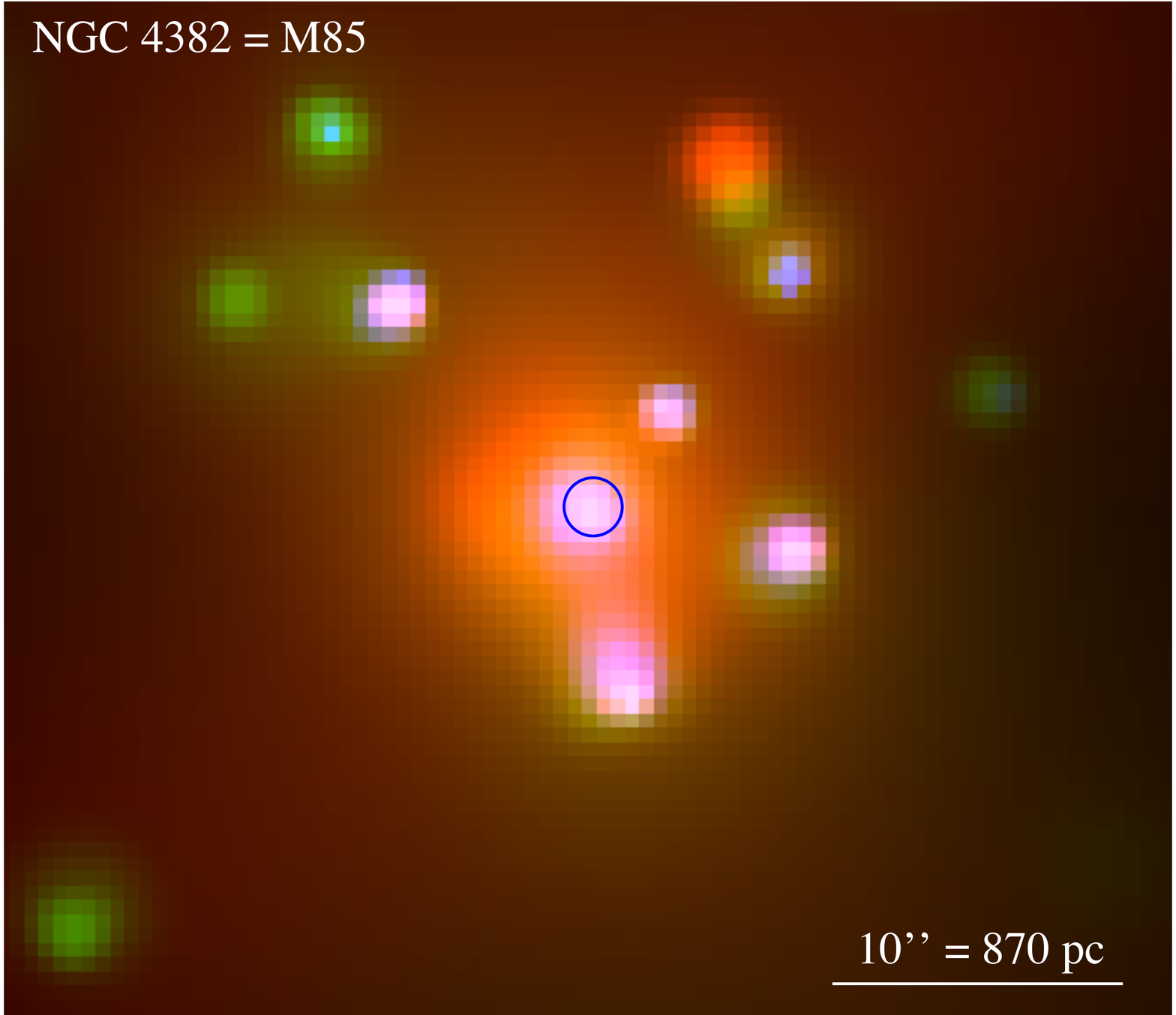} &
 \includegraphics[angle=90, trim=0cm 3.0cm 0.cm 3.0cm, height=5cm]{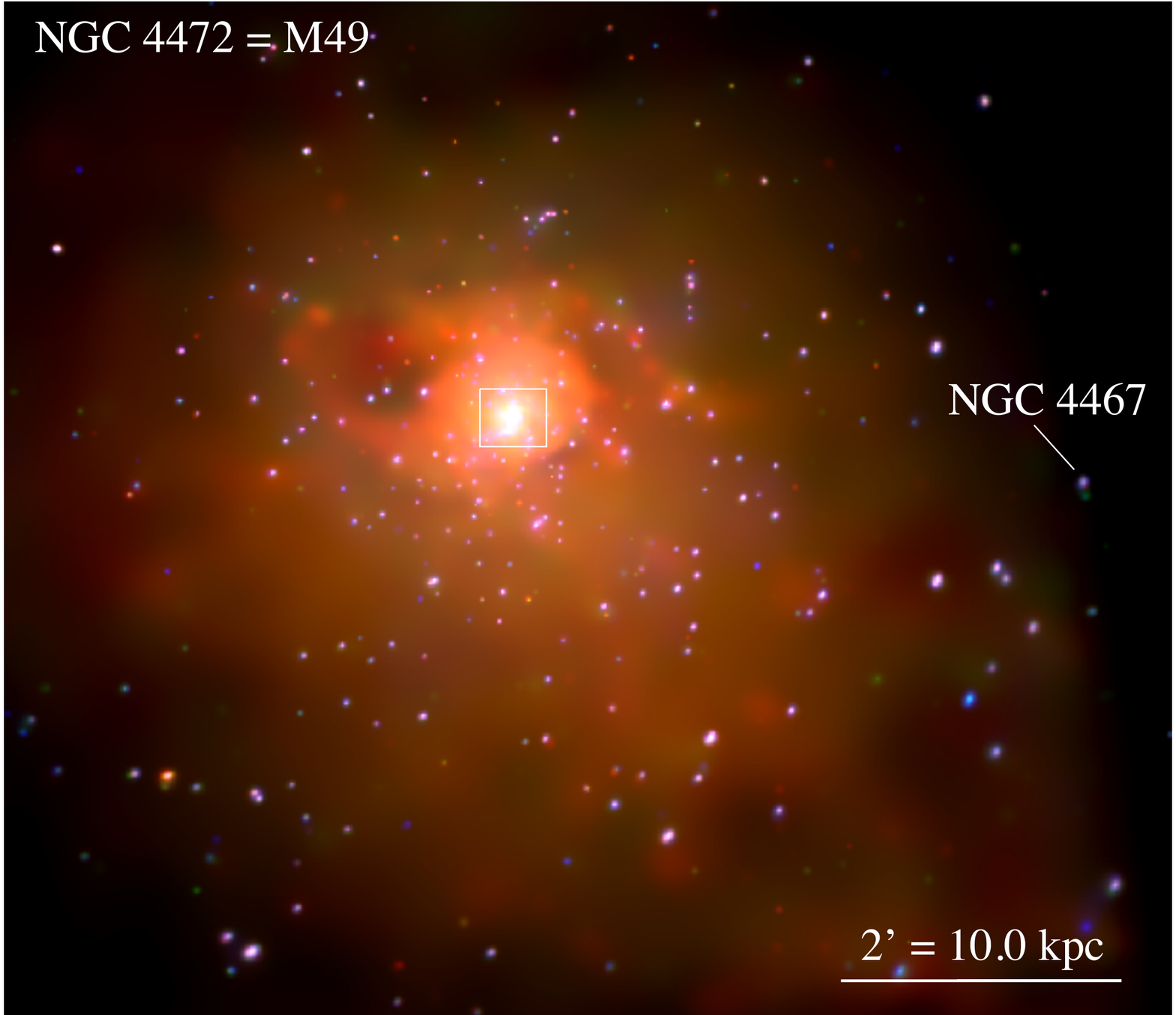} & 
 \includegraphics[angle=90, trim=0cm 2.5cm 0.cm 3.0cm, height=5cm]{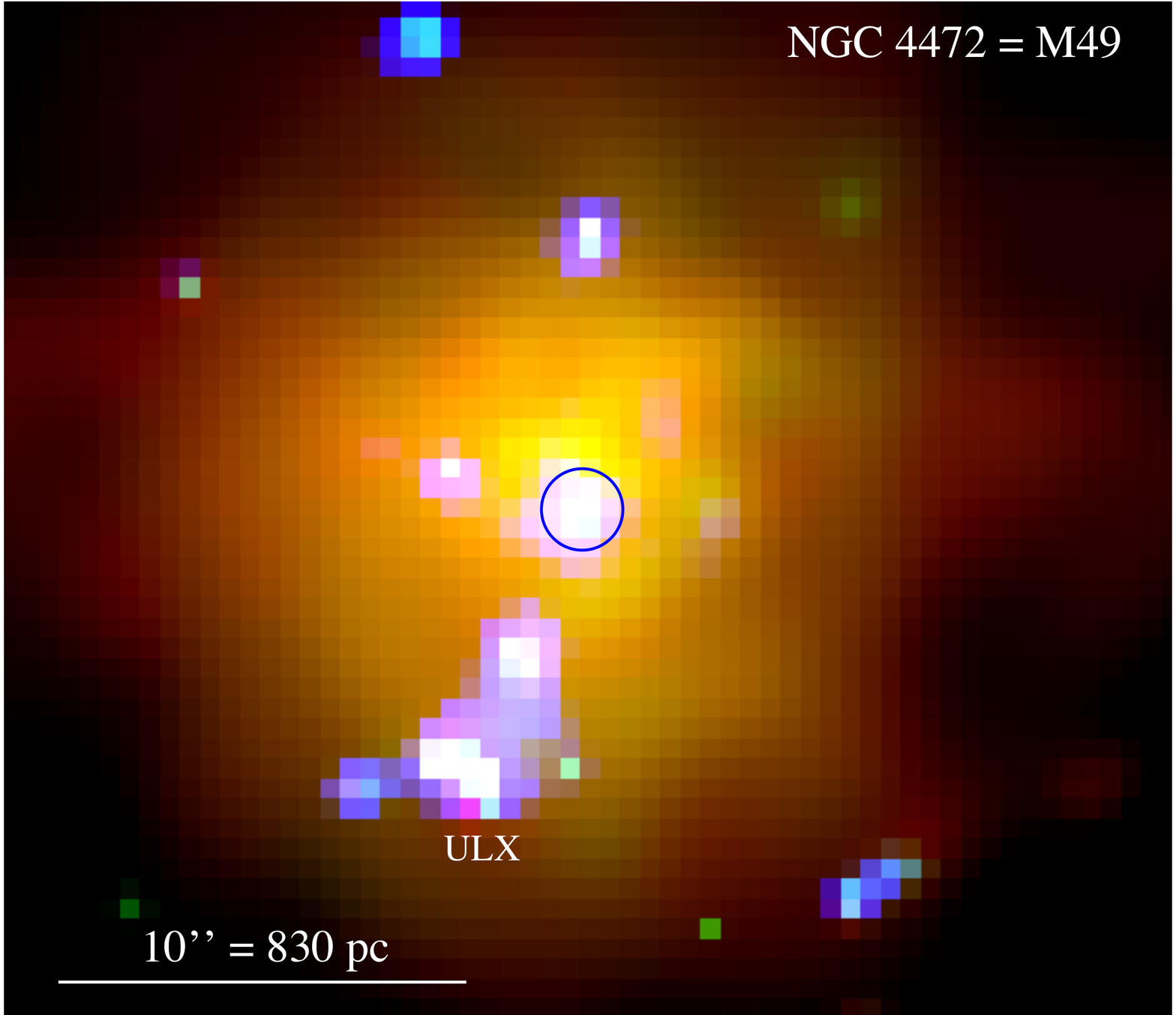} \\
\end{array}                                                         
$
    \caption{
Left panel: adaptively-smoothed {\it Chandra} image of the nuclear region of
NGC~4382 = M85; colours are: red = 0.3--1 keV; green = 1--2 keV; blue = 2--7
keV. Here and in all other panels, North is up and East to the left. The blue
circle represents the position of the optical/IR nucleus and coincides with a
point-like X-ray source, previously unreported. The hard nuclear source is
surrounded by softer diffuse emission (hot gas).
Middle panel: adaptively-smoothed {\it Chandra} image of NGC~4472 = M49, in
the same colour bands. The anisotropic distribution of the diffuse hot gas is
evidence of recent interactions. The central region inside the white box is
shown in more details in the right panel.  About 4$\arcmin$ to the east of the
nucleus of NGC~4472, an X-ray source is located at the nuclear position of NGC
4467. 
Right panel: zoomed-in view of the nuclear region of NGC~4472; the soft
(0.3--1 keV) and medium (1--2 keV) band images have been adaptively smoothed,
while the hard band (2--7 keV) has been Gaussian-smoothed (1$''$ core), to
better differentiate the harder point sources from the softer diffuse
emission. The blue circle represents the position of the optical/IR nucleus 
and coincides with a point-like X-ray source, previously unreported. The ULX
located south of the nucleus was reported in Plotkin et al.~(2014).
}
\label{Fig2b}
\end{figure*}

\begin{figure*}
\begin{center}
$
\begin{array}{cc}
\includegraphics[angle=90, trim=0cm 1.5cm 0cm 3.8cm, width=0.35\textwidth]{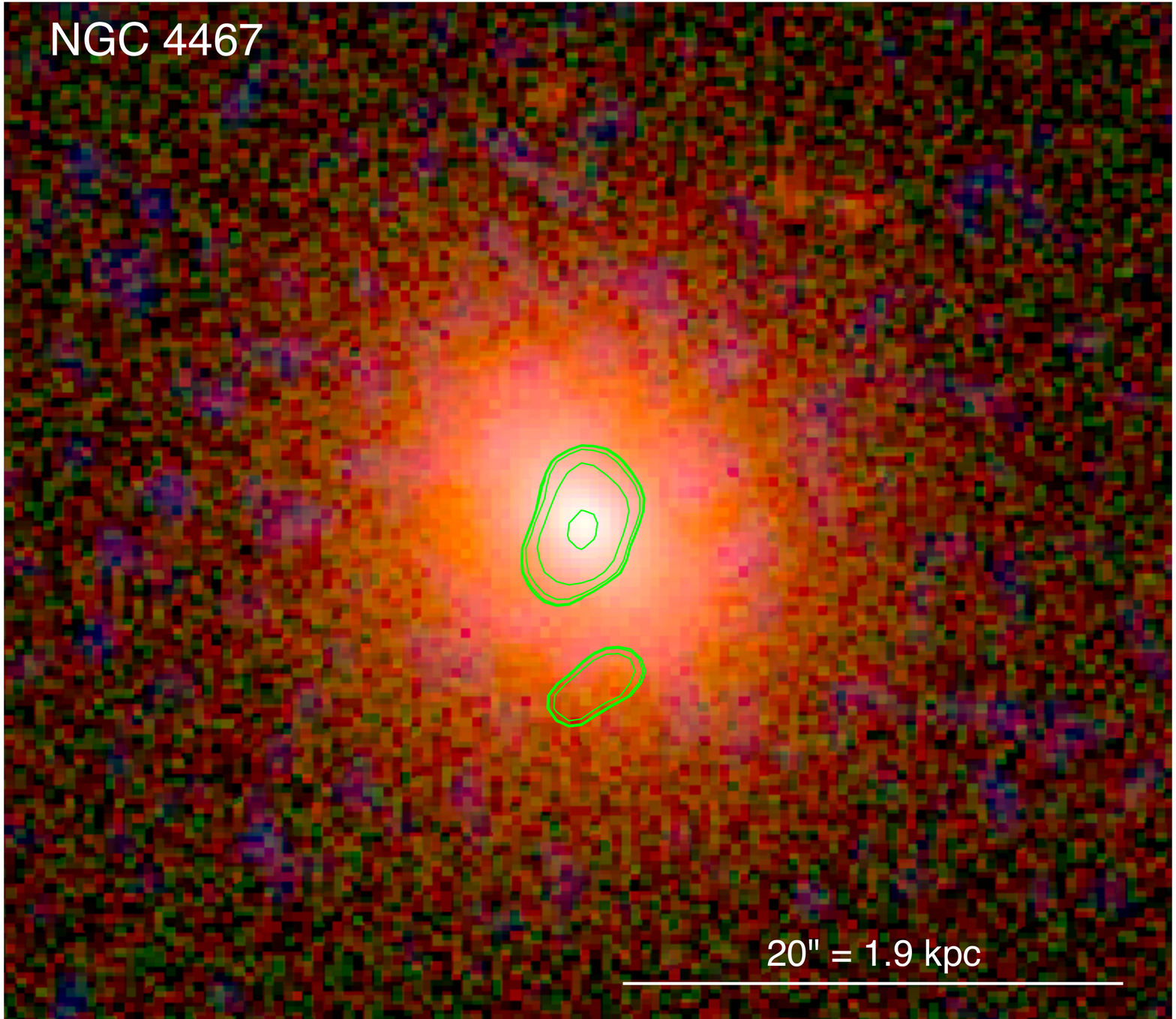} & 
\includegraphics[angle=90, trim=0cm 0.0cm 0cm 5.3cm, width=0.35\textwidth]{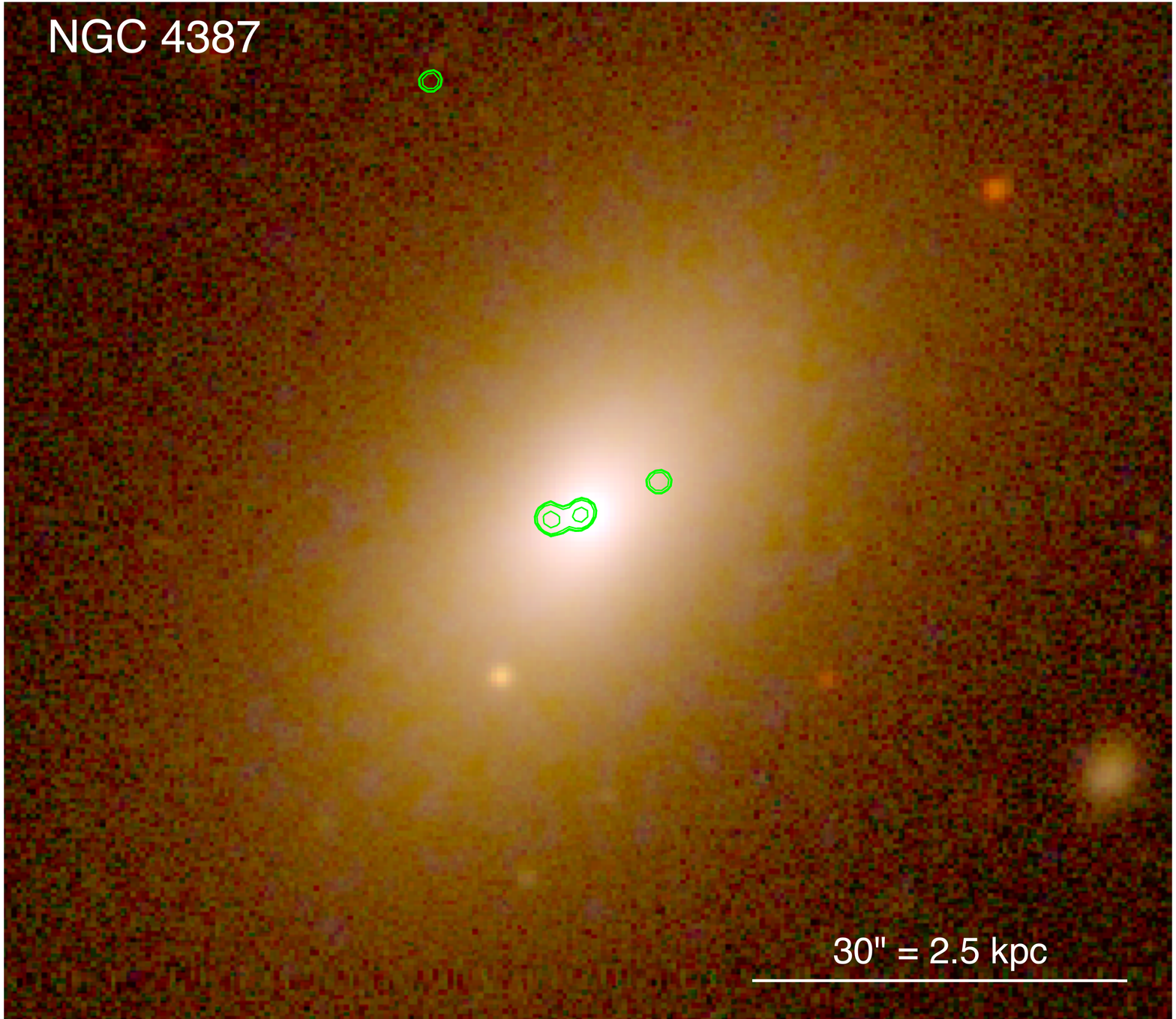} \\
\includegraphics[angle=90, trim=0cm 1.5cm 0cm 3.8cm, width=0.35\textwidth]{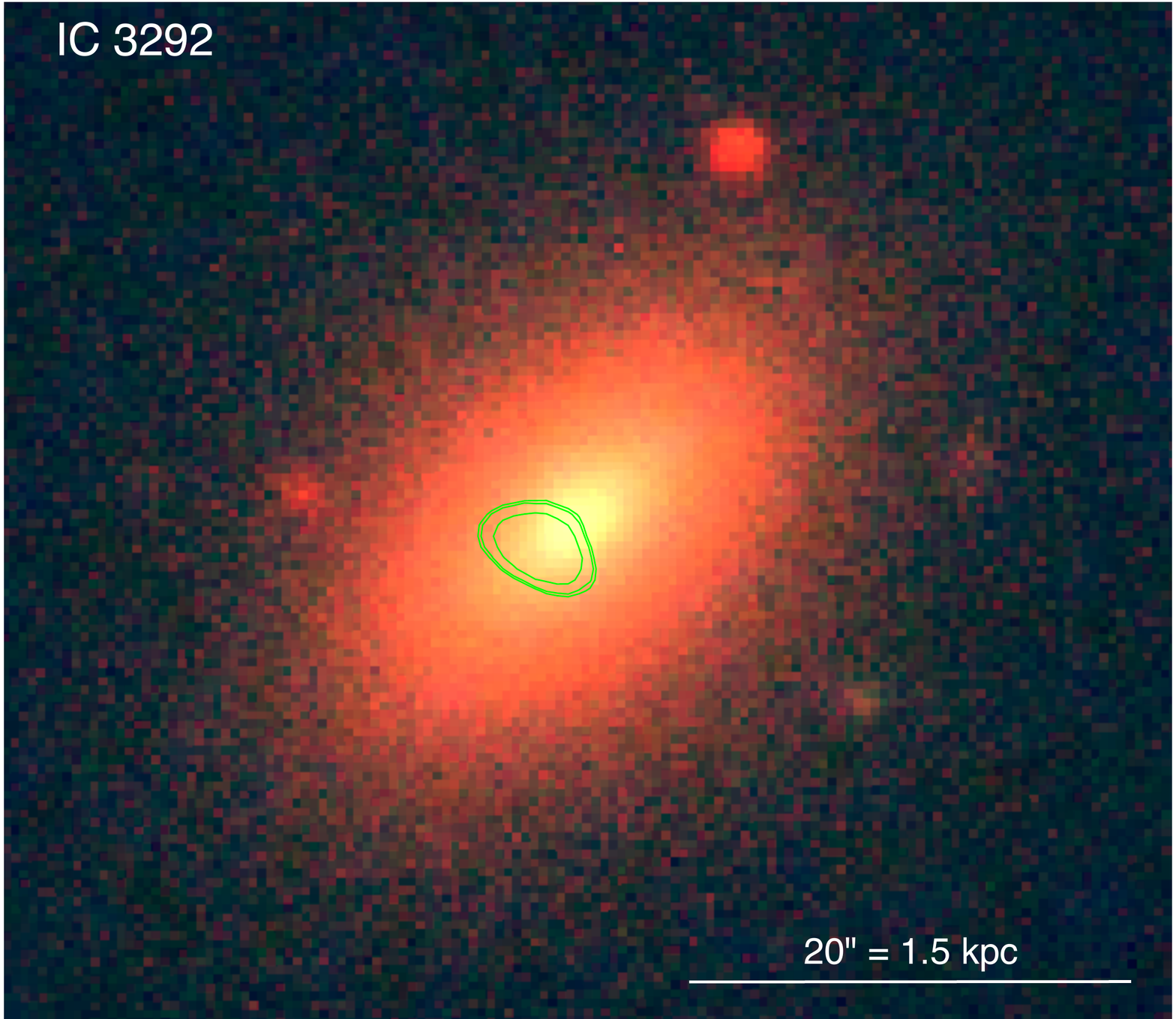} &
\includegraphics[angle=90, trim=0cm 0.0cm 0cm 5.3cm, width=0.35\textwidth]{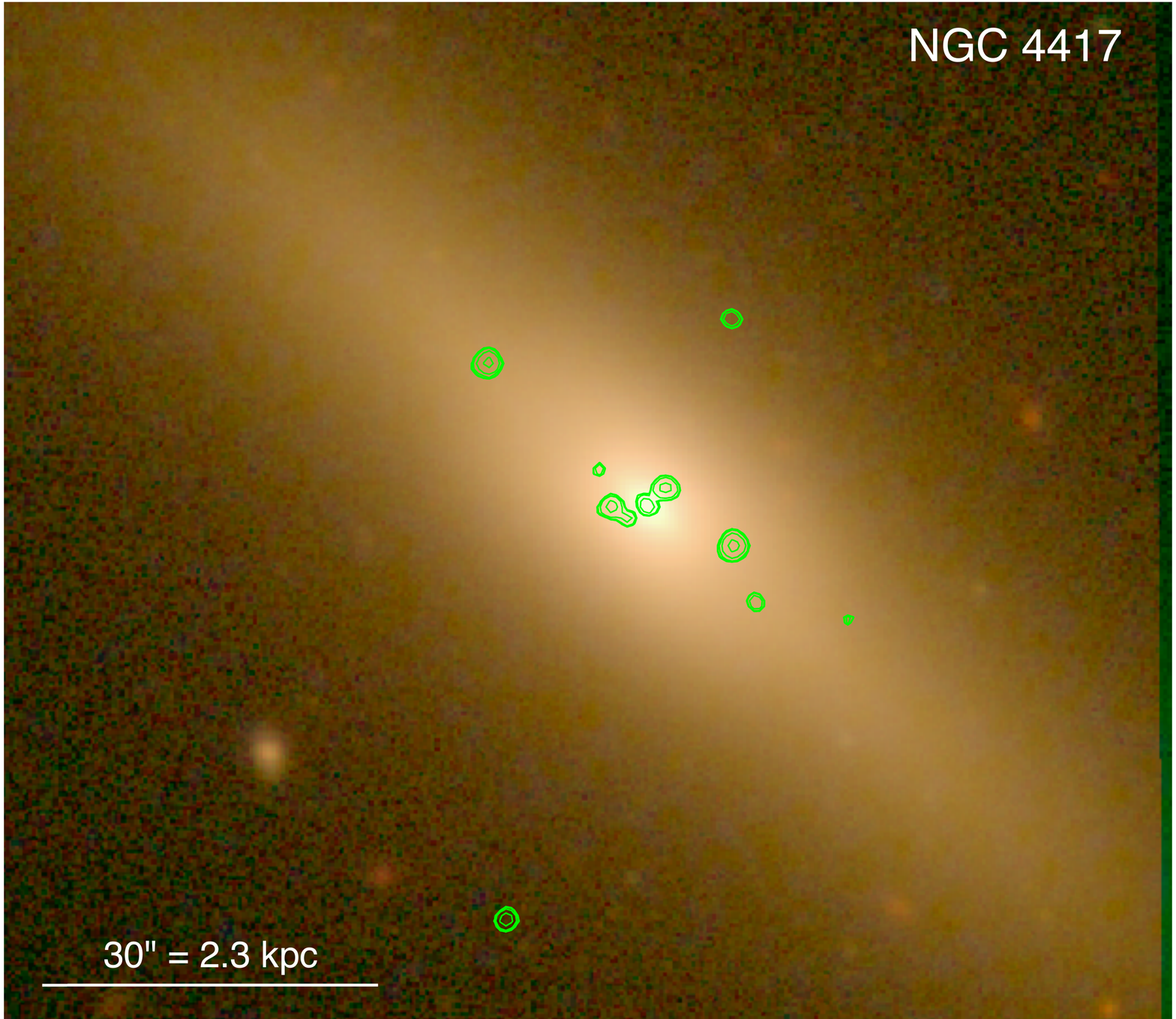} \\
\end{array}
$
\end{center}
\caption{
SDSS images with over-plotted {\it Chandra} contours (in green) for four
galaxies in the AMUSE sample for which X-ray emission in the nuclear region
was previously unreported (see Section 2.2). Colours are: red = $i^{\prime}$-band;
green = $g^{\prime}$-band; blue = $u^{\prime}$-band. North is up and East to
the left. 
In NGC~4467, NGC~4387, and NGC~4417, there is a point-like source located within
0$\arcsec$.5 of the optical nucleus. In IC~3292, the X-ray emission is offset from
the optical nucleus by $\approx$2$\arcsec$.
}
   \label{Fig2a}
\end{figure*}

\subsection{X-ray checkups}\label{Sec_X}

Gallo et al.\ (2010) reported centrally-located X-ray fluxes for 32 of the 100
galaxies.  We have re-examined the X-ray properties of the nuclear regions in
all those 100 galaxies, because some of those galaxies were covered by
additional {\it Chandra} observations after 2010, and more generally because
newer optical surveys such as the Sloan Digital Sky Survey (SDSS) Data Release
12 (Alam et al.\ 2015) and the {\it Gaia} Data Release 2 (Gaia Collaboration
et al.\ 2018) give us a chance to improve the astrometric association of the
X-ray and optical frames.

We reprocessed and analyzed the public archival data (pre- and post-2010) with
the Chandra Interactive Analysis of Observations ({\sc ciao}) software version
4.10 (Fruscione et al.\ 2006).  Specifically, we reprocessed the event files
with the {\it chandra\_repro} script.  We then reprojected and combined the
datasets with the task {\it reproject\_obs}, and created exposure-corrected
images in various energy bands with the task {\it flux\_obs}.  For galaxies
with multiple {\it Chandra} observations, we used the {\sc ciao} task {\it
  merge\_obs} to combine newer and older observations, in order to deepen the
exposures and improve the statistics on the fainter X-ray sources. We
identified the location of the point sources with the source-finding tool {\it
  wavdetect}. In some cases, point-like sources in the nuclear region are
embedded in softer, diffuse emission and do not stand out in the full 0.3--8
keV band; in those cases, we also ran {\it wavdetect} in the 2--7 keV band
alone, which resolves accreting point-like sources more effectively.  We used
the SAOImage {\sc DS9} visualization package (Joye \& Mandel 2003) to display
the X-ray and optical images, check their alignment, perform aperture
photometry on the X-ray images, and select the source and background regions
that we needed for spectral extraction, for sources with a sufficient number
of counts. We chose suitable sizes for the source extraction regions (between
1$''$.5 and 4$''$ in different cases) based on whether a source was observed
near the aimpoint of the ACIS detector or closer to the chip edge, and also in
a way that minimized contamination from nearby sources. Local background
regions were chosen as annuli around the source (whenever possible) or, in any
case, from the surrounding field at distances of $\approx$5$''$--10$''$, with
an area of about four times the area of the source extraction region. Spectra
and associated response files were extracted from individual exposures and
then stacked, with the {\sc ciao} task {\it specextract}.

For sources for which we extracted a spectrum, we used software from NASA's
High Energy Astrophysics Science Archive Center for spectral analysis. In
particular, we used the task {\it grppha} to group the spectra to 1 count per
bin for fitting based on maximum-likelihood statistics (Cash 1979) and to
$>15$ counts per bin for $\chi^2$ fitting. We then fitted the spectra with
{\sc xspec} Version 12.9.1 (Arnaud 1996).

We have found a point-like {\it Chandra} source near the nuclei of six
additional galaxies from the subsample of 68 galaxies for which Gallo et
al.\ reported only an upper limit to the nuclear X-ray luminosity.
We do, however, note that IC~3292 (VCC~751) might only house an X-ray binary near its nucleus
given the 2$\arcsec$ offset, although it has been speculated that our 
galaxy may harbour an offset IMBH (Oka et al.\ 2016, 2017; Tsuboi et
al.\ 2017; Ballone et al.\ 2018, but see Ravi et al.\ 2018).  
This increase of ($6-1=$) 5
represents a 16 per cent increase, from 32 to 37 galaxies, of 
the lower-limit to the percentage of galaxies that are likely hosting a
central black hole,
sometimes referred to as the `occupation fraction' (Zhang et al.\ 2009; Miller
et al.\ 2015).
The five new galaxies are:
\begin{itemize}
\item NGC~4382 = VCC~798 = M85, an S0,pec and suspected wet merger remnant (Ko
  et al.\ 2018);
\item NGC~4472 = VCC~1226 = M49, an interacting E2 galaxy (Arrigoni Battaia et
  al.\ 2012; $L_X=39.0$ erg s$^{-1}$)\footnote{The nuclear X-ray source is not
    the slightly off-nuclear X-ray source listed in Plotkin et al.\ 2014,
    their table~3, and located at $0.02\,R_{2\ 5}$.};
\item NGC~4467 = VCC~1192 (CXOJ122930.206+075934.48, $L_x=$(2.72-4.71)e+38,
  Liu 2011)\footnote{The X-ray emission only {\it appears} extended because it
    was observed off-axis.};
\item NGC~4387 = VCC~828 , ($L_x<$1.29e+39, Liu 2011);
and
\item NGC~4417 = VCC~944.
\end{itemize}
Our new detections arise from a combination of new observations (post
2010) plus a re-analysis of past data.  We present them in
Figures~(\ref{Fig2b}--\ref{Fig2a}) and provide further details below.
As we shall see, VCC~1192 is particularly interesting because it is a dwarf
early-type galaxy.

\subsubsection{NGC~4382}\label{Sec_4382}

The only {\it Chandra} data available for the Gallo et al.\ (2010) paper was a
40-ks ACIS-S observation from observing Cycle 2 (ObsID 2016). In recent years,
NGC~4382 has been observed once with ACIS-I in Cycle 16 (ObsID 16968) and four
times with ACIS-S in Cycle 18 (ObsIDs 19331, 20012, 20013, 20014).
The total exposure time in the stacked dataset (including both ACIS-S and
ACIS-I) at the nuclear position is $\approx$170 ks.

There are several point-like X-ray sources as well as soft diffuse emission in
the nuclear region. To determine whether one of the X-ray sources corresponds
to the nuclear BH, first, we need to discuss the poorly determined location of
the optical nucleus for this galaxy. The default coordinates adopted by the
NASA/IPAC Extragalactic Database (NED\footnote{https://ned.ipac.caltech.edu})
are very much an outlier (offset by about 2$\arcsec$) with respect to various
other recent determinations of the optical/IR nucleus; the default position
adopted by the Hyperleda database is also an outlier, in the opposite
direction (displacement of $\approx$5$\arcsec$ between the two positions). The
nuclear position fitted to the SDSS images (Ann et al.\ 2015) and that derived
from the {\it Two Micron All Sky Survey}
(2MASS\footnote{www.ipac.caltech.edu/2mass}, Jarrett et al.\ 2000) Redshift
Survey of Huchra et al.\ (2012) is closer to the median value of the most
recent estimates, are consistent with each other, and are essentially the same
value already adopted by Gallo et al.\ (2010).  Then, we improved the relative
astrometric position of the stacked X-ray image by selecting six relatively
strong, point-like X-ray sources with an unambiguous point-like counterpart in
the SDSS DR12 images. We noticed a systematic offset of $\approx$0$''$.2
between the two frames.  After correcting the astrometry of the stacked {\it
  Chandra} image for this offset, we found that the brightest X-ray point
source in the nuclear region (Figure~\ref{Fig2b}, left panel) does indeed
coincide with the optical position listed by the SDSS and the {\it Gaia}
catalogues within 0$''$.1, and is $\approx$0$''$.3 offset from the nuclear
position listed in the 2MASS catalog. This source is located roughly at the
peak of the soft, diffuse emission, but it has distinctly harder colours (as
we expect for an accreting nuclear black hole). Thus, we take this point
source as the most likely nuclear candidate.

We fitted the spectrum of the candidate nuclear source with a power-law model,
absorbed by cold material. We find (Table B1, and left panel of Figure B1)
that the column density of the absorber is consistent with the Galactic
line-of-sight value ($N_{\rm H} = 2.5 \times 10^{20}$ cm$^{-2}$); the
best-fitting photon index is $\Gamma = 1.4 \pm 0.2$.  We obtain a 0.3--10 keV
intrinsic luminosity $L_{\rm X} = \left(8.0^{+1.8}_{-1.4}\right) \times
10^{38}$ erg s$^{-1}$.  We also re-analyzed the original 40-ks observation
used by Gallo et al.\ (2010) and found that the nuclear emission is consistent
with this average luminosity also in that epoch, although the signal-to-noise
ratio from that observation alone is too poor to give a significant {\it
  wavdetect} detection in the full band, because the point-like source is
swamped by the diffuse soft emission. (A significant detection is obtained
instead when the image is filtered to the 2--7 keV band).  In addition to the
positional coincidence with the optical nucleus, the hard power-law spectrum
is a strong clue that the source is the nuclear black hole; stellar-mass X-ray
binaries with similar luminosities tend to have softer spectra (Remillard \&
McClintock 2006).

\subsubsection{NGC~4472}

The non-detection of a nuclear source by Gallo et al.~(2010) was based on a
33-ks observation from Cycle~1 (ObsID 321). Since then, NGC~4472 has been
observed with ACIS-S another nine times (ObsIDs 8095, 8107, 11274, 12888,
12889, 12978, 16260, 16261 and 16262) and once with ACIS-I (ObsID 322). We
downloaded, reprocessed, aligned, and stacked all of these datasets with the
{\sc ciao} tools, following the procedure described early in Section 2.2.  The
total exposure time in the nuclear region is $\approx$470 ks (most of it from
the two long observations in Cycle~12).

For this galaxy, the location of the optical/IR nucleus is accurately known 
and in agreement between NED and Hyperleda. We selected ten reasonably strong
X-ray sources with point-like counterparts in SDSS DR12 images. We determined
that there is no systematic offset (to better than 0$\arcsec$.1) between {\it
  Chandra} and SDSS astrometry; however, there is a random scatter of
$\approx$0$\arcsec$.2 between the fitted positions of corresponding sources in
the two bands. There is a point-like {\it Chandra} source (Figure~\ref{Fig2b})
located $\approx$0$\arcsec$.1 from both the SDSS DR12 and {\it Gaia} DR2
optical nuclear positions, and $\lesssim$0$\arcsec$.5 from the 2MASS position.
It is also coincident with the peak of
the diffuse emission, but stands out because of its harder colours.

There are other luminous point-like X-ray sources in the field: one of them,
located $\approx$7$\arcsec$ south-east of the nucleus, has an X-ray luminosity
of $\approx10^{39}$ erg s$^{-1}$ and was listed as a ULX candidate in Plotkin
et al.\ (2014). The nucleus is only slightly fainter. We extracted a stacked
spectrum and fitted it (Table B1 and middle panel of Figure B1) with a
power-law model plus thermal plasma emission (to account for a significant
soft excess below 1 keV). We obtain a photon index $\Gamma =
2.4^{+0.4}_{-0.3}$ and a temperature of the hot gas $kT = (0.6 \pm 0.1)$ keV
(Table A3).  The de-absorbed luminosity is $L_{\rm X} =
\left(9.2^{+1.0}_{-1.6}\right) \times 10^{38}$ erg s$^{-1}$; of this, the
contribution from the power-law component is $L_{\rm X} \approx 7 \times
10^{38}$ erg s$^{-1}$.
                                                                                                                    
We suggest that the thermal component comes from a stronger concentration of
hot gas in the inner arcsec, which is under-subtracted from the local
background annuli.  The slope of the power-law component is marginally steeper
than in typical low-luminosity AGN ($1.6 \lesssim \Gamma \lesssim 2$:
Terashima \& Wilson 2003), but it may be affected by contamination from the
thermal component; we do not have enough counts to attempt a meaningful fit
with a multi-temperature thermal plasma.  Our identification of this source as
the nucleus rather than an X-ray binary in the high/soft state is based mostly
on its perfect positional coincidence with the optical nucleus.

\subsubsection{NGC~4467}

The early-type galaxy NGC~4467 is located $\approx$4$\arcmin$.1 west of
NGC~4472: thus, it was serendipitously observed near the edge of the ACIS
chips in most of the observations centred on the latter galaxy. From our
stacked image and exposure map of the field (Section 2.2.2), we infer an
effective exposure time of $\approx$310 ks at the location of NGC~4467.  We
refer to the discussion in Sect 2.2.2 for the relative alignment of X-ray and
optical frames because we used the same datasets for the two galaxies.  There
is a moderately bright X-ray source coincident with the optical position of
this galaxy (Figures~\ref{Fig2b} and \ref{Fig2a}). More exactly, the
best-fitting centroid of the X-ray source is displaced by
$\approx$0$\arcsec$.5 from the SDSS, {\it Gaia} and 2MASS positions; however,
the X-ray centroid itself has an uncertainty of $\approx$0$\arcsec$.5 because
of the non-circular, large point spread function at such distances from the
telescope's central aim-point.  Thus, we can safely consider this source as
the candidate X-ray nucleus of NGC~4467. Following a similar procedure as
described before, we extracted a stacked spectrum.  We fitted it with a
power-law model (additional components do not improve the fit); the photon
index is $\Gamma = 1.4 \pm 0.2$ (Table B1 and right panel of Figure B1),
consistent with the expected spectrum of a nuclear black hole. The unabsorbed
0.3--10 keV luminosity is $L_{\rm X} = (6 \pm 1) \times 10^{38}$ erg s$^{-1}$.

\subsubsection{NGC~4387}

The nondetection in Gallo et al.~(2010) was based on a 5-ks ACIS-S
observation from Cycle 8 (ObsID 8056). More recently, NGC~4387 was observed
with ACIS-S in Cycle 15 (ObsID 16031), for 33.6 ks. We reprocessed and stacked
the two observations following the same procedure as described earlier.  To
improve the astrometry of the {\it Chandra} image, we matched the position of
six X-ray/optical point-like sources and corrected for a small systematic
offset of $\approx$0$\arcsec$.2 between {\it Chandra} and SDSS. We then found
two faint point-like X-ray sources in the nuclear region: one coincides with
the optical nucleus (within 0$\arcsec$.3 of the nuclear position listed in
SDSS-DR12, {\it Gaia} and 2MASS), the other is located $\approx$2$\arcsec$.5
($\approx$200 pc) to the east (Figure~\ref{Fig2a}). Both sources have
$\approx$15 net counts; however, the source to the east of the nuclear
position has harder X-ray colours, probably because of higher
absorption. Neither source has enough counts for spectral analysis; thus, we
used the Portable, Interactive Multi-Mission Simulator ({\sc
  pimms}\footnote{http://cxc.harvard.edu/toolkit/pimms.jsp}) conversion tool
(part of the {\it Chandra} Proposal Planning Toolkit) to derive a de-absorbed
flux from the observed count rate. For consistency with the luminosities of
the other nuclear sources listed in Gallo et al.\ (2010), we adopted the same
spectral model: a power-law with photon index $\Gamma = 2$, absorbed by cold
material with a column density of $2.5 \times 10^{20}$ cm$^{-2}$.  The
conversion factor depends on the observation cycle; we obtained a weighted-average
conversion factor, using the exposure times of the individual observations as
weights.  For the nuclear source, we estimate a de-absorbed 0.3--10 keV
luminosity $L_{\rm X} = (2 \pm 1) \times 10^{38}$ erg s$^{-1}$. For the
eastern source, we estimate $L_{\rm X} = (4 \pm 1) \times 10^{38}$ erg
s$^{-1}$.  Our identification of the candidate nuclear X-ray source is based
on the low probability of finding an unrelated X-ray binary by chance within
0$\arcsec$.3 of the optical nucleus, in a galaxy with only 4 point-like X-ray
sources inside its D25 ellipse (area of $\approx$4400 $\arcsec^2$).

\subsubsection{NGC~4417}

Gallo et al.\ (2010) detected no nuclear X-ray source in a 5-ks
observation from Cycle 8 (ObsID 8125). A new ACIS-S observation of NGC~4417 in
Cycle 14 (ObsID 14902) added another 30 ks. After reprocessing and stacking
the two datasets, we found several (faint) point-like X-ray sources in the
inner region of this galaxy (Figure~\ref{Fig2a}). To improve the {\it Chandra}
astrometric solution, we matched 6 X-ray sources (located outside the D25)
with their optical (SDSS) counterparts, and corrected a small systematic
offset of 0$\arcsec$.2 in the {\it Chandra} frame.  We then found that one
X-ray source coincides with the optical nucleus (within 0$\arcsec$.3 of the
nuclear position from SDSS-DR12, {\it Gaia} and 2MASS); three others are
within 3$\arcsec$ of the nuclear position and are likely X-ray binaries.
From its observed count rate ($\approx$4 $\times 10^{-4}$ ct s$^{-1}$ at
0.3--7 keV), using {\sc pimms} as described in Section 2.2.4, we estimate a
0.3--10 keV unabsorbed luminosity $L_{\rm X} = (1.0 \pm 0.3) \times 10^{38}$
erg s$^{-1}$ for the nuclear source. The two brightest non-nuclear sources in
this galaxy have $L_{\rm X} \approx 5 \times 10^{38}$ erg s$^{-1}$.

\subsubsection{IC 3292}\label{Sec_3292}

Two faint X-ray sources ($\approx$5 net counts each, corresponding to $L_{\rm
  X} \approx (2 \pm 1) \times 10^{38}$ erg s$^{-1}$) are detected near the
centre of the the 5-ks snapshot image (from 2007) available to Gallo et
al.\ (2010). The two sources are centred at $\approx$2$\arcsec$ and 3$\arcsec$
($\approx$ 150 and 230 pc) south-east of the optical nucleus (based on SDSS
and {\it Gaia} positions).  We used three X-ray/optical coincidences located
further out (outside the D25) to verify the {\it Chandra} astrometry, and
found that it coincides with the SDSS astrometry within 0$\arcsec$.2. Thus, an
offset of 2$\arcsec$ is too large to be explained by astrometric error.

Located 8$\arcmin$.4 west of NGC~4382, IC~3292 was additionally detected at
the edge of the ACIS-I chips in some of the observations aimed at NGC~4382
(Section 2.2.1).  This yields another 55 ks of exposure time in the stacked
dataset, but several arcmin off-axis, such that we cannot resolve the two
candidate point-like sources in this image.  However, from these 2015--2017
observations, using {\sc pimms}, we estimate a combined average unabsorbed
luminosity $L_{\rm X} = (2.0 \pm 0.5) \times 10^{38}$ erg s$^{-1}$, offset
slightly to the south-east of the nucleus (Figure~\ref{Fig2a}). The location
is consistent with that of the two resolved sources in the 2007 observation;
however, their combined luminosity appears to be a factor of 2 fainter.  Due
to the spatial offset, we speculate that this emission is due to X-ray
binaries rather than a massive black hole, unless the massive black hole is
displaced from this dwarf galaxy's optical centre --- a scenario that we do
not rule out.

\section{Predicting black hole masses}\label{sec_Param}

Of the 100 ETGs in the ACSVCS / AMUSE-Virgo sample, 11 have directly measured
black hole masses.  This count excludes NGC~4382 (G\"ultekin et al.\ 2011).  
The 11 galaxies are listed in Table~(1), and we do not
need/use predictions for the black hole masses in these 11 galaxies (which
were used to define the black hole scaling relations).

\begin{table}
\centering
\caption{Virgo ETGs with directly measured black hole masses.}
\begin{tabular}{lcl}
\hline\hline
Galaxy Id. &  $M_{\rm bh}$  &  Ref. \\
\hline
NGC~4374  (VCC~763) &  $9.0^{+0.9}_{-0.8}\times10^8 \,M_{\odot}$  & 1 \\
NGC~4434  (VCC~1025) & $7.0^{+2.0}_{-2.8}\times10^7 \,M_{\odot}$ & 2 \\
NGC~4473  (VCC~1231) & $1.2^{+0.4}_{-0.9}\times10^8 \,M_{\odot}$ & 3 \\ 
NGC~4486  VCC~1316) & $58^{+3.5}_{-3.5}\times10^8 \,M_{\odot}$ & 4 \\  
NGC~4486A (VCC~1327) & $1.3^{+0.8}_{-0.8}\times10^7 \,M_{\odot}$ & 5 \\
NGC~4486B (VCC~1297) & $6.0^{+3.0}_{-2.0}\times10^8 \,M_{\odot}$ & 6 \\
NGC~4552  (VCC~1632)& $4.7^{+0.5}_{-0.5}\times10^8 \,M_{\odot}$ &  7 \\
NGC~4578  (VCC~1720) & $1.9^{+0.6}_{-1.4}\times10^7 \,M_{\odot}$ & 2 \\
NGC~4621  (VCC~1903) & $3.9^{+0.4}_{-0.4}\times10^8 \,M_{\odot}$ & 7 \\ 
NGC~4649  (VCC~1978) & $47^{+10}_{-10}\times10^8 \,M_{\odot}$ & 8 \\
NGC~4762  (VCC~2095) & $2.3^{+0.9}_{-0.6}\times10^7 \, M_{\odot}$ & 2 \\
\hline
\end{tabular}

\label{Tab_Direct}
References: (1) Walsh et al.\ (2010); 
(2) Krajnovi\'c et al.\ (2018, who report 3$\sigma$
  uncertainties because there was often not a single well-defined minimum
  within the 1$\sigma$ levels of their minimisation routine); 
(3) G\"ultekin et al.\ (2009); 
(4) Gebhardt et al.\ (2011); 
(5) Nowak et al.\ 2007; 
(6) Kormendy et al.\ (1997); 
(7) Cappellari et al.\ (2008, a preliminary value determined by Hu
    2008 from Conf.\ Proc.\ figures of Cappellari et al.\ 2008); 
(8) Shen \& Gebhardt 2009. 
\end{table}

Our objective here is to predict the black hole masses for the remaining 89
galaxies while reconciling the three orders of magnitude differences reported
previously in the literature.

\subsection{Velocity Dispersion}\label{Sec_sig}

As noted in Section~\ref{Sec_mags}, homogenised velocity dispersions are
available in Hyperleda for 67 of the 100 Virgo galaxies.  These are
error-weighted values based upon re-normalised measurements from the
literature, building upon the principles of McElroy (1995).  We have assigned
a 10 per cent uncertainty to each of these values, which can be seen in the
Appendix Table.  For those interested in issues pertaining to the measurement
of stellar velocity dispersions, section 4.2.2 of Graham et al.\ (2011)
mentions a few concerns, such as gradients in the luminosity-weighted aperture
velocity dispersion profiles of bright ETGs, biases from rotating disks or
young stars, sigma drops from cold inner discs or bars, slit orientation, etc.
While our bright ETGs have velocity dispersion measurements based on ten or
more entries in Hyperleda, typically from long-slit spectra, the fainter
galaxies may have only one or two entries.  Given that any single study can be
prone to systematic biases, we acknowledge that there is room for more work to
be done here, but that is beyond the scope of the current investigation.

To predict the black hole masses from the velocity dispersions, Gallo et
al.\ (2008) used the following (symmetrical, bisector) linear regression,
which was determined by 
Ferrarese \& Ford (2005) using a sample of 25 (21 early-type + 4 late-type) galaxies. 
\begin{equation}
\log (M_{\rm bh}/M_{\odot}) = (8.22\pm0.24) + 
(4.86\pm0.43)\log \left( \sigma/ 200\, {\rm km\, s}^{-1} \right).
\label{eq1a}
\end{equation}
In this work, we use the latest relation, from Sahu et al.\ (2018, in
preparation), obtained from a bisector linear 
regression of $\sim$50 early-type galaxies with directly measured black hole masses
and without partially depleted cores. It is such that 
\begin{equation}
\log (M_{\rm bh}/M_{\odot}) = (8.16\pm0.06) + (4.76\pm0.46)\log \left( \sigma
/ 185\, {\rm km\, s}^{-1} \right). 
\label{eq1b}
\end{equation}
The root mean square (rms) scatter in the $\log M_{\rm bh}$ direction 
$\Delta_{\rm rms}=0.42$ dex, and here we adopt an intrinsic scatter (in the
$\log M_{\rm bh}$ direction) of 0.3 dex
for estimating the uncertainties on our predicted black hole masses (see
equation~4 in Graham et al.\ 2011). 
For the few core-S\'ersic galaxies in our sample, we use the following steeper relation 
from Sahu et al.\ (2018, in preparation).  Obtained from a bisector linear
regression of $\sim$30 early-type galaxies with directly measured black hole masses
and having partially depleted cores, it is such that 
\begin{equation}
\log (M_{\rm bh}/M_{\odot}) = (9.42\pm0.08) + (8.86\pm1.53)\log \left( \sigma
/ 282\, {\rm km\, s}^{-1} \right), 
\label{eq1c}
\end{equation}
with $\Delta_{\rm rms}=0.48$ dex, and here we adopt an intrinsic scatter (in
the $\log M_{\rm bh}$ direction) of 0.4 dex. 
The $\sim$30 core-S\'ersic galaxies that defined this relation have $\sigma \gtrsim 200$ km s$^{-1}$.

Modulo the updated and new velocity dispersions that we can use for our sample
of Virgo cluster galaxies, the 
similarity between equations~\ref{eq1a} and \ref{eq1b} dictates that the
(velocity dispersion)-based black hole masses predicted by Gallo et
al.\ (2008) will agree well with our predictions for the S\'ersic galaxies. 
Only 4 of the 7 ETGs with depleted cores do not have a directly measured black
hole mass.  
Due to the extrapolation of equation~\ref{eq1c} in order to predict the 
black hole mass for our core-S\'ersic galaxy NGC~4382 with the lowest
velocity dispersion (176 km s$^{-1}$) 
of our core-S\'ersic sample (see Figure~\ref{Fig1}), we caution that it 
may be preferable to refer to the (galaxy luminosity)-based black hole mass
for this galaxy with a nuclear X-ray source (section~\ref{Sec_4382}).

\subsection{Galaxy Magnitude}\label{Sec_mag}

Gallo et al.\ (2008) used the $B$-band $M_{\rm bh}$--$L$ relation from Ferrarese \&
Ford (2005), in which 
\begin{equation}
\log(M_{\rm bh}/M_{\odot}) = (8.37\pm0.11) - (0.419\pm0.085)[\mathfrak{M}_B +20], 
\nonumber 
\end{equation}
and where $\mathfrak{M}_B$ is the absolute $B$-band magnitude of early-type galaxies or the
bulges of late-type galaxies. Gallo et al.\ (2008) applied this to their Virgo cluster
sample's $B$-band magnitudes. 

We have used the bisector linear regressions between $\log M_{\rm bh}$ and 
the $B$-band spheroid magnitude from Graham \& Scott (2013, see their 
Figure~2 and Table~3).  
For spheroids with partially depleted cores, 
\begin{equation}
\log(M_{\rm bh}/M_{\odot}) = 9.03\pm0.09 - (0.54\pm0.12)[\mathfrak{M}_B +21], 
\label{eq3}
\end{equation}
and for spheroids without such cores, 
\begin{equation}
\log(M_{\rm bh}/M_{\odot}) = 7.37\pm0.15 - (0.94\pm0.16)[\mathfrak{M}_B +19].
\label{eq4}
\end{equation} 
Given that the majority of galaxies which went into the derivation of 
equation~\ref{eq3} were elliptical galaxies, the `spheroid' magnitude was the
`galaxy' magnitude. 
The majority of the galaxies used for equation~\ref{eq4} 
were (early-type) galaxies whose dust corrected spheroid magnitudes 
came close to matching the galaxy magnitudes.  It has since been determined 
(Graham et al.\ 2016) that the 
dust corrections from Driver et al.\ (2008) should have been
restricted to the late-type galaxies, as it was too large a correction for the early-type
galaxies.\footnote{We note that 
the more massive ETGs of the Virgo cluster do contain some dust (e.g.\ di 
  Serego Alighieri et al.\ 2013), while the dwarf ETGs will have their 
  dust removed by ram pressure stripping by the hot ICM 
(e.g.\ De Looze et al.\ 2010; Grossi et al.\ 2015).}  Given the apparent balancing act, 
we proceed by using the relations (equations~\ref{eq3} and \ref{eq4}) as they are. 
Given that they were established using RC3 magnitudes (Vega), for consistency,
i.e.\ to avoid bias, we apply them to the Virgo galaxy sample's RC3 $B$-band
magnitudes that have been tabulated in the Appendix for convenience.  For the
purpose of determining the uncertainty on the predicted black hole mass (see
section~3.3 in Graham \& Scott 2013), we assume an uncertainty on the magnitude of 0.25 mag.

Multiplying the slope $-$0.94 in equation~\ref{eq4} by $-$2.5, one has that
$M_{\rm bh} \propto L_B^{2.35}$, where $L_B$ is the $B$-band luminosity.  
This is the `super-quadratic' relation referred to in Graham \& Scott (2013)
because the exponent is slightly higher than 2.

\begin{figure}
        \includegraphics[trim=1cm 2cm 2cm 1.6cm, width=\columnwidth]{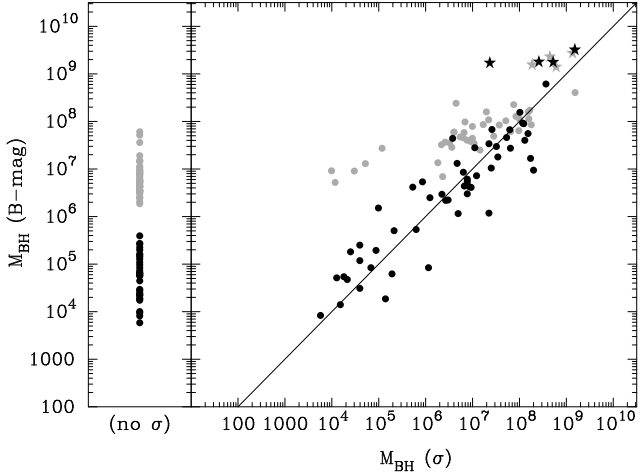}
    \caption{Predicted black hole masses in our Virgo galaxy sample, derived
      from both the galaxy's $B$-band magnitude, using equations~(\ref{eq3})
      and (\ref{eq4}), 
      and, when available, the velocity dispersion $\sigma$, using
      equations~(\ref{eq1b}) and (\ref{eq1c}).  The grey circles show the predictions 
      using the Gallo et al.\ (2008) data and their adopted scaling relations,
      while the black symbols use our data and relations (see section~\ref{sec_Param}).
      We have excluded the 11 galaxies with directly 
      measured black hole masses (Table~1). 
}
    \label{Fig5}
\end{figure}

\begin{figure}
        \includegraphics[trim=1cm 2.0cm 3cm 4.5cm, width=\columnwidth]{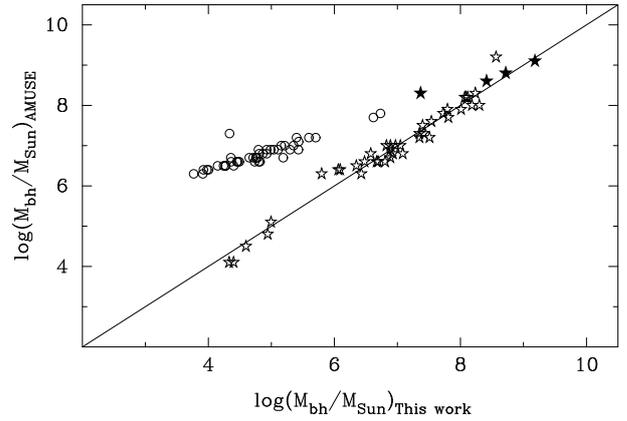}
    \caption{Comparison of our predicted black hole masses with those from Gallo et
      al.\ (2008).  Stars are based on velocity dispersions, and circles are
      based on $B$-band magnitudes.  The 4 filled stars are core-S\'ersic
      galaxies.} 
    \label{Fig6}
\end{figure}

Combining equations~(\ref{eq1c}) and (\ref{eq3}), to eliminate $M_{\rm bh}$, one
obtains the relation
\begin{equation}
-\mathfrak{M}_B/2.5 = (8.67\pm0.22) + (6.56\pm1.85)\log(\sigma/300\, {\rm km\, s}^{-1}) 
\label{eq5}
\end{equation}
for core-S\'ersic galaxies, 
while combining equations~(\ref{eq1b}) and (\ref{eq4}) yields the relation
\begin{equation}
 -\mathfrak{M}_B/2.5 = (7.94\pm0.27) + (2.02\pm0.40)\log(\sigma/185\, {\rm km\, s}^{-1})
\label{eq6}
\end{equation}
for the S\'ersic galaxies. 
The former equation, with a slope of $\sim$6.6$\pm$1.9, agrees well with the
known relation for luminous ETGs; Lauer et al.\ (2007) report $6.5\pm1.3$. 
The latter equation, with a slope of $\sim$2.0$\pm0.4$, also provides a good description for
the distribution of S\'ersic galaxies seen in Figure~\ref{Fig1}. 
In Figure~\ref{Fig5}, we plot our predicted black masses based on both the
velocity dispersion and the $B$-band magnitude of the 100 ETGs.  We
additionally show the predictions from Gallo et al.\ (2008).  To better
facilitate a comparison of our predictions with those from Gallo et
al.\ (2008), Figure~\ref{Fig6} plots these predictions against each other.  One can see
that the near-linear $M_{\rm bh}$--$L_B$ relation used by Gallo et al.\ (2008)
is the reason why they did not obtain consistent black hole masses with the
$M_{\rm bh}$--$\sigma$ relation.

Table~2 reveals the galaxy magnitudes and velocity dispersions
associated with a range of different black hole masses.
While the $L\propto \sigma^2$ relation seen in Figure~\ref{Fig1}, 
and revealed in equation~\ref{eq6}, can be seen here to  
apply to systems with velocity dispersions down to $\approx$25 km s$^{-1}$,
figure~17 in Toloba et al.\ (2014) reveals that it may roughly\footnote{From
 $\sigma =$ 5 to 10 km s$^{-1}$, dwarf spheroidal galaxies {\it may} follow the
  relation $M_{\rm *,gal}\propto \sigma^{3}$.} 
 hold all the way down to $\sigma \approx$9 km s$^{-1}$.  

\begin{table}
\centering
\caption{Black hole mass calibration points}
\begin{tabular}{ccc}
\hline
$M_{\rm bh}$   & $\mathfrak{M}_B$  & $\sigma$     \\
 $M_{\odot}$   & [mag]     & km s$^{-1}$  \\
\hline
$10^9$    &  $-$20.73  &  278    \\
$10^8$    &  $-$19.67  &  171    \\
$10^7$    &  $-$18.61  &  106    \\
$10^6$    &  $-$17.54  &   65    \\
$10^5$    &  $-$16.48  &   40    \\
$10^4$    &  $-$15.41  &   25    \\
$10^3$    &  $-$14.35  &   15    \\
$10^2$    &  $-$13.29  &    9    \\
\hline
\end{tabular}

\label{Tab_Calib}
Reversing equations~\ref{eq1b} and \ref{eq4} for S\'ersic galaxies,
i.e.\ those without partially depleted cores, we provide the $B$-band galaxy
magnitude and stellar velocity dispersion that corresponds to the black hole
masses listed in column~1.
\end{table}

\begin{table}
\centering
\caption{40 early-type galaxies with a potential IMBH} 
\begin{tabular}{llcc}
\hline
\multicolumn{2}{c}{Galaxy} & $M_{\rm bh}$ $(\sigma)$ &  $M_{\rm bh}$ ($\mathfrak{M}_B$ mag) \\
 VCC \#    &  Other Id.    &   $M_{\odot}$     &   $M_{\odot}$    \\
\hline
\multicolumn{4}{c}{11 galaxies: 2 estimates $\lesssim 10^5\,M_{\odot}$} \\
1049 & UGC~7580 &    1.3$\times10^4$  &    5.1$\times10^4$  \\
 856 & IC~3328  &    3.9$\times10^4$  &    1.2$\times10^5$  \\
1087 & IC~3381  &    8.7$\times10^4$  &    1.9$\times10^5$  \\
 543 & UGC~7436 &    2.5$\times10^4$  &    1.8$\times10^5$  \\
2019 & IC~3735  &    6.8$\times10^4$  &    8.4$\times10^4$  \\
1488 & IC~3487  &    2.1$\times10^4$  &    4.7$\times10^4$  \\
1075 & IC~3383  &    3.9$\times10^4$  &    3.1$\times10^4$  \\
2050 & IC~3779  &    1.4$\times10^5$  &    1.9$\times10^4$  \\
1407 & IC~3461  &    1.8$\times10^4$  &    5.4$\times10^4$  \\
1743 & IC~3602  &    1.5$\times10^4$  &    1.4$\times10^4$  \\
1826 & IC~3633  &    5.7$\times10^3$  &    8.3$\times10^3$  \\
\multicolumn{4}{c}{29 galaxies: 1 estimate $\lesssim 10^5\,M_{\odot}$}  \\
1910 & IC~809   &    ...  &    1.4$\times10^5$  \\
 140 & IC~3065  &    ...  &    1.6$\times10^5$  \\
1355$^a$ & IC~3442 &    ...  &    2.0$\times10^5$  \\
1861 & IC~3652  &    ...  &    9.7$\times10^4$  \\
1528 & IC~3501  &    ...  &    1.3$\times10^5$  \\
1833 &          &    ...  &    1.1$\times10^5$  \\
  33 & IC~3032  &    ...  &    6.6$\times10^4$  \\
 200 &          &    ...  &    8.4$\times10^4$  \\
  21 & I3025    &    ...  &    2.3$\times10^4$  \\
1779 & I3612    &    ...  &    5.7$\times10^4$  \\
1895 & UGC~7854 &    ...  &    2.9$\times10^4$  \\
1499$^a$ & IC~3492 &    ...  &    4.5$\times10^4$  \\
1545 & IC~3509  &    ...  &    7.3$\times10^4$  \\
1857 & IC~3647  &    ...  &    1.5$\times10^5$  \\
1948 &          &    ...  &    1.0$\times10^4$  \\
1627$^b$ &      &    ...  &    2.3$\times10^4$  \\
1440 & IC~798   &    ...  &    5.7$\times10^4$  \\
 230 & IC~3101  &    ...  &    1.7$\times10^4$  \\
1993 &          &    ...  &    9.5$\times10^3$  \\
 751$^a$ & IC~3292 &    ...  &    6.1$\times10^4$  \\
1828 & IC~3635   &    ...  &    6.3$\times10^4$  \\
 538 & NGC~4309A &    ... &    1.9$\times10^4$  \\
1886  &          &    ...  &    1.9$\times10^4$  \\
1199$^b$  &      &    ...  &    5.9$\times10^3$  \\
1539  &          &    ...  &    2.8$\times10^4$  \\
1185  &          &    ...  &    2.5$\times10^4$  \\
1512  &          &    ...  &    2.2$\times10^4$  \\
1489 & IC~3490   &    ...  &    8.2$\times10^3$  \\
1661 &           &    ...  &    6.4$\times10^4$  \\
\hline
\end{tabular}

\label{Tab_IMBH}
30 galaxies have $M_{\rm bh} \le 10^5\, M_{\odot}$, 10 galaxies have 
$10^5 < M_{\rm bh} / M_{\odot} < 2\times10^5$. 
$^a$ Nuclear X-ray emission has been detected in VCC~1355, VCC~1499 and VCC~751. 
$^b$ Possibly a stripped galaxy which may host a more massive black hole (see section~\ref{Sec_CM}). 
\end{table}

\section{IMBH targets of interest}\label{Sec_IMBH}

From the previous section, we have identified 40 targets of interest, of which
30 have a predicted black hole mass less than $10^5\, M_{\odot}$ and the
remaining 10 have a predicted black hole mass of (1--2)$\times 10^5\, 
M_{\odot}$ (see Table~3). 

Of these 40 IMBH candidates, 11 have {\it both} a magnitude and a velocity dispersion measurement 
suggestive of such an intermediate-mass black hole.  Among these 11, 
there are two targets immediately worthy of highlighting even though we have
not detected the (expectedly faint) X-ray emission from their nuclei. 
\begin{itemize}
\item IC~3602 = VCC~1743 is predicted to have a black hole mass of $\times10^4\,
M_{\odot}$ according to both its absolute magnitude and velocity dispersion.
\item IC~3633 = VCC~1826 is predicted to have a black hole mass of just 6 to 8 thousand
solar masses based on both its absolute magnitude and velocity dispersion. 
\end{itemize} 

Among these 40 galaxies, 
three additional targets of interest stand out, 
because X-ray activity has been detected in their core.  They are: 
\begin{itemize}
\item IC~3442 = VCC~1355 ($\log(M_{\rm bh,B-band}) = 5.30\pm0.87$), whose
nuclear X-ray point-source was reported by Gallo et al.\ (2010); 
\item IC~3492 = VCC~1499 ($\log(M_{\rm bh,B-band}) = 4.65\pm0.92$), seen in
Figure~\ref{Fig2a} and whose X-ray source was also noted by Gallo et al.\ (2010); 
and  
\item IC~3292 = VCC~751, see section~\ref{Sec_3292}, 
($\log(M_{\rm bh,B-band}) = 4.78\pm0.91$), whose 
central X-ray emission was previously unreported by Gallo et al.\ (2010).
\end{itemize} 
Unfortunately, no velocity dispersions are available for these three galaxies,
however, this is as expected given their low luminosities.  
IC~3442 (dE2,N) and IC~3292 (dS0,N) do, however, have nuclear star clusters, with 
$\log(M_{\rm nc}/M_{\odot})$ 
equal to 6.38 and 6.34 dex (Leigh et al.\ 2015)
based on the mass-to-light ratios provided by Bell et al.\ (2003). 
The optimal $M_{\rm bh}$--$M_{\rm nc}$ relation, extracted from Graham (2016b),
is such that 
\begin{eqnarray}
&&\log(M_{\rm nc}/M_{\odot}) = \\\nonumber
&&(0.40\pm0.13)\times\log(M_{\rm bh}/[10^{7.89}\,M_{\odot}]) + (7.64\pm0.25).
\label{eqNC} 
\end{eqnarray}
From this, we are able to predict that 
$\log (M_{\rm bh}/M_{\odot}) = 4.74$ and 4.64 dex, 
respectively.  That is, from the $\sim$2 million solar mass nuclear star clusters, we
expect a black hole mass of $\sim$40 thousand solar masses.  For reference,
the Milky Way has a nuclear star cluster that is some ten times more massive
than its $\sim 4 \times 10^6\, M_{\odot}$ black hole (e.g.\ Sch\"odel et
al.\ 2007; Graham \& Spitler 2009). 

\begin{figure}
\centering
  \includegraphics[angle=90, trim=0.3cm 2.6cm 0cm 2.6cm, width=0.8\columnwidth]{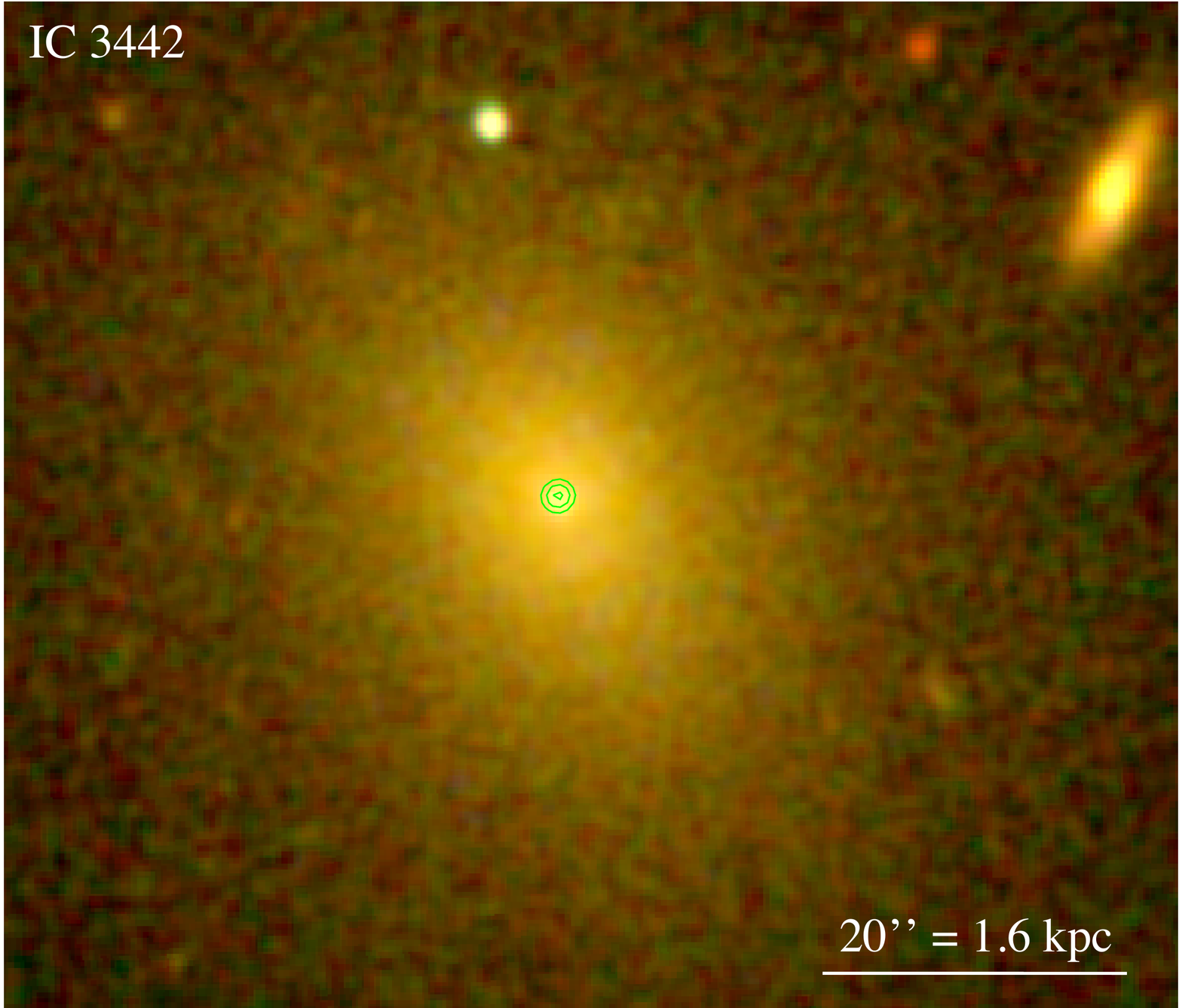}
    \caption{SDSS image of IC~3442 (VCC~1355). Colours are: red = $i^{\prime}$-band;
      green = $g^{\prime}$-band; blue = $u^{\prime}$-band. North is up and East to the left. A
      point-like X-ray source (green contours) coincides with the nuclear star
      cluster and is a prime candidate for an active IMBH.
}
\label{Fig_3442}
\end{figure}

To comment a little further on IC~3442, its main body has a red population 
while its nuclear star cluster is notably bluer. The {\it Chandra} source is
right on top of this nuclear cluster (Figure~\ref{Fig_3442}).  We cannot prove
it is not just a stellar-mass black hole, but it is worth further
investigation. For instance, the blue light might be direct optical emission
from the hot accretion disk around the nuclear black hole, analogous to the
scenario proposed by Soria et al.\ (2017) for the IMBH candidate HLX-1 in the
galaxy ESO 243-49, rather than coming from a younger star cluster.  Soria et
al.\ revealed that the blue light in HLX-1 originates from the IMBH disk
because it varies with the X-ray emission, while the red optical component
remains constant.  We will pursue this possibility for IC~3442 in future
work.

\begin{figure}
\centering
\includegraphics[angle=90, trim=0.3cm 2.6cm 0.0cm 2.6cm, width=0.8\columnwidth]{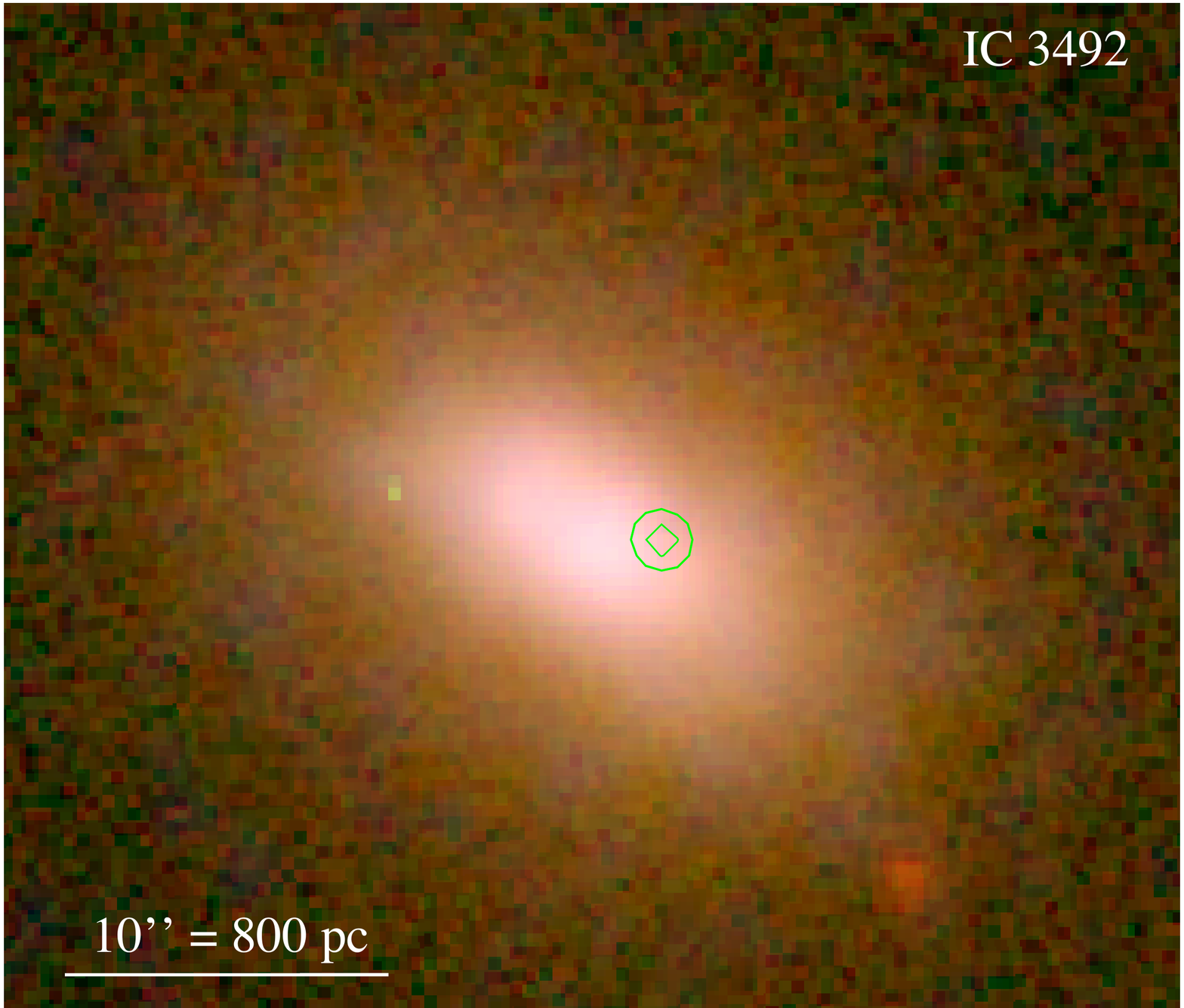}
    \caption{SDSS image of IC~3492 (VCC~1499), revealing the off-centre 
     {\it Chandra} X-ray point source (green contours). No nuclear star
     cluster is evident in {\it HST} images. 
}
\label{Fig_3492}
\end{figure}

As for the second target with an active X-ray nucleus (IC~3492), but without a
detected nuclear star cluster, we find that its {\it Chandra} source is
2$\arcsec$ west of the optical centre (Figure~\ref{Fig_3492}). There are only
two other point-like X-ray/optical associations in the field of view; based on
those two references points, we find that the {\it Chandra} astrometry
coincides with that of SDSS-12 and {\it Gaia} within an error of
$\approx$0$\arcsec$.2.  Thus, the offset of the {\it Chandra} source from the
optical nucleus is significantly larger than the relative positional
uncertainty.  It could be an X-ray binary, but we are not yet able to tell the
difference between this possibility and that of an offset IMBH in this small,
spheroid-shaped galaxy.  IC~3292 similarly had its X-ray point source
displaced by 2$\arcsec$, this time east of the optical nucleus (Section
2.2.6).

Returning to IC~3633 --- the galaxy predicted to have the smallest black hole
according to its magnitude and velocity dispersion --- the single 
stellar population models employed by Paudel et al.\ (2011) yielded an age of
$1.7^{+0.6}_{-0.2}$ Gyr for its nuclear component (with [Z/H]=$+0.13\pm0.17$
dex), within a host galaxy that is $11.4^{+1.7}_{-2.3}$ Gyr old (with
[Z/H]=$−0.90\pm0.15$ dex).  
Leigh et al.\ (2015) report that IC~3633 has a nuclear star
cluster with a mass of $(7.802\pm0.2947) \times 10^6\, M_{\odot}$, from which 
equation~\ref{eqNC} would predict a black hole mass of $10^6\, M_{\odot}$.
This is two
orders of magnitude larger than the predictions derived from the galaxy's
$B$-band magnitude and velocity dispersion.  However, IC~3633 has an
extended (radius $\approx 0\arcsec.3 \approx 24$ pc) 
and notably red ($g^{\prime} - z^{\prime}$)=1.5 (Ferrarese et al.\ 2006, their figure~99)
nuclear component which may be a nuclear disk rather than the types of smaller  
nuclear star clusters used to establish equation~\ref{eqNC}. 
We have therefore refrained from predicting black hole masses for all of the
Virgo ETGs reported to have nuclear star clusters (Ferrarese et al.\ 2006).
Many of these galaxies contain disks, and some also contain bars, rings and
ansae. However, the inclusion of such components in a detailed modelling of
the galaxy light, in order to better constrain the spheroid light and obtain a
more accurate nuclear star cluster luminosity, and nuclear disk / nuclear star cluster
separation (e.g.\ Balcells et al.\ 2007), is beyond the scope of this paper.

\section{Eddington ratios}\label{Sec_Edd}

\begin{figure}
        \includegraphics[trim=1.0cm 2.0cm 1.0cm 1.0cm, width=\columnwidth]{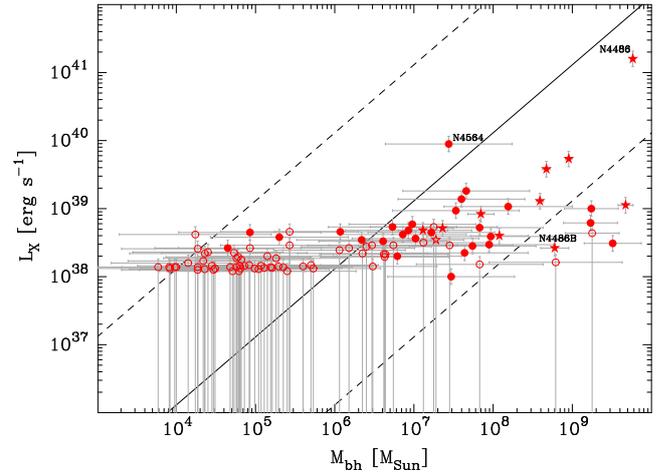}
    \caption{Nuclear X-ray luminosities (erg s$^{-1}$), for X-rays with
      energies between 0.3 and 10 keV (filled symbols), and upper limits (open
      symbols), plotted against 11 directly measured black hole masses (stars)
      and 89 predicted black hole masses (circles) based on the host galaxy's
      RC3 $B$-band luminosity.  
      The solid line shows an Eddington ratio of $10^{-6}$, while the dashed
      lines show Eddington ratios of $10^{-4}$ and $10^{-8}$. 
}
    \label{FigX}
\end{figure}

Eddington ratios, defined here as $L_X/L_{\rm Edd}$, with $L_{\rm Edd} =
1.3\times10^{38} M_{\rm bh}/M_{\odot}$ erg s$^{-1}$, of course vary with time,
depending on the phase of the AGN duty cycle, i.e.\ the black hole's sporadic
accretion rate (Czerny 2013).  The Galactic Centre has long been known to be
abuzz with signs of past activity (Chevalier 1992; Morris \& Serabyn 1996;
Yusef-Zadeh, Melia \& Wardle 2000; Cheng et al.\ 1997; Crocker et al.\ 2015).
The $\sim$200~pc, out-of-plane, Galactic Centre radio lobes (Sofue \& Handa
1984; Tsuboi et al.\ 1985), also detected as a limb-brightened bipolar
structure at mid-infrared wavelengths (8.3 $\mu$m) using the Midcourse Space
Experiment (Bland-Hawthorn \& Cohen 2003), and extending to much higher
latitudes in the radio (Sofue 1977, 2000), ROSAT 1.5 keV X-ray band (Bland-Hawthorn \&
Cohen 2003) and at Fermi gamma-ray wavelengths (Su et al.\ 2010), provides compelling
evidence that our Galaxy had a variable AGN over the last several million
years (Bland-Hawthorn 2015).  The multiple, nested bubbles observed in other galaxies
(e.g.\ McNamara \& Nulsen 2007; Gendron-Marsolais et al.\ 2016; 
see also Russell et al.\ 2017) further reveal the semi-cyclical nature of AGN activity.

No doubt due to the limited 90 minute exposure times in the AMUSE-Virgo X-ray
survey, no X-ray photons were observed to emanate from the centers of many
galaxies (see Section~\ref{Sec_X}). 
The X-ray survey was designed to be sensitive enough to provide a 3$\sigma$
detection of a 3 $M_{\odot}$ black hole accreting at the Eddington rate, or a
3 million solar mass black hole that is accreting at an Eddington ratio of $10^{-6}$.
For reference, the Milky Way's 4 million solar mass black hole is currently
accreting at a rate less than $10^{-8}$ of the theoretical Eddington
accretion limit, and elliptical galaxies have been reported to have a typical
$L_X/L_{\rm Edd}$ ratio of $10^{-8}$ (Zhang et al.\ 2009).  
In Figure~\ref{FigX}, we see that many of the more massive Virgo ETGs have ratios between
$10^{-8}$ and $10^{-6}$. 
IMBHs could, therefore, readily have gone undetected at X-ray wavelengths within the AMUSE-VIRGO
survey.  The X-ray detection of 3 galaxies out of the 40 suspected to have
IMBHs (Table~3) may be revealing a $7.5\pm4.3$ per cent `on' / 92.5 per cent
`off' AGN duty cycle in IMBHs, where `on' means an Eddington ratio greater
than $\sim 10^{-5}$.

In total, 62 (63 if we include IC~3292 from section~\ref{Sec_3292}, or 64 if
we also include IC~3492 from section~\ref{Sec_IMBH}) of the 100 Virgo ETGs
have only an upper limit to their X-ray luminosity arising from
non-detections.  While one might speculate on the absence of a black hole, for
many galaxies their AGN are likely to be too faint and a longer
exposure time is required.  Observing their 24 $\mu$m images, in a search for
dust obscured nuclear emission, Leipski et al.\ (2012) concluded that the
quiescent ETGs in the sample have `low bolometric Eddington ratios arising
from low accretion rates and/or highly radiatively inefficient accretion'.
This creates an artificial boundary in the $L_x$--$M_{\rm bh}$ diagram
(Figure~\ref{FigX}).  To minimise the biasing influence that this would have
on the linear regression between $\log L_X$ and $\log M_{\rm bh}$, for the 37
galaxies (including IC~3492) with measurements, we minimised their offsets in
the ($\log\, M_{\rm bh}$)-direction (see Lynden-Bell et al.\ 1988 for a
discourse on this approach, and their Figure~10).  This is sometimes referred
to as an ordinary least squares regression of the $X$ variable on the $Y$
variable, OLS($X|Y$), or as an `inverse' regression.  Using the BCES routine
from Akritas \& Bershady (1996) we obtained
\begin{equation} 
\log (L_X/10^{38} {\rm erg\, s}^{−1}) = (1.26\pm0.16) + (0.99\pm0.23)\log(M_{\rm
  bh}/10^8\, M_{\odot}). \nonumber 
\end{equation}
This is consistent with a slope of unity, suggesting that there is no
dependence of $L_{\rm Edd}$ ($\propto M_{\rm bh}$) on $M_{\rm bh}$.  
Although, this result is perhaps 
overly influenced by the two galaxies with the highest $L_X$ values.  
Excluding NGC~4486 and NGC~4584, to acquire what may be a more robust
regression with a sample size of 35 galaxies, we obtain 
\begin{equation}
\log (L_X/10^{38} {\rm erg\, s}^{−1})  =  (1.05\pm0.11) +
(0.64\pm0.19)\log(M_{\rm bh}/10^8\, M_{\odot}). 
\label{eqX}
\end{equation}
This relation is steeper than that from Gallo et al.\ (2010), who
reported a similar intercept at $M_{\rm bh}=10^8\, M_{\odot}$ 
of $1.0\pm0.1$ and a slope of $0.38^{+0.13}_{-0.12}$; although, the $1\sigma$
uncertainties on these two slopes do overlap with each other. 

Equation~\ref{eqX} can be rewritten in terms of the Eddington ratio, such that 
$(L_X/L_{\rm Edd} \propto M_{\rm bh}^{-0.36\pm0.19}$, suggesting that higher
mass black holes are not emitting as close to their Eddington limit as the
low-mass black holes do.  
According to equation~\ref{eqX}, at 
$M_{\rm bh}=3\times 10^4\, M_{\odot}$, the average Eddington ratio is
$1.6\times10^{-6}$, and if one extrapolates down to   
$M_{\rm bh}=3 M_{\odot}$, the average Eddington ratio is $4.4\times10^{-5}$,
i.e.\ 27.5 times larger. 

\begin{figure}
        \includegraphics[trim=1cm 2.0cm 3cm 4.5cm, width=\columnwidth]{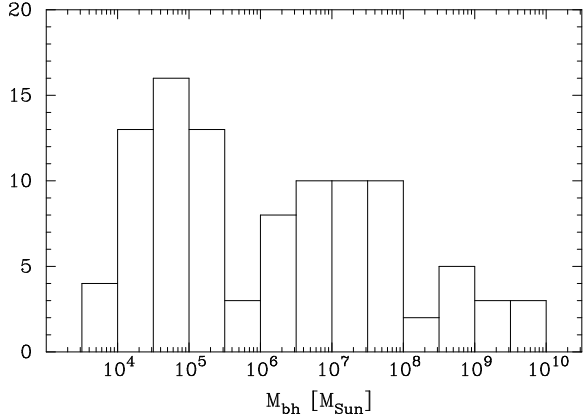}
    \caption{Histogram of expected black hole masses, based upon the host
      galaxy's $B$-band magnitude. 
}
    \label{Fig_Hist}
\end{figure}

It may be that equation~\ref{eqX} represents a biased (by observational
limits) upper envelope to the full distribution of black holes in
Figure~\ref{FigX}; that is, equation~\ref{eqX} may not trace the actual
relation.  The sampling of the 100 galaxies was such that more low-luminosity
galaxies, compared to high-luminosity galaxies, were included (see the
histogram in Figure~\ref{Fig_Hist}), thereby increasing the
chances\footnote{The more galaxies one includes, the more 2- and 3-$sigma$
  outliers one will detect. Also, the chances of capturing an X-ray
  photon, during a 90 minute exposure, from a low-mass black hole with a low
  Eddington ratio will also increase as more attempts are made to detect such
  a photon by sampling a greater number of low-luminosity galaxies.} of
detecting some low mass ($< 10^6 M_{\odot}$) black holes in an active state,
while still missing the majority of those with Eddington ratios $\le 10^{-6}$.  As
such, the X-ray detections from this galaxy sample with a non-uniform
luminosity function provides a somewhat biased census of (low-level)
supermassive black hole activity in the Virgo cluster, and consequently our
$M_{\rm bh}$--$L_X$ relation in equation~\ref{eqX} is arguably too shallow.

Probing to fainter X-ray luminosities might be rewarding.  For $M_{\rm
  bh}=3\times10^4\, M_{\odot}$, and an Eddington ratio of $10^{-5}$, this will
require exposure times that are ten times longer than the 90 minutes used in
the {\it Chandra} Cycle~8 Project; that is, 15 hours of integration would be
required for a 3-sigma detection.  However, this calculation is based on the
sensitivity of Chandra around the time when the AMUSE survey was performed
(2007-2008). Since then, the sensitivity has decreased by a factor of 2 due to
contaminants on the detector. If the same sources were to be observed next
year, it would take twice as long for a 3-sigma detection as the above
calculation indicates, i.e.\ 30 hours.

\section{The colour-magnitude diagram and implications for the $M_{\rm
    bh}$--$M_{*,gal}$ relation}\label{Sec_CM}

\begin{figure}
 \includegraphics[trim=5.8cm 2.2cm 1.0cm 1.0cm, width=\columnwidth]{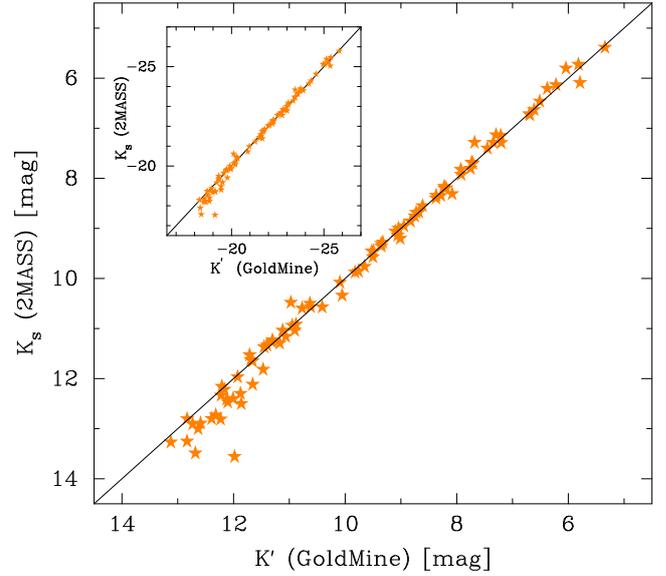}
    \caption{2MASS $K_s$-band {\it apparent} magnitudes versus the GOLD Mine 
      $K^{\prime}$-band {\it apparent} magnitudes. The inset panel 
      shows the {\it absolute} magnitudes. Both are on the Vega magnitude system and
      are corrected here for Galactic extinction.  At the faint end, some of the 2MASS
      magnitudes underestimate the galaxy light.} 
    \label{Fig_Comp}
\end{figure}

We have obtained the 
2MASS $K_s$-band (2.2 $\mu$m) total apparent magnitudes, $K_{\rm tot}$, 
for the sample of 100 Virgo cluster ETGs. 
In Figure~\ref{Fig_Comp}, we compare these with the 
$K^{\prime}$-band (2.2 $\mu$m) magnitudes available in the 
GOLD Mine\footnote{http://goldmine.mib.infn.it/} database (Gavazzi et al.\ 2003).
The agreement is good until we reach 2MASS magnitudes fainter than $K_s = 12$
mag.  2MASS is, however, known to underestimate the luminosity of ETGs with
faint mean effective surface brightnesses (e.g.\ Kirby et al.\ 2008, their
figure~13), and thus ETGs galaxies with faint magnitudes given the relation
between $\langle \mu \rangle_e$ and absolute magnitude (e.g.\ Graham \&
Guzm\'an 2003, their figure~9a). 
We, therefore, proceed using the GOLD Mine $K^{\prime}$-band magnitudes, albeit 
subject to a small correction at bright magnitudes, noted below. 

2MASS magnitudes are known to underestimate the total galaxy
flux in luminous ETGs with high S\'ersic indices due to the long tail in these
galaxies' surface brightness profiles, as discussed in Scott et al.\ (2013).  Given the
relative agreement seen in Figure~\ref{Fig_Comp} among the luminous ETGs, we
have applied the following 2MASS correction from Scott et al.\ (2013, see
their figures~1 and 2) to the GOLD Mine $K^{\prime}$-band absolute magnitudes
brighter than $-$22 mag:
\begin{equation}
K_{s,ARCH} = (1.070 \pm 0.030)K_{s,2MASS} + (1.527 \pm 0.061). 
\label{eq_Scott} 
\end{equation} 
This correction should bring the bright 2MASS, and thus the bright GOLD Mine,
magnitudes inline with the {\sc archangel} photometry (Schombert \& Smith
2012).  The size of these corrections were small for our sample: 0.24--0.28
mag for the brightest seven galaxies, and less than 0.2 mag for the other galaxies
with $M_K < -22$~mag, making their B-K value larger by this amount.  
This adjustment is not applied, and thus has no affect, on the faint galaxies 
where IMBHs may reside.

\begin{figure}
 \includegraphics[trim=2.5cm 2.cm 1.0cm 1.0cm, width=\columnwidth]{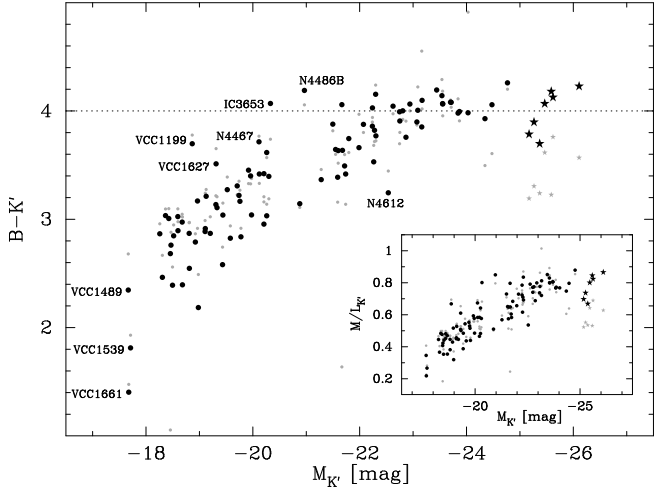}
    \caption{Colour-magnitude diagram for the 100 Virgo ETGs using GOLD Mine
      $K^{\prime}$-band magnitudes (see Section~\ref{Sec_CM} for
      details). Gallo et al.\ $B$-band data (grey points); RC3 $B$-band data
      (black points).  Core-S\'ersic galaxies (stars); S\'ersic galaxies
      (circles).  Inset panel: $K^{\prime}$-band (stellar mass)-to-(stellar
      light) ratio as a function of the $K^{\prime}$-band magnitude
      (Eq.~\ref{eq_Bell}).  The offset `faint' galaxies are predicted to have
      over-massive black holes (section~\ref{Sec_strip}).}
    \label{Fig2}
\end{figure}

In Figure~\ref{Fig2} we plot the $B-K^{\prime}$ colour against the 
$K^{\prime}$-band absolute magnitude using the corrected (i.e.\ slightly
brightened) $K^{\prime}$ GOLD Mine magnitudes.  At the luminous end, one can see that
the Gallo et al.\ (2008) $B$-band magnitudes are brighter than the RC3
$B$-band magnitudes.  At the faint end of the colour-magnitude diagram,
i.e.\ the faint-end of the so-called `red-sequence' for ETGs, the bluer
$B-K^{\prime}$ colours (also seen in Forbes et al.\ 2008, their figure~1) are 
considered to be real; it is also evident in the ($g-z$)
versus $g$-band colour-magnitude diagram for this galaxy sample 
(Smith Castelli et al.\ 2013).   Indeed, the general trend seen in Figure~\ref{Fig2}
is nothing new, and many other studies have revealed that the red sequence flattens at
bright magnitudes (e.g.\ Metcalfe, Godwin \& Peach 1994; 
Secker et al.\ 1997; Terlevich et al.\ 2001; Tremonti et al.\ 2004; Boselli et al.\ 2008; 
Ferrarese et al.\ 2006; Misgeld et al.\ 2008; Janz et al.\ 2009, Chen et
al. 2010; Lieder et al.\ 2012). 
Jim\'enez et al.\ (2011) report how the bright-end 
can be reconciled with dry galaxy mergers preserving the colour but increasing
the luminosity. 

We can use Figure~\ref{Fig2} to predict how the $B$-band $M_{\rm bh}$--$L$
relations from section~\ref{Sec_mag}, in particular equations~\ref{eq3} and \ref{eq4}, 
 will map into the $K$-band. 
Brighter than a $K^{\prime}$-band absolute magnitude of approximately $-$23
mag in Figure~\ref{Fig2}, the $B-K^{\prime}$ colour is roughly constant, at a value of around 4 or
slightly greater.  Therefore, the $K^{\prime}$-band $M_{\rm
  bh}$--$L_{K^{\prime}}$ relation should have the same slope as the $B$-band
$M_{\rm bh}$--$L_{B}$ relation for $B$-band magnitudes brighter than $\approx
-$19 mag.  That is, the bright ETGs, many of which are 
core-S\'ersic galaxies thought to have been built
from dry merger events, should have a near-infrared $M_{\rm
  bh}$--$\mathfrak{M}_{K^{\prime},galaxy}$ relation with a slope of $-0.54$
(equation~\ref{eq3}).  Given that the magnitude $\mathfrak{M}$ 
equals $-2.5\log L$, we have that 
$M_{\rm bh} \propto L_{B,galaxy}^{1.35}$, and thus 
$M_{\rm bh} \propto L_{K^{\prime},galaxy}^{1.35}$. 

In contrast to this, the `red-sequence' will ensure that the ETGs fainter than 
$\mathfrak{M}_{K^{\prime},galaxy} \approx -23\pm1$~mag, i.e.\ the S\'ersic
ETGs, will have a different $M_{\rm bh}$--$L$ slope in the
near-infrared than they do in the $B$-band.  Given that the $B-K^{\prime}$
colour changes from $\approx$2.5 at $K^{\prime} = -18$ mag (see also Forbes et
al.\ 2008) to $\approx$4 at $K^{\prime} = -23$ mag, one can deduce that the
$K^{\prime}$-band equivalent of equation~\ref{eq4} should have a slope of
$-0.66$. From the above, the red sequence is such that $B-K \propto 
(-1.5/5)K$, which can be rearranged to give $B \propto (3.5/5)K$.  Given that
equation~\ref{eq4} reports that $\log M_{\rm bh} \propto -0.94B$, we have that 
$\log M_{\rm bh} \propto -0.66K$.  Finally, because the $K$ band magnitude equals
$-2.5$ times the logarithm of the $K$-band luminosity, we have that 
$M_{\rm bh} \propto L_{K^{\prime},galaxy}^{1.65}$.

Sahu et al.\ (2018, in preparation) has derived the 3.6 $\mu$m $M_{\rm
  bh}$--$L_{galaxy}$ relation for $\sim$30 core-S\'ersic ETGs and $\sim$50
S\'ersic ETGs.  Their work supercedes the $K$-band relation from Graham \&
Scott (2013), whose smaller sample size suffered from a ($B-K$
colour)-magnitude relation that does not represent the ETG population at
large.  Sahu et al.\ find an exponent on the luminosity term of $1.48\pm0.18$ 
and $1.64\pm0.17$, respectively (cf.\ $1.35\pm0.30$ and $1.65\pm0.30$ predicted
above), providing additional confidence in equations~\ref{eq3} and \ref{eq4}
that have been used in this paper to derive the photometry-based black hole masses. 

Taking things further regarding the ETGs, we can convert the galaxy 
stellar luminosities into stellar masses
using the relation from Bell \& de Jong (2001) for the $K$-band\footnote{Here
  we set $K=K^{\prime}$.} 
(stellar mass)-to-(stellar light) ratio, $M/L_K$, which is such that 
\begin{equation}
\log M/L_{K} = 0.2119(B-K)-0.9586.
\label{eq_Bell}
\end{equation}
The inset panel in Figure~\ref{Fig2} shows the $M/L_{K}$ ratios for our sample
plotted against their $K$-band absolute magnitude.  
Brighter than $-$23 to $-$24 mag, the ratios are roughly
constant at $\sim$0.8.  As a consequence, the $M_{\rm bh}$--$M_{\rm *,gal}$
relation for luminous ETGs should have a slope of $1.35\pm0.30$.  
Given that the $M/L_{K}$ ratio drops from $\sim$0.8 at $K^{\prime} = -23$ mag to
$\sim$0.4 at $K^{\prime} = -18$ mag,
one can similarly deduce that the $M_{\rm bh}$--$M_{\rm *,gal}$
relation for the less luminous ETGs should have a slope of 1.43,
consistent with that at the high-mass end.  That is, a single (black
hole)-galaxy mass relation appears to unite the dwarf, ordinary, and bright ETG
sequence (Savorgnan et al.\ 2016; Sahu et al.\ 2018, in preparation). 

The above applies to gas-poor early-type galaxies; it should not be
assumed that the same scalings apply to late-type galaxies.  Savorgnan et
al.\ (2016) have provided evidence that the late-type galaxies follow a
steeper relation than ETGs in the $M_{\rm bh}$--$M_{\rm *,bulge}$ diagram, and
given the systematic change in the bulge-to-total flux ratio among late-type
galaxies, the $M_{\rm bh}$--$M_{\rm *,galaxy}$ relation for late-type galaxies
will be even steeper than their $M_{\rm bh}$--$M_{\rm *,bulge}$ relation.
This was recently shown to be the case by Davis et al.\ (2018a,b), who
presents the $M_{\rm bh}$--$M_{\rm *,bulge}$ and $M_{\rm bh}$--$M_{\rm
  *,galaxy}$ relation for 40 spiral galaxies with directly measured black hole
masses.

\subsection{Faint outliers from the colour-magnitude diagram}\label{Sec_strip}

Regarding the red outliers in Figure~\ref{Fig2}, we have identified 
the following galaxies.
\begin{itemize}
\item NGC~4612 = VCC~1883.  The galaxy image contains a bright foreground star which {\it may}
  have brightened the published $B$-band magnitude, although this requires
  investigating. 
\item NGC~4486B = VCC~1297.  It is a `compact elliptical' galaxy, thought to
  be the remnant of a more luminous, tidally stripped galaxy, possibly
  explaining both the offset in Figure~\ref{Fig1} and \ref{Fig2}, and its apparent
  overly massive black hole (Kormendy et al.\ 1997). 
\item NGC~4467 = VCC~1192.  Faint outlier from the $L_B$--$\sigma$ diagram
  (Figure~\ref{Fig1}), and
  reported being rather bright and compact (Chilingarian \& Mamon 2008) 
\item IC~3653 = VCC~1871.  Faint outlier from the $L_B$--$\sigma$ diagram and
  reported being rather bright and compact (Chilingarian \& Mamon 2008). 
\item VCC~1627. Reported being rather bright and compact (Chilingarian \& Mamon 2008). 
\item VCC~1199.  Extreme outlier in the $(g-z)$--$M_g$ colour-magnitude diagram of 
Smith Castelli et al.\ (2013), who identify a `red group' containing this
galaxy plus VCC~1192, VCC~1297, VCC~1627, and VCC~1871 (plus possibly
VCC~1327). 
\end{itemize}
Although a few `red outliers', relative to their current magnitude, are
evident in Figure~\ref{Fig2}, it may be more accurate to refer to them as
`faint outliers' relative to their colour.

It is not unreasonable to speculate that if NGC~4486B has an unusually over-massive
black hole for its $B$-band luminosity (Kormendy et al.\ 1997), 
then so may some of the other galaxies
listed above if they too have been stripped of their outer stellar population
like NGC~4486B is thought to be. 
That is, they may not be overly red galaxies but are instead now under-luminous in both the
$B$- and $K$-bands.  NGC~4486B is similar to M32, the prototype for the
`compact elliptical' galaxy class.  Thought to have previously been a
lenticular disk galaxy (Graham 2002), much of M32's disk has been removed by
M31, leaving behind a bulge with structural properties that are similar to the
bulges of other disk galaxies (see figure~1 in Graham 2013).  While these
galaxies are rare (Chilingarian et al.\ 2009; Chilingarian \& Zolotukhin 2015) --- by matched luminosity,
just 0.5 percent of 
dwarf galaxies are compact elliptical galaxies (Chilingarian 2018,
priv.\ comm.) --- they are often over-represented in local
scaling diagrams, sometimes leading to a skewed/biased view of reality. 
However, partially-stripped galaxies which have not (yet) become compact
elliptical galaxies may be less rare, and we might be seeing some of these in our
Virgo cluster sample (especially at $\mathfrak{M}_{K^{\prime}} = -20\pm1$ mag
in Figure~\ref{Fig2}).  If
correct, this would imply that our ($B$-band magnitude)-based predictions for
the black hole masses in these partially-stripped galaxies may be too low, because
the stellar luminosities are too low due to the partial removal of their disk
and outer layers.  It would be prudent to exclude such
galaxies when constructing the $M_{\rm bh}$--$L$ scaling relations.  Indeed,
the inclusion of M32, when constructing the $M_{\rm bh}$--$L$ relation in
Graham (2007), resulted in a biased relation with a near-linear slope. 
If these galaxies' stellar velocity dispersions have not been significantly reduced,
then these galaxies will also appear to have low-luminosities in the
luminosity-(velocity dispersion) diagram (see Figure~\ref{Fig1}), as indeed
some do.

Two of the above mentioned outlying galaxies (VCC~1627 and VCC~1199) are listed in
Table~3 as hosting an IMBH based upon their $B$-band luminosity.  It
may well be that they instead host an `over-massive' black hole relative to
their luminosity.

If the offset galaxies in the colour-magnitude diagram are due
to stripping events, then one could assume that the unstripped galaxy 
would have roughly the same colour as it does now, 
and use the offset from the colour-magnitude relation to estimate the original
galaxy magnitude.  Of course, the existence of radial colour gradients within
an individual galaxy, coupled with the stripping of the outer stellar material, 
does make the recovery of the original galaxy magnitude somewhat uncertain,
and we leave this exercise for a later study.

\section{Discussion: Past and Future}\label{Sec_Disc}

Graham (2007, see the final paragraph
of his Appendix) proposed that the $M_{\rm bh}$--$L_{\rm galaxy}$ relation may
have two slopes: a linear slope at the high mass end where bright ETGs reside, 
and a quadratic relation such that $M_{\rm bh} \propto L_{\rm
  galaxy}^2$ at lower masses.\footnote{Note: Graham (2007) contains an obvious
  typographical error, and mistakingly reads $M_{\rm bh} \propto L^{0.5}$
  rather than $M_{\rm bh} \propto L^{1/0.5}$.}  
Bernardi et al.\ (2007) and Graham \& Driver (2007) also remarked on the
inconsistency of a single log-linear relation.  It was not, however, until
Graham (2012) that systematic departures from the linear $M_{\rm bh}$--$M_{\rm
  dyn,galaxy}$ relation were clearly pointed out and again discussed in this
context of a broken relation.  A more detailed historical telling of
the literary evolution of the $M_{\rm bh}$--$M_{\rm galaxy}$ relation is
provided in Graham (2016a), where it can be seen that Laor (1998, 2001) and
Salucci et al.\ (2000) had correctly noted that the $M_{\rm bh}$--$L$ relation 
is not linear.  While they did not 
mention the possibility of a broken or curved relation, their single
log-linear relation had an exponent greater than 1 as their sample probed both
the bright (near-linear) and faint (near-quadratic) regions of the $B$- and $V$-band 
$M_{\rm bh}$--$L$ diagram.  Despite this pioneering work from two decades ago, most previous
studies had not predicted an abundance of IMBHs --- reviewed by 
Koliopanos (2017) and Mezcua (2017)  --- because they extrapolated the
near-linear $M_{\rm bh}$--$M_{\rm galaxy}$ relation to low masses, resulting
in inconsistent black hole masses with the $M_{\rm bh}$--$\sigma$ relation. 

It may be worthwhile pursuing the most promising IMBH targets in the Virgo
cluster with high 
spatial resolution radio/sub-millimetre interferometry in case any have maser
emission coming from a rotating disc surrounding the black hole (e.g.\ Miyoshi
et al.\ 1995; Greenhill et al.\ 2003).  
Long XMM observations, acquiring a few $10^2$ counts, may also be useful for
investigating the time-scale of any X-ray variability, and the X-ray spectra can provide further
insight.  For instance, if the spectrum is a hard power-law and the luminosity
is $\sim$$10^{38}$--$10^{39}$ erg s$*{-1}$, the radiating source is more
likely to be associated with an IMBH/SMBH because an X-ray binary
would have a softer thermal spectrum at those luminosities.  
In the absence of directly
measured constraints on the black hole mass --- which are challenging due to
the small angular extent of their sphere of gravitational influence --- one
can turn to an array of indirect mass estimates.

While the luminosity of the X-ray photons coming from 
the hot accretion discs around IMBHs is not a reliable proxy for black hole mass, when
coupled with the AGN's radio luminosity, it can be used in the
`fundamental plane of black hole activity' (Merloni et al.\ 2003; Falcke et
al.\ 2004) to estimate black hole masses. 
High spatial resolution radio images can also be used to determine if the
radio emission is concentrated in a point-source, and/or originates from a
jet, or instead comes from an extended, more diffuse star formation region.  
In addition, the radio brightness 
temperature can be used to distinguish between thermal and non-thermal
emission in supermassive black holes (e.g.\ Reines \& Deller 2012).  
The X-ray spectrum can also reveal clues as to the mass of the black
hole, although this requires the collection of more than just a few photons.

Another approach which can be used to provide an estimate of the black hole
masses is `reverberation mapping' (e.g.\ Bahcall et al.\ 1972; Netzer \&
Peterson 1997).  Broad and narrow line signatures in optical spectra can
reveal an AGN, as can a galaxy's location in the `BPT diagram' (Baldwin et
al.\ 1981; Kewley et al.\ 2001).  However, while an ensemble {\it average}
virial $f$-factor can be established for deriving black hole masses {\it en
  masse} --- using $M_{\rm bh} = f \times r \Delta V^2/G$, where $\Delta V$ is
the Doppler-broadened width of the emission lines in an individual galaxy, $r$
is the associated size of the `broad line region' in that galaxy, and $G$ is
the gravitational constant (e.g.\ Onken et al.\ 2004; Graham et al.\ 2011),
--- the (typically unknown) individual virial factors can vary widely from
galaxy to galaxy due to different, and time-varying, geometries of the `broad
line region' surrounding the black hole.  This undermines attempts using
reverberation mapping to obtain an individual galaxy's BH mass more accurately
than already achieved in section~\ref{sec_Param}.  Nonetheless, we do note
that Chilingarian et al.\ (2018) has used the width and luminosity of the
H$\alpha$ emission line to identify 305 IMBH candidates, with $0.3\times 10^5
< M_{\rm bh}/M_{\odot} < 2\times 10^5$, using the $M_{\rm bh}$-$H\alpha$
relation previously calibrated against virial black hole mass estimates.  Ten
of these galaxies display nuclear X-ray emission signalling the presence of a
black hole, whose masses could be estimated using the techniques employed here
and in Paper~II for spiral galaxies.  This could also be done for the IMBH
candidate in NGC~3319 (Jiang et al.\ 2018), and perhaps for some of the 40 low-mass
AGN out to $z=2.4$ in Mezcua et al.\ (2018a) which have so far had their black
hole masses estimated using a near-linear, $z=0$, $M_{\rm bh}$--$M_{*,galaxy}$
relation. 

The merging of black holes is now quite famously (Abbott et al.\ 2016,2017)
known to produce gravitational radiation during their orbital decay
(Poincar\'e 1905; Einstein 1916, 1918; see Barack et al.\ 2018 and
Caballero-Garcia et al.\ 2018 for a modern and relevant overview).  The
inspiral of compact stellar-mass objects, such as neutron stars and black
holes, around IMBHs (e.g.\ Amaro-Seoane et al. 2007; Mapelli et al.\ 2012) ---
referred to as extreme mass-ratio inspirals (EMRIs) --- may one day be
detectable by the gravitational radiation they emit on their inward journey.
In addition, the merging of galaxies containing IMBHs should result in the
eventual merging of these black holes.  The long wavelengths associated with
their large orbits are currently beyond the reach of our ground-based
facilities, ruling out any chance of measuring IMBH masses this way in the
near future.  The Kamioka Gravitational Wave Detector (KAGRA: Aso et
al.\ 2013) will be a 3 km long underground interferometer in Japan capable of
detecting the gravitational radiation emanating from 
collisions involving
black holes with masses up to 200 $M_{\odot}$ (T\'apai et al.\ 2015).  The 
planned Deci-Hertz Interferometer Gravitational wave Observatory (DECIGO:
Kawamura et al.\ 2011) and the European, Laser Interferometer Space Antenna
(LISA) Pathfinder mission\footnote{\url{http://sci.esa.int/lisa-pathfinder/}}
(Anza et al.\ 2005; McNamara 2013), with their greater separation of mirrors,
will be able to detect longer wavelength gravitational waves, and thus better
reach into the domain of intermediate-mass and supermassive black hole
mergers, the latter of which are currently being searched for via `pulsar
timing arrays' (PTAs) (e.g.\ Hobbs et al.\ 2010; Kramer \& Champion 2013;
Shannon et al.\ 2015).  It is reasonable to expect that the densely packed
nuclear star clusters, which coexist with low-mass SMBHs (e.g.\ Gonz{\'a}lez
Delgado et al.\ 2008; Seth et al.\ 2008; Graham \& Spitler 2009), will
similarly surround many IMBHs.  Gravitational radiation, and the gravitational
tidal disruption of ill-fated white dwarf stars that venture too close to
these black holes (e.g.\ Komossa 2015; Shen 2018), is therefore expected from these
astrophysical entities.

Now, a decade on from the {\it Chandra} Cycle~8 Project `The Duty Cycle
of Supermassive Black Holes: X-raying Virgo', a Cycle 18 Large Project ({\it
  Chandra} Proposal Id.18620568, P.I.\ Soria), has been awarded 559 ks (155
hours) of {\it Chandra X-ray Observatory} time to image 52 spiral galaxies in
the Virgo cluster with the Advanced CCD Imaging Spectrometer (ACIS-S
detector).  This will be combined with suitably-deep archival {\it Chandra}
images for an additional 22 spiral galaxies in the Virgo cluster, and the
results presented in a separate series of papers.  Given the low degree of
scatter about the $M_{\rm bh}$--$|\phi|$ relation for spiral galaxies (Seigar
et al.\ 2008; Davis et al.\ 2017), the spiral arm pitch angle $\phi$ appears
to be the most accurate predictor of black hole mass in spiral galaxies.  We
shall use this on the (52+22=) 74 Virgo spiral galaxies.  In future work, we
intend to additionally identify non-Virgo late-type spiral galaxies with open,
loosely-wound spiral arms, i.e.\ those expected to have the lowest mass black
holes at their centre, and then check for the X-ray signature of a hot accretion
disk heralding the presence of potentially further intermediate-mass black
holes.

For now, we conclude having identified 30 (40) ETGs in the Virgo cluster with
predicted black hole masses less than $10^5\, M_{\odot}$ ($2\times 
10^5\, M_{\odot}$).  Based on the velocity dispersion, Gallo et al.\ (2008,
their figure~3) 
had also predicted black hole masses of less than $10^5\, M_{\odot}$ for 
VCC~1488 (IC~3487), VCC~543 (UGC~7436), and VCC~856 (IC~3328), but this was
inconsistent with their predictions based upon the absolute magnitude.  
Of particular note in the full sample 
are IC~3602 with $M_{\rm bh} \approx 10^4\, M_{\odot}$, and IC~3633
with $M_{\rm bh} \approx$ (6--8)$\times 10^3\, M_{\odot}$.  
Two (three) of these 30 (40) galaxies has had a {\it Chandra} point 
source detected near the centre of their optical image.
Given that 
most IMBHs will have low Eddington ratios at any given time during their AGN
duty cycle,  
longer exposure times are required if one is to see X-rays from more IMBHs. 
We additionally note that two of these three detections may be due to X-ray binaries as
they occur $\approx 2\arcsec$ ($\approx 200$ pc) from the precise optical
centre.  Alternatively, they may be offset AGN given that the gravitational
gradient is not steep near the bottom of the 
gravitational potential well in dwarf galaxies with low S\'ersic indices. 
For example, Binggeli et al.\ (2000) report that ($10^5$--$10^6\,M_{\odot}$) nuclear star
clusters are offset by $\approx 100$ parsec 20 per cent of the time. 

We have resolved the three orders of magnitude difference in predicted black
hole mass that was reported by Gallo et al.\ (2008, their figure~4)
for the 100 Virgo cluster ETGs in the Advanced Camera for
Surveys Virgo Cluster Survey (C\^ot\'e et al.\ 2004).  
It is plausible that we have discovered the whereabouts of some of the previously thought to be
missing population of IMBHs. 
Thought to be inhabiting the centres of
low-luminosity ETGs, we predict black hole masses down to 6--8
thousand solar masses. 
Additional IMBHs may reside at the centres of
low-luminosity late-type galaxies, see Paper~II of this series (Graham et
al.\ 2018), and we will further explore this possibility
through the {\it Chandra X-ray Observatory} Cycle 18 Large Project 
`Spiral galaxies of the Virgo cluster' (Proposal Id.18620568, P.I.\ Soria).

\section*{Acknowledgements}

This research was supported under the Australian Research
Council's funding scheme DP17012923. 
Part of this research was conducted within the Australian Research Council's 
  Centre of Excellence for Gravitational Wave Discovery (OzGrav), through
  project number CE170100004. 
Support for this work was provided by the National Aeronautics and Space
Administration through Chandra Award Number 18620568.
AWG is grateful to the National Astronomical Observatories at the Chinese Academy of
Sciences in Beijing for his two-week visit in 2017. 
This research has made use of the NASA/IPAC Extragalactic Database (NED).
This publication makes use of data products from the Two Micron All Sky Survey.
We acknowledge use of the HyperLeda database (http://leda.univ-lyon1.fr).
This research has made use of the GOLD Mine Database.

\appendix
\section{Galaxy sample and predicted black hole masses} 

\begin{table*}
	\centering
	\caption{Galaxy parameters and predicted black hole masses} 
	\begin{tabular}{rlc|rc|cc|c} 
	\hline
VCC \# & Other  &  Dist. &  $\sigma$   & $\log M_{\rm bh}(\sigma)$ & $\mathfrak{M}_B$ &    $\log M_{\rm bh}(\mathfrak{M}_B)$  &  $\log L_X$  \\
       &        &  [Mpc] & km s$^{-1}$ &         [dex]          &    [mag]  &    [dex]                  &  [dex]     \\  
 1 & 2 & 3 & 4 & 5 & 6 & 7 & 8 \\
	\hline
 1226 & NGC~4472$^a$ & 17.14 & 282$\pm$28  &   9.18$\pm$0.56    &    $-$21.88  &   9.51$\pm$0.36   &   $ $38.49$^*$  \\
 1316 & NGC~4486$^a$ & 17.22 & 323$\pm$32  &  (9.70$\pm$0.56)   &    $-$21.67  &  (9.39$\pm$0.35)  &   $ $41.20      \\
 1978 & NGC~4649$^a$ & 17.30 & 330$\pm$33  &  (9.79$\pm$0.56)   &    $-$21.49  &  (9.30$\pm$0.35)  &   $ $39.05      \\
  881 & NGC~4406$^a$ & 16.83 & 231$\pm$23  &   8.41$\pm$0.59    &    $-$21.41  &   9.25$\pm$0.34   &   $<$38.64      \\
  798 & NGC~4382$^a$ & 17.86 & 176$\pm$17  &   7.37$\pm$0.66$^b$ &   $-$21.37  &   9.23$\pm$0.34   &   $ $38.79$^*$  \\
  763 & NGC~4374$^a$ & 18.45 & 278$\pm$27  &  (9.13$\pm$0.56)   &    $-$21.39  &  (9.24$\pm$0.34)  &   $ $39.73      \\
  731 & NGC~4365$^a$ & 23.33 & 250$\pm$25  &   8.72$\pm$0.57    &    $-$21.40  &   9.24$\pm$0.34   &   $ $39.00      \\
 1535 & NGC~4526   & 16.50  &  225$\pm$22  &   8.56$\pm$0.37    &    $-$20.51  &   8.79$\pm$0.84   &   $<$38.21      \\
 1903 & NGC~4621   & 14.93  &  228$\pm$22  &  (8.59$\pm$0.37)   &    $-$20.42  &  (8.70$\pm$0.83)  &   $ $39.11      \\
 1632 & NGC~4552   & 15.85  &  250$\pm$25  &  (8.78$\pm$0.37)   &    $-$20.42  &  (8.70$\pm$0.83)  &   $ $39.58      \\
 1231 & NGC~4473   & 15.28  &  179$\pm$17  &  (8.09$\pm$0.37)   &    $-$19.86  &  (8.18$\pm$0.81)  &   $ $38.60      \\
 2095 & NGC~4762   & 16.50  &  141$\pm$14  &  (7.60$\pm$0.37)   &    $-$20.05  &  (8.35$\pm$0.82)  &   $ $38.71      \\
 1154 & NGC~4459   & 16.07  &  172$\pm$17  &   8.01$\pm$0.37    &    $-$19.87  &   8.19$\pm$0.81   &   $ $39.03      \\
 1062 & NGC~4442   & 15.28  &  179$\pm$17  &   8.09$\pm$0.37    &    $-$19.62  &   7.95$\pm$0.81   &   $ $38.47      \\
 2092 & NGC~4754   & 16.14  &  177$\pm$17  &   8.07$\pm$0.37    &    $-$19.63  &   7.97$\pm$0.81   &   $ $38.59      \\
  369 & NGC~4267   & 15.85  &  150$\pm$15  &   7.73$\pm$0.37    &    $-$19.31  &   7.66$\pm$0.80   &   $ $39.26      \\
  759 & NGC~4371   & 16.98  &  129$\pm$12  &   7.41$\pm$0.38    &    $-$19.49  &   7.83$\pm$0.80   &   $<$38.18      \\
 1692 & NGC~4570   & 17.06  &  187$\pm$18  &   8.18$\pm$0.37    &    $-$19.40  &   7.74$\pm$0.80   &   $ $38.45      \\
 1030 & NGC~4435   & 16.75  &  155$\pm$15  &   7.79$\pm$0.37    &    $-$19.49  &   7.83$\pm$0.80   &   $ $38.72      \\
 2000 & NGC~4660   & 15.00  &  192$\pm$19  &   8.24$\pm$0.37    &    $-$18.84  &   7.22$\pm$0.80   &   $ $38.65      \\
  685 & NGC~4350   & 16.50  &  181$\pm$18  &   8.11$\pm$0.37    &    $-$19.25  &   7.60$\pm$0.80   &   $ $39.14      \\
 1664 & NGC~4564   & 15.85  &  156$\pm$15  &   7.81$\pm$0.37    &    $-$19.07  &   7.44$\pm$0.80   &   $ $39.95      \\
  654 & NGC~4340   & 16.50  &  108$\pm$10  &   7.05$\pm$0.38    &    $-$19.08  &   7.45$\pm$0.80   &   $<$38.46      \\
  944 & NGC~4417   & 16.00  &  135$\pm$13  &   7.51$\pm$0.37    &    $-$19.11  &   7.47$\pm$0.80   &   $ $38.00$^*$  \\
 1938 & NGC~4638   & 17.46  &  125$\pm$12  &   7.35$\pm$0.38    &    $-$19.17  &   7.53$\pm$0.80   &   $ $38.97      \\
 1279 & NGC~4478   & 16.98  &  137$\pm$13  &   7.54$\pm$0.37    &    $-$18.88  &   7.25$\pm$0.80   &   $<$38.73      \\
 1720 & NGC~4578   & 16.29  &  112$\pm$11  &  (7.12$\pm$0.38)   &    $-$18.75  &  (7.14$\pm$0.80)  &   $<$38.54      \\
  355 & NGC~4262   & 15.42  &  198$\pm$19  &   8.30$\pm$0.37    &    $-$18.58  &   6.98$\pm$0.80   &   $ $38.77      \\
 1619 & NGC~4550   & 15.49  &   96$\pm$ 9  &   6.80$\pm$0.39    &    $-$18.53  &   6.93$\pm$0.80   &   $ $38.68      \\
 1883 & NGC~4612   & 16.60  &   86$\pm$ 8  &   6.58$\pm$0.40    &    $-$19.29  &   7.64$\pm$0.80   &   $ $38.35      \\
 1242 & NGC~4474   & 15.56  &   90$\pm$ 9  &   6.67$\pm$0.40    &    $-$18.73  &   7.12$\pm$0.80   &   $<$38.50      \\
  784 & NGC~4379   & 15.85  &  110$\pm$11  &   7.09$\pm$0.38    &    $-$18.46  &   6.86$\pm$0.80   &   $ $38.62      \\
 1537 & NGC~4528   & 15.85  &  104$\pm$10  &   6.97$\pm$0.39    &    $-$18.20  &   6.61$\pm$0.81   &   $ $38.52      \\
  778 & NGC~4377   & 17.78  &  128$\pm$12  &   7.40$\pm$0.38    &    $-$18.63  &   7.02$\pm$0.80   &   $ $38.56      \\
 1321 & NGC~4489   & 15.42  &   57$\pm$ 5  &   5.73$\pm$0.44    &    $-$18.20  &   6.62$\pm$0.81   &   $<$38.33      \\
  828 & NGC~4387   & 17.95  &  100$\pm$10  &   6.89$\pm$0.39    &    $-$18.38  &   6.79$\pm$0.81   &   $ $38.30$^*$  \\
 1250 & NGC~4476   & 17.62  &   63$\pm$ 6  &   5.93$\pm$0.43    &    $-$18.32  &   6.73$\pm$0.81   &   $ $38.73      \\
 1630 & NGC~4551   & 16.14  &  102$\pm$10  &   6.93$\pm$0.39    &    $-$18.21  &   6.63$\pm$0.81   &   $<$38.29      \\
 1146 & NGC~4458   & 16.37  &   97$\pm$ 9  &   6.83$\pm$0.39    &    $-$18.23  &   6.64$\pm$0.81   &   $<$38.33      \\
 1025 & NGC~4434   & 22.44  &  116$\pm$11  &  (7.20$\pm$0.38)   &    $-$18.81  &  (7.19$\pm$0.80)  &   $ $38.92      \\
 1303 & NGC~4483   & 16.75  &  100$\pm$10  &   6.89$\pm$0.39    &    $-$18.05  &   6.48$\pm$0.81   &   $<$38.15      \\
 1913 & NGC~4623   & 17.38  &   77$\pm$ 7  &   6.35$\pm$0.41    &    $-$18.04  &   6.47$\pm$0.81   &   $<$38.46      \\
 1327 & NGC~4486A  & 18.28  &  131$\pm$13  &  (7.45$\pm$0.38)   &    $-$18.14  &  (6.57$\pm$0.81)  &   $ $38.68      \\
 1125 & NGC~4452   & 16.50  &  100$\pm$10  &   6.89$\pm$0.39    &    $-$18.33  &   6.74$\pm$0.81   &   $<$38.46      \\
 1475 & NGC~4515   & 16.60  &   82$\pm$ 8  &   6.48$\pm$0.40    &    $-$17.91  &   6.35$\pm$0.82   &   $<$38.34      \\
 1178 & NGC~4464   & 15.85  &  125$\pm$12  &   7.35$\pm$0.38    &    $-$17.62  &   6.07$\pm$0.83   &   $ $38.66      \\
 1283 & NGC~4479   & 17.38  &   80$\pm$ 8  &   6.43$\pm$0.41    &    $-$17.91  &   6.34$\pm$0.82   &   $ $38.54      \\
 1261 & NGC~4482   & 18.11  &   40$\pm$ 4  &   4.99$\pm$0.48    &    $-$17.73  &   6.18$\pm$0.83   &   $<$38.42      \\
  698 & NGC~4352   & 18.71  &   68$\pm$ 6  &   6.09$\pm$0.42    &    $-$17.96  &   6.40$\pm$0.82   &   $<$38.44      \\
	\hline
	\end{tabular}

Column~1: Virgo Cluster Catalog number (Binggeli et al.\ 1985).
Column~2: Other identification.  $^a$ The first seven galaxies have partially
depleted cores (see Section~\ref{Sec_mags}). 
Column~3: Distance from Mei et al.\ (2007) via  Gallo et al.\ (2010). 
Column~4: Velocity dispersion from HyperLeda.
Column~5: Predicted black hole mass (in units of solar masses) based on the
velocity dispersion (equation~\ref{eq1b} and \ref{eq1c}). 
The 11 galaxies with directly measured black hole masses (available from 
Table~1) have had their predicted masses placed in parenthesis. 
$^b$ Likely to be an underestimate (see the final paragraph of section~\ref{Sec_sig}). 
Column~6: (Galactic extinction)-corrected $B$-band absolute magnitude derived from the
RC3's apparent magnitude. 
Refer to section~\ref{Sec_strip} for six potentially stripped galaxies whose magnitudes, and in
turn predicted black hole masses, may be low. 
Column~7: Predicted black hole mass (in units of solar masses) based on the
$B$-band absolute magnitude (equations~\ref{eq3} and \ref{eq4}).  
Column~8: Central $X$-ray luminosity between 0.3 and 10 keV, in ergs per
second and corrected for absorption.  $^c$ Two galaxies have X-ray source $\approx 2\arcsec$
from their optical centre. 
All nuclear point-source luminosities have been taken from Table~1 of Gallo et al.\ 2010, except
for 7 galaxies where an asterisk indicates either a new X-ray detection or an
updated value from our work (sections~\ref{Sec_X} and \ref{Sec_IMBH}).
	\label{Tab_App1}
\end{table*}

\begin{table*}
	\centering
	\caption{Continued.} 
	\begin{tabular}{rlc|rc|cc|c} 
	\hline
VCC \# & Other    &  Dist.  &  $\sigma$   & $\log M_{\rm bh}(\sigma)$ & $\mathfrak{M}_B$ &    $\log M_{\rm bh}(\mathfrak{M}_B)$  &  $\log L_X$  \\
       &          &  [Mpc]  & km s$^{-1}$ &         [dex]          &    [mag]  &    [dex]                  &  [dex]     \\  
 1 & 2 & 3 & 4 & 5 & 6 & 7 & 8 \\
	\hline
 1422 & IC~3468   & 15.35   &   33$\pm$ 3  &   4.60$\pm$0.50    &    $-$16.90  &   5.40$\pm$0.87   &   $<$38.08      \\
 2048 & IC~3773   & 16.50   &   59$\pm$ 5  &   5.80$\pm$0.43    &    $-$17.25  &   5.73$\pm$0.85   &   $<$38.12      \\
 1871 & IC~3653   & 15.49   &   46$\pm$ 4  &   5.28$\pm$0.46    &    $-$16.26  &   4.79$\pm$0.91   &   $<$38.08      \\
    9 & IC~3019   & 17.14   &      ...     &      ...           &    $-$17.11  &   5.60$\pm$0.86   &   $<$38.15      \\
  575 & NGC~4318   & 22.08  &   91$\pm$ 9  &   6.69$\pm$0.40    &    $-$17.61  &   6.06$\pm$0.83   &   $<$38.39      \\
 1910 & IC~809    & 16.07   &      ...     &      ...           &    $-$16.64  &   5.15$\pm$0.88   &   $<$38.30      \\
 1049 & U7580   & 16.00     &   26$\pm$ 2  &   4.10$\pm$0.54    &    $-$16.17  &   4.71$\pm$0.92   &   $<$38.08      \\
  856 & IC~3328   & 16.83   &   33$\pm$ 3  &   4.60$\pm$0.50    &    $-$16.56  &   5.07$\pm$0.89   &   $<$38.16      \\
  140 & IC~3065   & 16.37   &      ...     &      ...           &    $-$16.70  &   5.21$\pm$0.88   &   $<$38.13      \\
 1355 & IC~3442   & 16.90   &      ...     &      ...           &    $-$16.79  &   5.30$\pm$0.87   &   $ $38.58      \\
 1087 & IC~3381   & 16.67   &   39$\pm$ 3  &   4.94$\pm$0.48    &    $-$16.79  &   5.29$\pm$0.88   &   $<$38.15      \\
 1297 & NGC~4486B  & 16.29  &  166$\pm$16  &  (7.94$\pm$0.37)   &    $-$16.78  &  (5.28$\pm$0.88)  &   $ $38.42      \\
 1861 & IC~3652   & 16.14   &      ...     &      ...           &    $-$16.47  &   4.99$\pm$0.90   &   $<$38.12      \\
  543 & U7436   & 15.70     &   30$\pm$ 3  &   4.40$\pm$0.52    &    $-$16.75  &   5.26$\pm$0.88   &   $<$38.27      \\
 1431 & IC~3470   & 16.14   &      ...     &      ...           &    $-$16.94  &   5.43$\pm$0.87   &   $<$38.66      \\
 1528 & IC~3501   & 16.29   &      ...     &      ...           &    $-$16.59  &   5.11$\pm$0.89   &   $<$38.13      \\
 1695 & IC~3586   & 16.52   &      ...     &      ...           &    $-$16.85  &   5.35$\pm$0.87   &   $<$38.14      \\
 1833 & ...     & 16.22     &      ...     &      ...           &    $-$16.52  &   5.04$\pm$0.89   &   $<$38.11      \\
  437 & U7399A  & 17.14     &   47$\pm$ 4  &   5.33$\pm$0.46    &    $-$17.23  &   5.70$\pm$0.85   &   $<$38.17      \\
 2019 & IC~3735   & 17.06   &   37$\pm$ 3  &   4.83$\pm$0.49    &    $-$16.40  &   4.92$\pm$0.90   &   $<$38.17      \\
   33 & IC~3032   & 15.07   &      ...     &      ...           &    $-$16.29  &   4.82$\pm$0.91   &   $<$38.25      \\
  200 & ...     & 18.20     &      ...     &      ...           &    $-$16.40  &   4.93$\pm$0.90   &   $<$38.42      \\
  571 & ...     & 23.77     &      ...     &      ...           &    $-$16.94  &   5.43$\pm$0.87   &   $<$38.46      \\
   21 & IC~3025   & 16.50   &      ...     &      ...           &    $-$15.79  &   4.36$\pm$0.95   &   $<$38.11      \\
 1488 & IC~3487   & 16.50   &   29$\pm$ 2  &   4.33$\pm$0.52    &    $-$16.13  &   4.68$\pm$0.92   &   $<$38.14      \\
 1779 & IC~3612   & 16.50   &      ...     &      ...           &    $-$16.22  &   4.76$\pm$0.92   &   $<$38.14      \\
 1895 & U7854   & 15.85     &      ...     &      ...           &    $-$15.91  &   4.47$\pm$0.94   &   $<$38.10      \\
 1499 & IC~3492   & 16.50   &      ...     &      ...           &    $-$16.11  &   4.65$\pm$0.92   &   $ $38.42$^c$  \\
 1545 & IC~3509   & 16.83   &      ...     &      ...           &    $-$16.33  &   4.86$\pm$0.91   &   $<$38.16      \\
 1192 & NGC~4467   & 16.50  &   67$\pm$ 6  &   6.06$\pm$0.42    &    $-$16.40  &   4.93$\pm$0.90   &   $ $38.65$^*$  \\
 1857 & IC~3647   & 16.50   &      ...     &      ...           &    $-$16.68  &   5.19$\pm$0.88   &   $<$38.14      \\
 1075 & IC~3383   & 16.14   &   33$\pm$ 3  &   4.60$\pm$0.50    &    $-$15.94  &   4.49$\pm$0.94   &   $<$38.12      \\
 1948 & ...     & 16.50     &      ...     &      ...           &    $-$15.42  &   4.00$\pm$0.98   &   $<$38.14      \\
 1627 & ...     & 15.63     &      ...     &      ...           &    $-$15.80  &   4.36$\pm$0.95   &   $<$38.34      \\
 1440 & IC~798    & 16.00   &      ...     &      ...           &    $-$16.22  &   4.76$\pm$0.92   &   $<$38.27      \\
  230 & IC~3101   & 17.78   &      ...     &      ...           &    $-$15.67  &   4.24$\pm$0.96   &   $<$38.62      \\
 2050 & IC~3779   & 15.78   &   43$\pm$ 4  &   5.14$\pm$0.47    &    $-$15.70  &   4.27$\pm$0.96   &   $<$38.10      \\
 1993 & ...     & 16.52     &      ...     &      ...           &    $-$15.39  &   3.98$\pm$0.99   &   $<$38.14      \\
  751 & IC~3292   & 15.78   &      ...     &      ...           &    $-$16.25  &   4.78$\pm$0.91   &   $ $38.30$^{*,c}$  \\
 1828 & IC~3635   & 16.83   &      ...     &      ...           &    $-$16.26  &   4.80$\pm$0.91   &   $<$38.16      \\
  538 & NGC~4309A  & 22.91  &      ...     &      ...           &    $-$15.70  &   4.27$\pm$0.96   &   $<$38.41      \\
 1407 & IC~3461   & 16.75   &   28$\pm$ 2  &   4.26$\pm$0.53    &    $-$16.20  &   4.73$\pm$0.92   &   $<$38.35      \\
 1886 & ...     & 16.50     &      ...     &      ...           &    $-$15.71  &   4.27$\pm$0.96   &   $<$38.14      \\
 1199 & ...     & 16.50     &      ...     &      ...           &    $-$15.17  &   3.77$\pm$1.01   &   $<$37.10$^*$  \\
 1743 & IC~3602   & 17.62   &   27$\pm$ 2  &   4.18$\pm$0.53    &    $-$15.57  &   4.15$\pm$0.97   &   $<$38.20      \\
 1539 & ...     & 16.90     &      ...     &      ...           &    $-$15.90  &   4.45$\pm$0.94   &   $<$38.16      \\
 1185 & ...     & 16.90     &      ...     &      ...           &    $-$15.84  &   4.40$\pm$0.95   &   $<$38.36      \\
 1826 & IC~3633   & 16.22   &   22$\pm$ 2  &   3.76$\pm$0.56    &    $-$15.33  &   3.92$\pm$0.99   &   $<$38.12      \\
 1512 & ...     & 18.37     &      ...     &      ...           &    $-$15.77  &   4.34$\pm$0.95   &   $<$38.23      \\
 1489 & IC~3490   & 16.50   &      ...     &      ...           &    $-$15.32  &   3.91$\pm$0.99   &   $<$38.14      \\
 1661 & ...     & 15.85     &      ...     &      ...           &    $-$16.27  &   4.81$\pm$0.91   &   $<$38.12      \\
	\hline
	\end{tabular}
	\label{Tab_App2}
\end{table*}


\section{X-ray spectral properties of three nuclear sources}

\begin{table*}
\centering
\caption{Best-fitting parameters for the nuclear X-ray sources in NGC~4382,
  NGC~4472 and NGC~4467. The model is {\it tbabs} $\times$ {\it tbabs}
  $\times$ {\it powerlaw} for NGC~4382 and NGC~4467, and {\it tbabs} $\times$
  {\it tbabs} $\times$ ({\it powerlaw} $+$ {\it mekal}) for NGC~4472.}
\begin{tabular}{lccc}
\hline
Parameter  & \multicolumn{3}{c}{Galaxy} \\
           &     NGC~4382 & NGC~4472 &  NGC~4467   \\
 1 & 2 & 3 & 4  \\ 
        \hline
 $N_{\rm {H, Gal}}$ ($10^{20}$ cm$^{-2}$) &  [2.5] & [1.6] &  [1.6] \\[3pt]
 $N_{\rm {H, int}}$ ($10^{20}$ cm$^{-2}$) &  $< 9.4$ & $2.1^{+6.3}_{-2.1}$ & $< 7.7$ \\[3pt]
 $\Gamma$    & $1.43^{+0.23}_{-0.23}$ & $2.39^{+0.37}_{-0.29}$ & $1.44^{+0.18}_{-0.18}$ \\[3pt]
 $N_{\rm {pl}}$ ($10^{-6}$ photons keV$^{-1}$ cm$^{-2}$ s$^{-1}$ at 1 keV) & $2.3^{+0.5}_{-0.4}$ & $4.3^{+1.8}_{-0.9}$  &  $2.1^{+0.3}_{-0.3}$ \\[3pt]
 $kT$ (keV) & ... & $0.59^{+0.12}_{-0.14}$ &  ... \\[3pt]
 $N_{\rm {mek}}^a$ $\left(10^{-6}\right)$ & ... & $1.8^{+0.7}_{-0.7}$ & ... \\[3pt]
 \hline
 C-stat  & 0.69 (77.7/112) & 0.98 (199.8/203) &  0.77 (139.3/181) \\[3pt]
  $F_{0.3-10}$ ($10^{-14}$ erg cm$^{-2}$ s$^{-1}$) & $2.0^{+0.5}_{-0.4}$ & $2.2^{+0.3}_{-0.2}$  &  $1.8^{+0.3}_{-0.2}$ \\[3pt]
  $L_{0.3-10}$ ($10^{38}$ erg s$^{-1}$) & $8.0^{+1.8}_{-1.4}$ & $9.2^{+1.0}_{-1.6}$ &  $6.1^{+1.0}_{-0.8}$ \\[3pt]
 \hline
\end{tabular}

$^a$ $N_{\rm {mek}} = 10^{-14}/(4\pi d^2) \int n_e n_{\rm H} dV$ where $d$ is
    the distance to the source (cm), and $n_e$ and $n_H$ are the electron and
    H densities (cm$^{-3}$)
        \label{Tab_App3}
\end{table*}

\begin{figure*}
\includegraphics[angle=270, trim=0cm 2.2cm 0.cm 2.5cm, width=5.8cm]{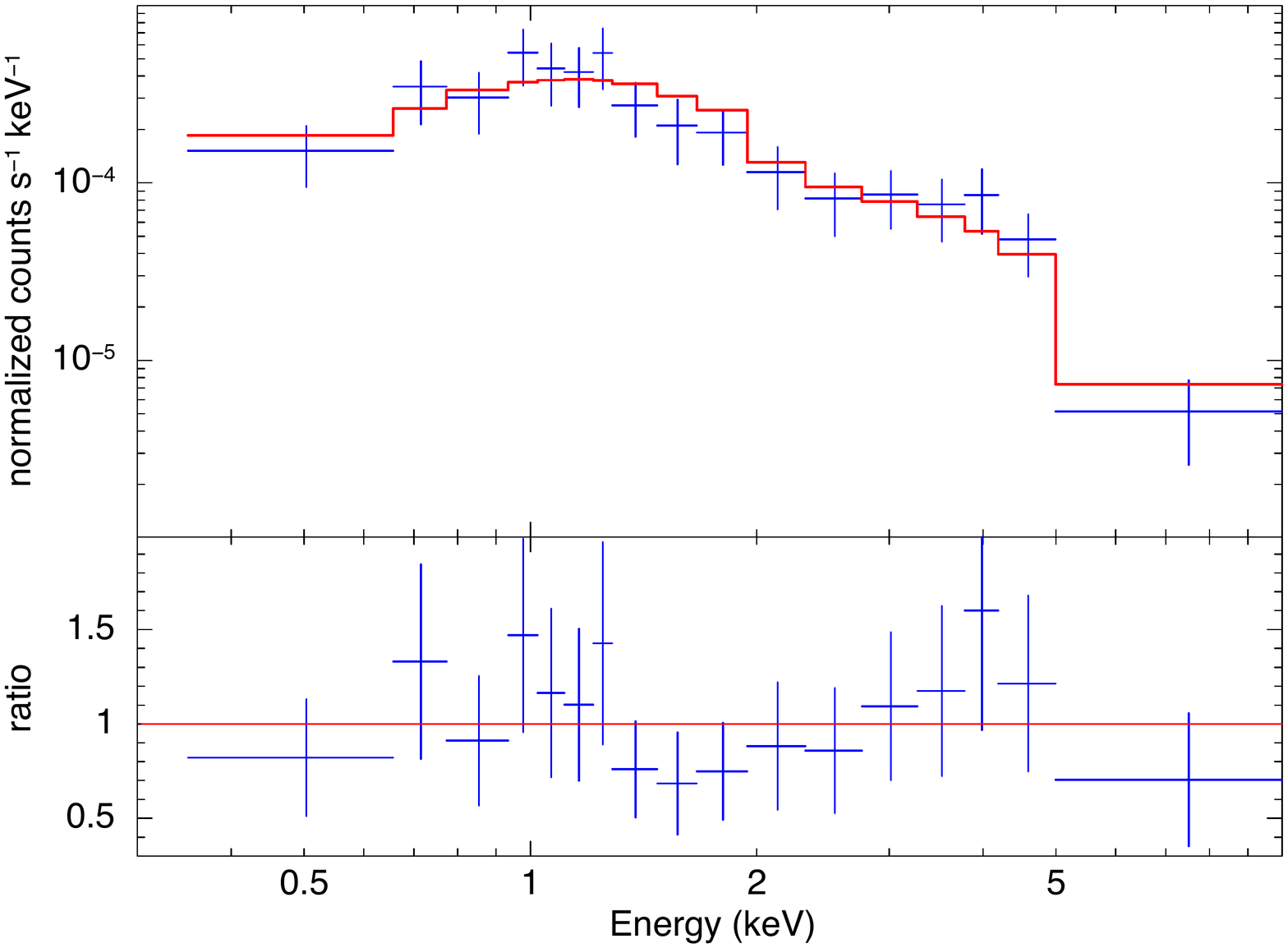}
\includegraphics[angle=270, trim=0cm 2.2cm 0.cm 2.5cm, width=5.8cm]{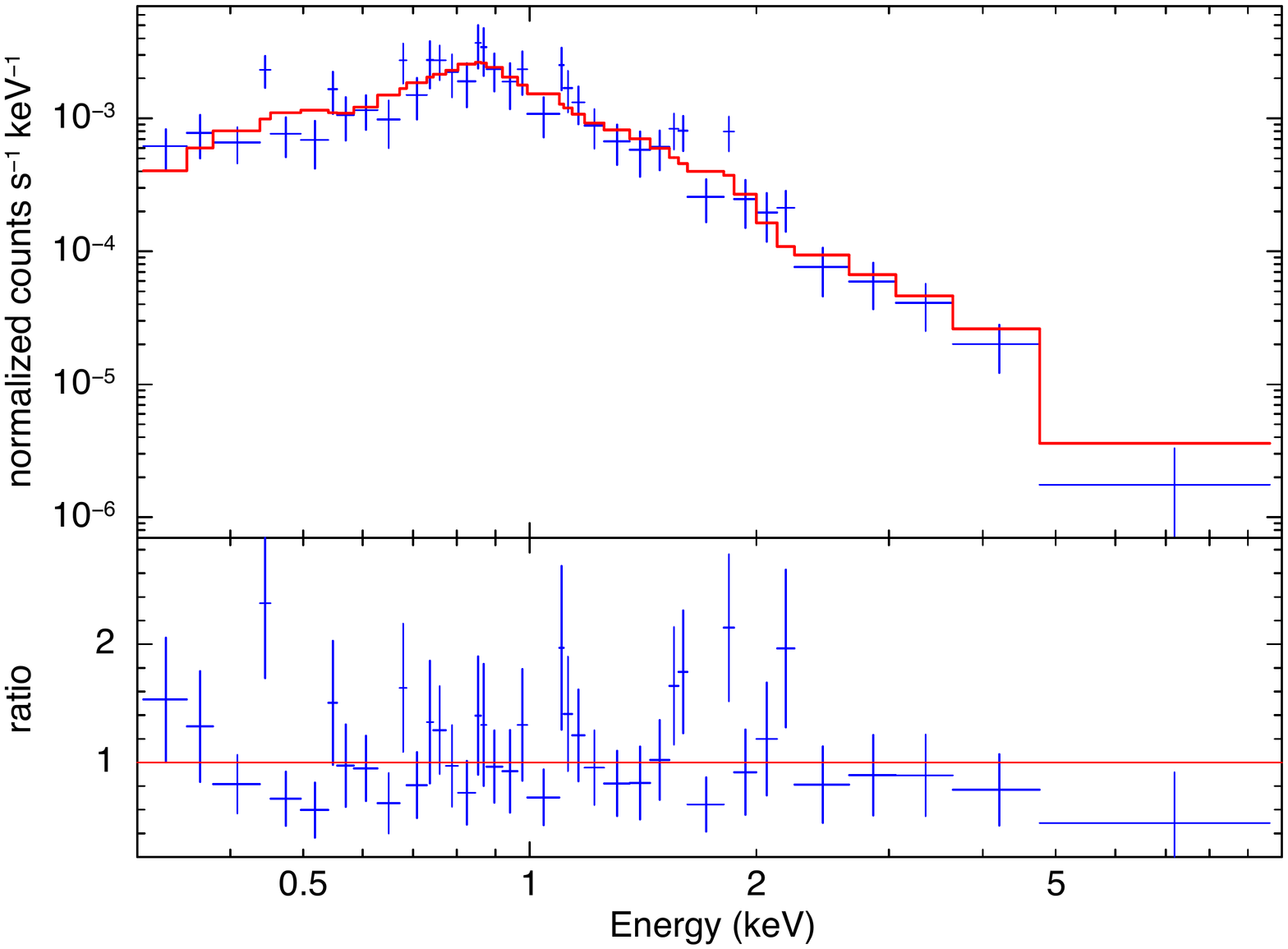}
\includegraphics[angle=270, trim=0cm 2.2cm 0.cm 2.5cm, width=5.8cm]{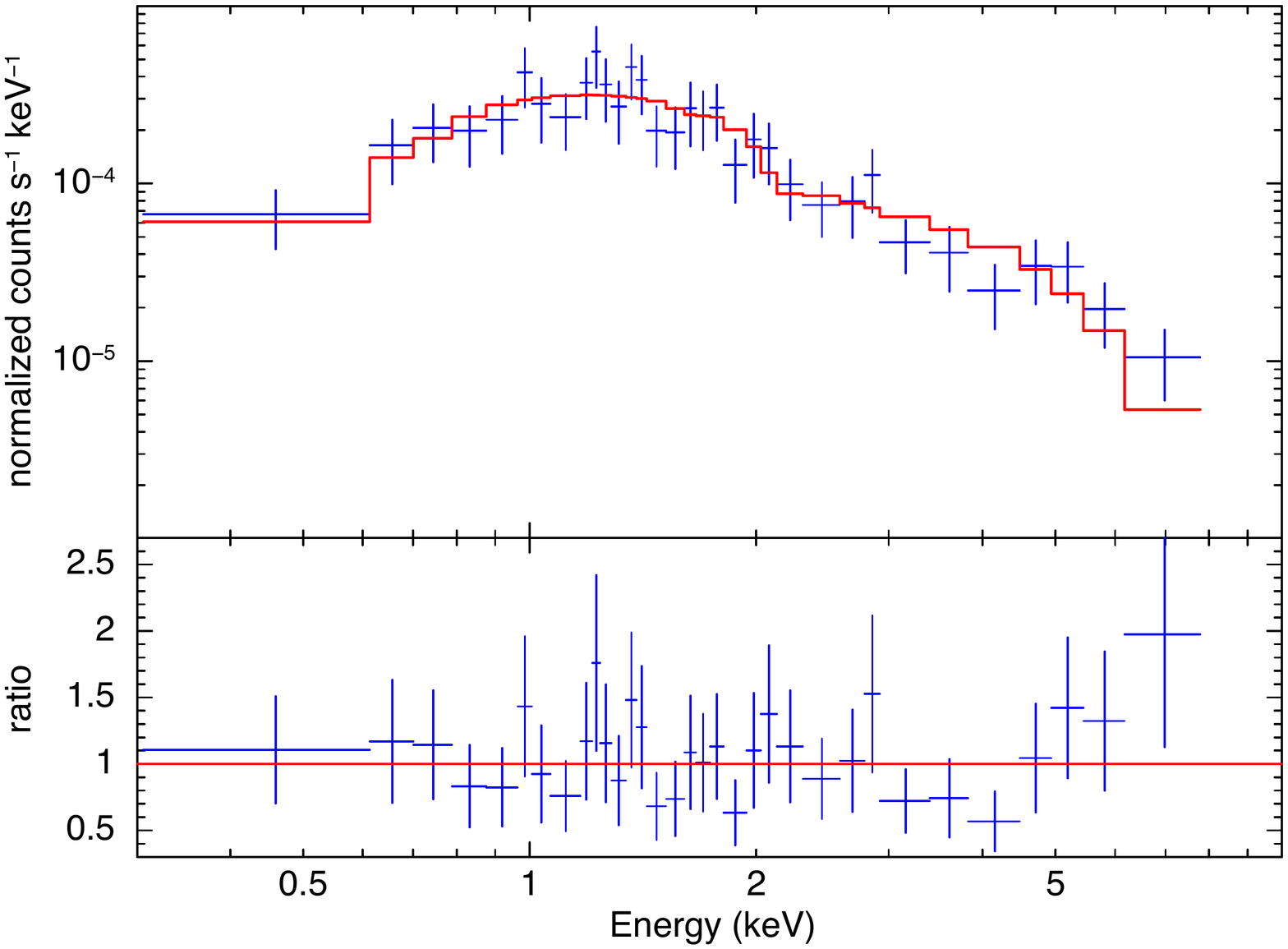}
    \caption{ Left panel: {\it Chandra}/ACIS-S spectrum of the nuclear source
      in NGC~4382, fitted with an absorbed power-law model, and corresponding
      residuals. See Table~\ref{Tab_App3} for the fit parameters. The data
      have been binned to a minimum signal-to-noise ratio of 2.5 for display
      purposes only. Middle panel: similar to the left panel, but for the
      nuclear source in NGC~4472; the model is an absorbed power-law plus
      optically thin thermal plasma. Right panel: similar to the left panel,
      but for the nuclear source in NGC~4467, fit with a power-law model.
    }
\label{FigA1}
\end{figure*}


\bsp	
\label{lastpage}

\begin{thebibliography}{999}

  \bibitem[Abbott et al.(2016)]{2016PhRvL.116f1102A} Abbott, B.~P., Abbott, R.,
  Abbott, T.~D., et al.\ 2016, Physical Review Letters, 116, 061102 
 \bibitem[Abbott et al.(2017)]{2017PhRvL.118v1101A} Abbott, B.~P., Abbott, R.,
  Abbott, T.~D., et al.\ 2017, Physical Review Letters, 118, 221101 
 \bibitem[Akritas \& Bershady(1996)]{AaB96}Akritas, M.G., Bershady, M.A.\ 1996, ApJ, 470, 706
 \bibitem[Alam et al.(2015)]{2015ApJS..219...12A} Alam, S., Albareti, F.~D.,
  Allende Prieto, C., et al.\ 2015, \apjs, 219, 12
 \bibitem[Amaro-Seoane et al.(2007)]{2007CQGra..24R.113A} Amaro-Seoane, P.,
  Gair, J.~R., Freitag, M., et al.\ 2007, Classical and Quantum Gravity, 24,
  R113 
 \bibitem[Anza et al.(2005)]{Anza:2005} Anza, S., Armano, M., Balaguer, E., et
   al.\ 2005, Classical and Quantum Gravity, 22, 125
\bibitem[Arnaud(1996)]{1996ASPC..101...17A} Arnaud K.A.\ 1996, ASPC, 101, 17
 \bibitem[Arrigoni Battaia et al.(2012)]{2012A&A...543A.112A} Arrigoni Battaia,
  F., Gavazzi, G., Fumagalli, M., et al.\ 2012, \aap, 543, A112 
 \bibitem[Aso et al.(2013)]{2013PhRvD..88d3007A} Aso, Y., Michimura, Y.,
  Somiya, K., et al.\ 2013, Physical Review D, 88, 043007 
 \bibitem[Bahcall et a.(1972)]{BKS72}Bahcall J.N., Kozlovsky B.Z., Salpeter E.E., 1972, ApJ, 171, 467
 \bibitem[Balcells et al.(2007)]{2007ApJ...665.1084B} Balcells, M., Graham,
  A.~W., \& Peletier, R.~F.\ 2007, \apj, 665, 1084 
 \bibitem[Baldassare et al.(2015)]{2015ApJ...809L..14B} Baldassare, V.~F.,
   Reines, A.~E., Gallo, E., \& Greene, J.~E.\ 2015, \apjl, 809, L14
 \bibitem[Baldwin et al.(1981)]{1981PASP...93....5B} Baldwin, J.~A., Phillips,
   M.~M., \& Terlevich, R.\ 1981, \pasp, 93, 5 
 \bibitem[Ballone et al.(2018)]{2018MNRAS.480.4684B} Ballone, A., Mapelli, M.,
  \& Pasquato, M.\ 2018, \mnras, 480, 4684 
 \bibitem[Barack et al.(2018)]{2018arXiv180605195B} Barack, L., Cardoso, V.,
   Nissanke, S., et al.\ 2018, arXiv:1806.05195 
 \bibitem[Belczynski et al.(2010)]{2010ApJ...714.1217B} Belczynski, K., Bulik,
  T., Fryer, C.~L., et al.\ 2010, \apj, 714, 1217 
 \bibitem[Bell \& de Jong(2001)]{2001ApJ...550..212B} Bell, E.~F., \& de Jong,
   R.~S.\ 2001, \apj, 550, 212 
 \bibitem[Bell et al.(2003)]{2003ApJS..149..289B} Bell, E.~F., McIntosh, D.~H.,
  Katz, N., \& Weinberg, M.~D.\ 2003, \apjs, 149, 289 
 \bibitem[Bernardi et al.(2007)]{2007AJ....133.1741B} Bernardi, M., Hyde,
   J.~B., Sheth, R.~K., Miller, C.~J., \& Nichol, R.~C.\ 2007, \aj, 133, 1741 
 \bibitem[Binggeli et al.(2000)]{2000A&A...359..447B} Binggeli, B., Barazza,
   F., \& Jerjen, H.\ 2000, \aap, 359, 447 
 \bibitem[Bland-Hawthorn \& Cohen(2003)]{2003ApJ...582..246B} Bland-Hawthorn,
  J., \& Cohen, M.\ 2003, \apj, 582, 246 
 \bibitem[Bland-Hawthorn(2015)]{2015IAUGA..2301410B} Bland-Hawthorn, J.\ 2015,
  IAU General Assembly Meeting 29, id.2301410 
 \bibitem[Bonfini et al.(2018)]{2018MNRAS.478.1161B} Bonfini, P.,
  Gonz{\'a}lez-Mart{\'{\i}}n, O., Fritz, J., et al.\ 2018, \mnras, 478, 1161 
 \bibitem[Boselli et al.(2008)]{Bos09}Boselli, A., Boissier, S., Cortese, L.,
   Gavazzi, G.\ 2008, A\&A, 489, 1015
 \bibitem[Caballero-Garcia et al.(2018)]{2018arXiv180207149C}
   Caballero-Garcia, M.~D., Fabrika, S., Castro-Tirado, A.~J., et al.\ 2018,
   arXiv:1802.07149
 \bibitem[Cappellari et al.(2008)]{2008IAUS..245..215C} Cappellari, M., Bacon,
   R., Davies, R.~L., et al.\ 2008, Formation and Evolution of Galaxy Bulges,
   245, 215
\bibitem[Cash(1979)]{1979ApJ...228..939C} Cash W.\ 1979, \apj, 228, 939
 \bibitem[Chen et al.(2010)]{2010ApJS..191....1C} Chen, C.-W., C{\^o}t{\'e},
   P., West, A.~A., Peng, E.~W., \& Ferrarese, L.\ 2010, \apjs, 191, 1 
 \bibitem[Cheng et al.(1997)]{1997ApJ...481L..43C} Cheng, L.~X., Leventhal,
   M., Smith, D.~M., et al.\ 1997, \apjl, 481, L43 
 \bibitem[Chevalier(1992)]{1992ApJ...397L..39C} Chevalier, R.~A.\ 1992, \apjl,
   397, L39 
 \bibitem[Chilingarian et al.(2009)]{2009Sci...326.1379C} Chilingarian, I.,
  Cayatte, V., Revaz, Y., et al.\ 2009, Science, 326, 1379 
 \bibitem[Chilingarian et al.(2018)]{2018ApJ...863....1C} Chilingarian, I.~V.,
   Katkov, I.~Y., Zolotukhin, I.~Y., et al.\ 2018, \apj, 863, 1 
 \bibitem[Chilingarian \& Mamon(2008)]{2008MNRAS.385L..83C} Chilingarian,
  I.~V., \& Mamon, G.~A.\ 2008, \mnras, 385, L83 
\bibitem[Chilingarian \& Zolotukhin(2015)]{2015Sci...348..418C} Chilingarian,
  I., \& Zolotukhin, I.\ 2015, Science, 348, 418 
 \bibitem[Colbert \& Mushotzky(1999)]{1999ApJ...519...89C} Colbert, E.~J.~M.,
   \& Mushotzky, R.~F.\ 1999, \apj, 519, 89
 \bibitem[C{\^o}t{\'e} et al.(2004)]{2004ApJS..153..223C} C{\^o}t{\'e}, P.,
   Blakeslee, J.~P., Ferrarese, L., et al.\ 2004, \apjs, 153, 223 
\bibitem[Coziol et al.(2011)]{Coziol2011} Coziol, R., Torres-Papaqui, J.P.,
  Plauchu-Frayn, I., et al. 2011, RMxAA, 47, 361
 \bibitem[Crocker et al.(2015)]{2015ApJ...808..107C} Crocker, R.~M., Bicknell,
  G.~V., Taylor, A.~M., \& Carretti, E.\ 2015, \apj, 808, 107 
 \bibitem[Crowther et al.(2010)]{2010MNRAS.408..731C} Crowther, P.~A., Schnurr,
  O., Hirschi, R., et al.\ 2010, \mnras, 408, 731 
 \bibitem[Cseh et al.(2015)]{2015MNRAS.446.3268C} Cseh, D., Webb, N.~A., Godet,
   O., et al.\ 2015, \mnras, 446, 3268 
 \bibitem[Czerny et al.(2013)]{2013A&A...555A..97C} Czerny, B., Kunneriath,
   D., Karas, V., \& Das, T.~K.\ 2013, \aap, 555, A97 
 \bibitem[Davis et al.(2018a)]{2018aDavis} Davis, B.~L., Graham,
  A.~W., \& Cameron, E.\ 2018a, ApJ, in press, arXiv:1810.04887
 \bibitem[Davis et al.(2018b)]{2018bDavis} Davis, B.~L., Graham,
  A.~W., \& Cameron, E.\ 2018b, ApJ, in press, arXiv:1810.04888
 \bibitem[Davis et al.(2017)]{2017MNRAS.471.2187D} Davis, B.~L., Graham, A.~W.,
  \& Seigar, M.~S.\ 2017, \mnras, 471, 2187 
 \bibitem[Davies et al.(1983)]{1983ApJ...266...41D} Davies, R.~L., Efstathiou,
   G., Fall, S.~M., Illingworth, G., \& Schechter, P.~L.\ 1983, \apj, 266, 41
 \bibitem[De Looze et al.(2010)]{2010A&A...518L..54D} De Looze, I., Baes, M.,
  Zibetti, S., et al.\ 2010, \aap, 518, L54 
 \bibitem[de Rijcke et al.(2005)]{2005A&A...438..491D} de Rijcke, S.,
   Michielsen, D., Dejonghe, H., Zeilinger, W.~W., \& Hau, G.~K.~T.\ 2005,
   \aap, 438, 491 
 \bibitem[de Vaucouleurs et al.(1991)]{deV91} de Vaucouleurs, G., de
   Vaucouleurs, A., Corwin, H.~G.~Jr, Buta R.~J., Paturel G., \& Fouque
   P.\ 1991, Third Reference Catalogue of Bright Galaxies. Springer-Verlag,
   Berlin (RC3) 
 \bibitem[di Serego Alighieri et al.(2013)]{2013A&A...552A...8D} di Serego
  Alighieri, S., Bianchi, S., Pappalardo, C., et al.\ 2013, \aap, 552, A8 
 \bibitem[Dressler(1989)]{1989IAUS..134..217D} Dressler, A.\ 1989, Active
  Galactic Nuclei, IAU Symp.\ 134, 217
 \bibitem[Driver et al.(2008)]{2008ApJ...678L.101D} Driver, S.~P., Popescu,
   C.~C., Tuffs, R.~J., et al.\ 2008, \apjl, 678, L101 
 \bibitem[Einstein(1916)]{1916SPAW.......688E} Einstein, A., 1916, Preuss. Akad. Wiss. Berlin, Sitzungsber., 688
 \bibitem[Einstein(1918)]{1918SPAW.......154E} Einstein, A., 1918, Preuss. Akad. Wiss. Berlin, Sitzungsber., 154
 \bibitem[Faber \& Jackson(1976)]{1976ApJ...204..668F} Faber, S.~M., \&
   Jackson, R.~E.\ 1976, \apj, 204, 668 
 \bibitem[Falcke et al.(2004)]{2004A&A...414..895F} Falcke, H., K{\"o}rding,
  E., \& Markoff, S.\ 2004, \aap, 414, 895
 \bibitem[Farrell et al.(2009)]{Far09}Farrell, S.A., Webb, N.A., Barret, D.,
   Godet, O., \& Rodrigues, J.M.\ 2009, Nature, 460, 73
 \bibitem[Farrell et al.(2014)]{2014MNRAS.437.1208F} Farrell, S.~A.,
   Servillat, M., Gladstone, J.~C., et al.\ 2014, \mnras, 437, 1208
 \bibitem[Feng \& Soria(2011)]{feng11} Feng H., Soria R.\ 2011, NewAR, 55, 166 
 \bibitem[Ferrarese et al.(2006)]{2006ApJS..164..334F} Ferrarese, L.,
  C{\^o}t{\'e}, P., Jord{\'a}n, A., et al.\ 2006, \apjs, 164, 334 
 \bibitem[Ferrarese \& Ford(2005)]{2005SSRv..116..523F} Ferrarese, L., \&
   Ford, H.\ 2005, \ssr, 116, 523 
 \bibitem[Ferrarese \& Merritt(2000)]{2000ApJ...539L...9F} Ferrarese, L., \&
   Merritt, D.\ 2000, \apjl, 539, L9 
 \bibitem[Forbes et al.(2008)]{2008MNRAS.389.1924F} Forbes, D.~A., Lasky, P.,
  Graham, A.~W., \& Spitler, L.\ 2008, \mnras, 389, 1924 
 \bibitem[Fragione et al.(2018)]{2018arXiv180608385F} Fragione, G., Leigh, N.,
   Ginsburg, I., \& Kocsis, B.\ 2018, arXiv:1806.08385 
 \bibitem[Frank \& Rees(1976)]{1976MNRAS.176..633F} Frank, J., \& Rees,
  M.~J.\ 1976, \mnras, 176, 633 
\bibitem[Fruscione et al.(2006)]{2006SPIE.6270E..1VF} Fruscione, A., McDowell,
  J.~C., Allen, G.~E., et al.\ 2006, \procspie, 6270, 62701V
\bibitem[Gaia Collaboration et al.(2018)]{2018A&A...616A...1G} Gaia
  Collaboration, Brown, A.~G.~A., Vallenari, A., et al.\ 2018, \aap, 616, A1
 \bibitem[Gallo et al.(2008)]{2008ApJ...680..154G} Gallo, E., Treu, T., Jacob,
   J., et al.\ 2008, \apj, 680, 154
 \bibitem[Gallo et al.(2010)]{2010ApJ...714...25G} Gallo, E., Treu, T.,
   Marshall, P.~J., et al.\ 2010, \apj, 714, 25 
 \bibitem[Gavazzi et al.(2003)]{2003A&A...400..451G} Gavazzi, G., Boselli, A.,
   Donati, A., Franzetti, P., \& Scodeggio, M.\ 2003, \aap, 400, 451
 \bibitem[Gebhardt et al.(2011)]{Geb11}Gebhardt, K., Adams, J., Richstone, D., et al.\ 2011, ApJ, 729, 119
 \bibitem[Gebhardt et al.(2000)]{2000ApJ...539L..13G} Gebhardt, K., Bender, R.,
  Bower, G., et al.\ 2000, \apjl, 539, L13 
 \bibitem[Gendron-Marsolais et al.(2017)]{2017ApJ...848...26G}
  Gendron-Marsolais, M., Kraft, R.~P., Bogdan, A., et al.\ 2017, \apj, 848, 26 
 \bibitem[Gonz{\'a}lez Delgado et al.(2008)]{2008AJ....135..747G} Gonz{\'a}lez
  Delgado, R.~M., P{\'e}rez, E., Cid Fernandes, R., \& Schmitt, H.\ 2008, \aj,
  135, 747 
 \bibitem[Graham(2002)]{2002ApJ...568L..13G} Graham, A.~W.\ 2002, \apjl, 568,
  L13 
 \bibitem[Graham(2007)]{2007MNRAS.379..711G} Graham, A.~W.\ 2007, \mnras, 379,
   711 
 \bibitem[Graham(2012)]{2012ApJ...746..113G} Graham, A.~W.\ 2012, \apj, 746,
  113 
 \bibitem[Graham(2013)]{2013pss6.book...91G}Graham, 2013, in ``Planets, Stars
   and Stellar Systems'', Volume 6, p.91-140, T.D.Oswalt \& W.C.Keel (Eds.),
   Springer Publishing (arXiv:1108.0997)
 \bibitem[Graham(2016a)]{2016ASSL..418..263G} Graham, A.~W.\ 2016a, in Galactic
  Bulges, E. Laurikainen, R.F. Peletier, and D.A. Gadotti (eds.), Springer
  Publishing, Astrophysics and Space Science Library, 2016, v.418, p.263-313 
 \bibitem[Graham et al.(2016)]{2016ApJ...818..172G} Graham, A.~W., Ciambur,
  B.~C., \& Soria, R.\ 2016, \apj, 818, 172 
 \bibitem[Graham et al.(2018)]{2018arXiv181103232G} Graham, A.~W., Soria, R.,
   \& Davis, B.~L.\ 2018, arXiv:1811.03232 
 \bibitem[Graham \& Driver(2007)]{2007ApJ...655...77G} Graham, A.~W., \&
   Driver, S.~P.\ 2007, \apj, 655, 77 
 \bibitem[Graham et al.(2003)]{2003AJ....125.2951G} Graham, A.~W., Erwin, P.,
   Trujillo, I., \& Asensio Ramos, A.\ 2003, \aj, 125, 2951 
 \bibitem[Graham et al.(2011)]{2011MNRAS.412.2211G} Graham, A.~W., Onken,
  C.~A., Athanassoula, E., \& Combes, F.\ 2011, \mnras, 412, 2211 
 \bibitem[Graham \& Scott(2013)]{2013ApJ...764..151G} Graham, A.~W., \& Scott,
   N.\ 2013, \apj, 764, 151 
 \bibitem[Graham \& Scott(2015)]{2015ApJ...798...54G} Graham, A.~W., \& Scott,
  N.\ 2015, \apj, 798, 54 
 \bibitem[Graham \& Spitler(2009)]{2009MNRAS.397.2148G} Graham, A.~W., \&
  Spitler, L.~R.\ 2009, \mnras, 397, 2148 
\bibitem[Greenhill et al.(2003)]{2003ApJ...590..162G} Greenhill, L.~J., Booth,
  R.~S., Ellingsen, S.~P., et al.\ 2003, ApJ, 590, 162
 \bibitem[Grossi et al.(2015)]{2015A&A...574A.126G} Grossi, M., Hunt, L.~K.,
  Madden, S.~C., et al.\ 2015, \aap, 574, A126 
 \bibitem[G\"ultekin et al.(2009)]{Guilty}G\"ultekin K., Richstone, D.~O., Gebhardt, K., et al.\ 2009, ApJ, 698, 198
 \bibitem[G{\"u}ltekin et al.(2011)]{2011ApJ...741...38G} G{\"u}ltekin, K.,
  Richstone, D.~O., Gebhardt, K., et al.\ 2011, \apj, 741, 38 
 \bibitem[Held et al.(1992)]{Hel92}Held, E.V., de Zeeuw, T., Mould, J., Picard, A.\ 1992, AJ, 103, 851
 \bibitem[Hills(1975)]{1975Natur.254..295H} Hills, J.~G.\ 1975, \nat, 254, 295 
 \bibitem[Hils \& Bender(1995)]{1995ApJ...445L...7H} Hils, D., \& Bender,
   P.~L.\ 1995, \apjl, 445, L7 
 \bibitem[Hobbs et al.(2010)]{2010CQGra..27h4013H} Hobbs, G., Archibald, A.,
  Arzoumanian, Z., et al.\ 2010, Classical and Quantum Gravity, 27, 084013
 \bibitem[Hu(2008)]{2008MNRAS.386.2242H} Hu, J.\ 2008, \mnras, 386, 2242 
 \bibitem[Janz \& Lisker(2009)]{2009ApJ...696L.102J} Janz, J., \& Lisker,
  T.\ 2009, \apjl, 696, L102 
 \bibitem[Jarrett et al.(2000)]{2MASS}Jarrett, T.H., Chester, T., Cutri, R., et al.\ 2000, AJ, 119, 2498 (2MASS)
 \bibitem[Jiang et al.(2011)]{2011ApJ...737L..45J} Jiang, Y.-F., Greene, J.~E.,
  \& Ho, L.~C.\ 2011, \apjl, 737, L45 
 \bibitem[Jiang et al.(2018)]{2018arXiv181010283J} Jiang, N., Wang, T., Zhou,
   H., et al.\ 2018, ApJ, in press (arXiv:1810.10283) 
 \bibitem[Jim{\'e}nez et al.(2011)]{2011MNRAS.417..785J} Jim{\'e}nez, N.,
   Cora, S.~A., Bassino, L.~P., Tecce, T.~E., \& Smith Castelli, A.~V.\ 2011,
   \mnras, 417, 785
 \bibitem[Joye \& Mandel(2003)]{2003ASPC..295..489J} Joye, W.~A., \& Mandel,
   E.\ 2003, Astronomical Data Analysis Software and Systems XII, 295, 489 
 \bibitem[Kaaret \& Feng(2013)]{2013ApJ...770...20K} Kaaret, P., \& Feng,
  H.\ 2013, \apj, 770, 20
 \bibitem[Kaaret et al.(2017)]{kaaret17} Kaaret P., Feng, H., Roberts, T.P.,
  2017, ARA\&A, 55, 303
 \bibitem[Kawamura et al.(2011)]{2011CQGra..28i4011K} Kawamura, S., Ando, M.,
  Seto, N., et al.\ 2011, Classical and Quantum Gravity, 28, 094011 
 \bibitem[Kewley et al.(2001)]{2001ApJ...556..121K} Kewley, L.~J., Dopita,
   M.~A., Sutherland, R.~S., Heisler, C.~A., \& Trevena, J.\ 2001, \apj, 556, 121
 \bibitem[Kirby et al.(2008)]{2008AJ....136.1866K} Kirby, E.~M., Jerjen, H.,
  Ryder, S.~D., \& Driver, S.~P.\ 2008, \aj, 136, 1866 
 \bibitem[Ko et al.(2018)]{2018ApJ...859..108K} Ko, Y., Lee, M.~G., Park,
   H.~S., et al.\ 2018, \apj, 859, 108 
 \bibitem[Koliopanos(2017)]{2017mbhe.confE..51K} Koliopanos, F.\ 2017,
   Proceedings of the XII Multifrequency Behaviour of High Energy Cosmic
   Sources Workshop, Proc.\ of Sci., 306, 51 
 \bibitem[Komossa(2015)]{2015JHEAp...7..148K} Komossa, S.\ 2015, Journal of
  High Energy Astrophysics, 7, 148 
 \bibitem[Komossa \& Bade(1999)]{1999A&A...343..775K} Komossa, S., \& Bade,
  N.\ 1999, \aap, 343, 775 
 \bibitem[Kormendy et al.(1997)]{1997ApJ...482L.139K} Kormendy, J., Bender, R.,
  Magorrian, J., et al.\ 1997, \apjl, 482, L139 
 \bibitem[Kormendy \& Richstone(1995)]{1995ARA&A..33..581K} Kormendy, J., \&
  Richstone, D.\ 1995, ARA\&A, 33, 581
 \bibitem[Kourkchi et al.(2012)]{2012MNRAS.420.2819K} Kourkchi, E.,
   Khosroshahi, H.~G., Carter, D., et al.\ 2012, \mnras, 420, 2819 
 \bibitem[Krajnovi{\'c} et al.(2018)]{2018arXiv180308055K} Krajnovi{\'c}, D.,
  Cappellari, M., McDermid, R.~M., et al.\ 2018, arXiv:1803.08055 
 \bibitem[Kramer \& Champion(2013)]{2013CQGra..30v4009K} Kramer, M., \&
  Champion, D.~J.\ 2013, Classical and Quantum Gravity, 30, 224009
 \bibitem[Laor(1998)]{1998ApJ...505L..83L} Laor, A.\ 1998, ApJ, 505, L83
 \bibitem[Laor(2001)]{2001ApJ...553..677L} Laor, A.\ 2001, ApJ, 553, 677
 \bibitem[Lauer et al.(2007)]{2007ApJ...662..808L} Lauer, T.~R., Faber, S.~M.,
   Richstone, D., et al.\ 2007, \apj, 662, 808 
 \bibitem[Leigh et al.(2015)]{2015MNRAS.451..859L} Leigh, N.~W.~C., Georgiev,
  I.~Y., B{\"o}ker, T., Knigge, C., \& den Brok, M.\ 2015, \mnras, 451, 859 
 \bibitem[Leipski et al.(2012)]{2012ApJ...744..152L} Leipski, C., Gallo, E.,
  Treu, T., et al.\ 2012, \apj, 744, 152 
 \bibitem[Lieder et al.(2012)]{2012A&A...538A..69L} Lieder, S., Lisker, T.,
   Hilker, M., Misgeld, I., \& Durrell, P.\ 2012, \aap, 538, A69 
\bibitem[Lin et al.(2018)]{2018NatAs.tmp...73L} Lin, D., Strader, J.,
  Carrasco, E.~R., et al.\ 2018, Nature Astronomy,  
 \bibitem[Liu(2011)]{2011ApJS..192...10L} Liu, J.\ 2011, \apjs, 192, 10 
 \bibitem[Liu et al.(2012)]{2012ApJ...745...89L} Liu, J., Orosz, J., \&
  Bregman, J.~N.\ 2012, \apj, 745, 89
 \bibitem[Lynden-Bell et al.(1988)]{1988ApJ...326...19L} Lynden-Bell, D.,
   Faber, S.~M., Burstein, D., et al.\ 1988, \apj, 326, 19 
 \bibitem[Magorrian et al.(1998)]{1998AJ....115.2285M} Magorrian, J.,
   Tremaine, S., Richstone, D., et al.\ 1998, \aj, 115, 2285 
 \bibitem[Malumuth \& Kirshner(1981)]{1981ApJ...251..508M} Malumuth, E.~M., \&
   Kirshner, R.~P.\ 1981, \apj, 251, 508 
 \bibitem[Mapelli et al.(2012)]{2012A&A...542A.102M} Mapelli, M., Ripamonti,
  E., Vecchio, A., Graham, A.~W., \& Gualandris, A.\ 2012, \aap, 542, A102 
 \bibitem[{{Marconi} \& {Hunt}(2003)}]{Marconi:2003} {Marconi}, A., \& {Hunt},
  L.~K. 2003, ApJ Lett, 589, L21
 \bibitem[Matkovi{\'c} \& Guzm{\'a}n(2005)]{2005MNRAS.362..289M} Matkovi{\'c},
   A., \& Guzm{\'a}n, R.\ 2005, \mnras, 362, 289 
 \bibitem[MacArthur et al.(2008)]{2008ApJ...680...70M} MacArthur, L.~A.,
   Ellis, R.~S., Treu, T., et al.\ 2008, \apj, 680, 70 
\bibitem[McElroy(1995)]{1995ApJS..100..105M} McElroy, D.~B.\ 1995, \apjs, 100,
  105 
 \bibitem[McLure \& Dunlop(2002)]{2002MNRAS.331..795M} McLure, R.~J., \&
  Dunlop, J.~S.\ 2002, MNRAS, 331, 795
 \bibitem[McNamara(2013)]{McNamara:2013} McNamara, P.~W.\ 2013, International
   Journal of Modern Physics D, 22, 41001
 \bibitem[McNamara \& Nulsen(2007)]{2007ARA&A..45..117M} McNamara, B.~R., \&
  Nulsen, P.~E.~J.\ 2007, \araa, 45, 117 
 \bibitem[Mei et al.(2007)]{2007ApJ...655..144M} Mei, S., Blakeslee, J.~P.,
  C{\^o}t{\'e}, P., et al.\ 2007, \apj, 655, 144 
 \bibitem[Merloni et al.(2003)]{2003MNRAS.345.1057M} Merloni, A., Heinz, S., \&
  di Matteo, T.\ 2003, MNRAS, 345, 1057 
 \bibitem[Merritt \& Ferrarese(2001)]{2001ApJ...547..140M} Merritt, D., \&
   Ferrarese, L.\ 2001, \apj, 547, 140 
 \bibitem[Metcalfe et al.(1994)]{MGP94}Metcalfe, N., Godwin, J.G., Peach, J.V.\ 1994, MNRAS, 267, 431
\bibitem[Mezcua(2017)]{2017IJMPD..2630021M} Mezcua, M.\ 2017, International
  Journal of Modern Physics D, 26, 1730021 
\bibitem[Mezcua et al.(2018a)]{Mezcua2018a} Mezcua, M., Kim, M., Ho,
  L.~C., \& Lonsdale, C.~J.\ 2018a, \mnras, 478, 2576
\bibitem[Mezcua et al.(2018b)]{2018MNRAS.480L..74M} Mezcua, M., Kim, M., Ho,
  L.~C., \& Lonsdale, C.~J.\ 2018b, \mnras, 480, L74 
 \bibitem[Mezcua et al.(2015)]{2015MNRAS.448.1893M} Mezcua, M., Roberts,
  T.~P., Lobanov, A.~P., \& Sutton, A.~D.\ 2015, \mnras, 448, 1893
 \bibitem[Miller et al.(2015)]{2015ApJ...799...98M} Miller, B.~P., Gallo, E.,
  Greene, J.~E., et al.\ 2015, \apj, 799, 98 
 \bibitem[Miller et al.(2013)]{2013:Miller} Miller, J. M., Walton, D. J., King,
  A. L., et al., 2013, ApJL, 776, L36
 \bibitem[Minkowski(1962)]{1962IAUS...15..112M} Minkowski, R.\ 1962, Problems
  of Extra-Galactic Research, 15, 112 
 \bibitem[Misgeld et al.(2008)]{2008A&A...486..697M} Misgeld, I., Mieske, S.,
   \& Hilker, M.\ 2008, \aap, 486, 697 
\bibitem[Miyoshi et al.(1995)]{1995Natur.373..127M} Miyoshi, M., Moran, J.,
  Herrnstein, J., et al.\ 1995, Nature, 373, 127
 \bibitem[Morris \& Serabyn(1996)]{1996ARA&A..34..645M} Morris, M., \&
   Serabyn, E.\ 1996, \araa, 34, 645 
 \bibitem[Netzer \& Peterson(1997)]{NaP97}Netzer H., Peterson B.M., 1997, in
  Astronomical Time Series, ed.\ D.\ Maoz, A.\ Sternberg, \& E.M.\ Leibowitz (Dordrecht: Kluwer), 85
 \bibitem[Nguyen et al.(2017)]{2017ApJ...836..237N} Nguyen, D.~D., Seth, A.~C.,
  den Brok, M., et al.\ 2017, \apj, 836, 237 
 \bibitem[Nguyen et al.(2018)]{2018ApJ...858..118N} Nguyen, D.~D., Seth,
   A.~C., Neumayer, N., et al.\ 2018, \apj, 858, 118 
 \bibitem[Nowak et al.(2007)]{Nowak}Nowak, N., Saglia, R.P., Thomas, J., et al.\ 2007, MNRAS, 379, 909
 \bibitem[Onken et al.(2004)]{2004ApJ...615..645O} Onken, C.~A., Ferrarese, L.,
  Merritt, D., et al.\ 2004, \apj, 615, 645 
 \bibitem[Oka et al.(2015)]{Oka:2015} Oka, T., Mizuno, R., Miura, K.,
   Takekawa, S.\ 2016, ApJ, 816, L7
 \bibitem[Oka et al.(2016)]{2016ApJ...816L...7O} Oka, T., Mizuno, R., Miura,
   K., \& Takekawa, S.\ 2016, \apjl, 816, L7 
 \bibitem[Oka et al.(2017)]{2017NatAs...1..709O} Oka, T., Tsujimoto, S.,
   Iwata, Y., Nomura, M., \& Takekawa, S.\ 2017, Nature Astronomy, 1, 709 
 \bibitem[Paudel et al.(2011)]{2011MNRAS.413.1764P} Paudel, S., Lisker, T., \&
  Kuntschner, H.\ 2011, \mnras, 413, 1764 
 \bibitem[Pasham et al.(2015)]{2015:Pasham} Pasham, D.R., Cenko, S.B., Zoghbi,
  A., Mushotzky, R.F., Miller, J., Tombesi, F.\ 2015, ApJL, 811, L11
 \bibitem[Paturel et al.(2003)]{Pat03}Paturel G., Petit C., Prugniel P., Theureau G., Rousseau J.,
  Brouty M., Dubois P., \& Cambr{\'e}sy L.\ 2003, A\&A\, 412, 45
 \bibitem[Peterson \& Ferland(1986)]{1986Natur.324..345P} Peterson, B.~M., \&
  Ferland, G.~J.\ 1986, \nat, 324, 345 
 \bibitem[Plotkin et al.(2014)]{2014ApJ...780....6P} Plotkin, R.~M., Gallo,
   E., Miller, B.~P., et al.\ 2014, \apj, 780, 6 
 \bibitem[Poincar\'e(1905)]{Poincare} Poincar\'e H., 1905, C.R.\ Acad.\ Sci.\ 140, 1504
 \bibitem[Ptak \& Griffiths(1999)]{1999ApJ...517L..85P} Ptak, A., \& Griffiths,
  R.\ 1999, \apjl, 517, L85 
 \bibitem[Ravi et al.(2018)]{2018MNRAS.478L..72R} Ravi, V., Vedantham, H., \&
   Phinney, E.~S.\ 2018, \mnras, 478, L72 
 \bibitem[Rees(1988)]{1988Natur.333..523R} Rees, M.~J.\ 1988, \nat, 333, 523 
 \bibitem[Reines \& Deller(2012)]{2012ApJ...750L..24R} Reines, A.~E., \&
  Deller, A.~T.\ 2012, \apjl, 750, L24 
\bibitem[Reines et al.(2013)]{2013ApJ...775..116R} Reines, A.~E., Greene,
  J.~E., \& Geha, M.\ 2013, \apj, 775, 116 
\bibitem[Remillard \& McClintock(2006)]{2006ARA&A..44...49R} Remillard, R.A.,
  McClintock, J.E.\ 2006, \araa, 44, 49
 \bibitem[Russell et al.(2017)]{2017MNRAS.472.4024R} Russell, H.~R., McNamara,
  B.~R., Fabian, A.~C., et al.\ 2017, \mnras, 472, 4024 
 \bibitem[Ryan et al.(2007)]{2007ApJ...654..799R} Ryan, C.~J., De Robertis,
   M.~M., Virani, S., Laor, A., \& Dawson, P.~C.\ 2007, ApJ, 654, 799
 \bibitem[Salucci et al.(2000)]{2000MNRAS.317..488S} Salucci, P., Ratnam, C.,
   Monaco, P., \& Danese, L.\ 2000, \mnras, 317, 488 
 \bibitem[Savorgnan et al.(2016)]{2016ApJ...817...21S} Savorgnan, G.~A.~D.,
  Graham, A.~W., Marconi, A., \& Sani, E.\ 2016, \apj, 817, 21 
 \bibitem[Schechter(1980)]{1980AJ.....85..801S} Schechter, P.~L.\ 1980, \aj,
   85, 801 
 \bibitem[Schlafly \& Finkbeiner(2011)]{2011ApJ...737..103S} Schlafly, E.~F.,
   \& Finkbeiner, D.~P.\ 2011, \apj, 737, 103 
 \bibitem[Sch{\"o}del et al.(2007)]{2007A&A...469..125S} Sch{\"o}del, R.,
  Eckart, A., Alexander, T., et al.\ 2007, \aap, 469, 125 
 \bibitem[Schombert \& Smith(2012)]{2012PASA...29..174S} Schombert, J., \&
   Smith, A.~K.\ 2012, \pasa, 29, 174 
 \bibitem[Scott et al.(2013)]{2013ApJ...768...76S} Scott, N., Graham, A.~W.,
   \& Schombert, J.\ 2013, \apj, 768, 76 
 \bibitem[Secker et al.(1997)]{1997PASP..109.1377S} Secker, J., Harris, W.~E.,
   \& Plummer, J.~D.\ 1997, \pasp, 109, 1377 
 \bibitem[Secrest et al.(2012)]{2012ApJ...753...38S} Secrest, N.~J., Satyapal,
  S., Gliozzi, M., et al.\ 2012, \apj, 753, 38
 \bibitem[Seigar et al.(2008)]{2008ApJ...678L..93S} Seigar, M.~S., Kennefick,
   D., Kennefick, J., \& Lacy, C.~H.~S.\ 2008, \apjl, 678, L93 
 \bibitem[S{\'e}rsic(1963)]{1963BAAA....6...41S} S{\'e}rsic, J.~L.\ 1963,
  Boletin de la Asociacion Argentina de Astronomia La Plata Argentina, 6, 41 
 \bibitem[Seth et al.(2008)]{2008ApJ...678..116S} Seth, A., Ag{\"u}eros, M.,
   Lee, D., \& Basu-Zych, A.\ 2008, \apj, 678, 116-130 
 \bibitem[Shannon et al.(2015)]{2015Sci...349.1522S} Shannon, R.~M., Ravi, V.,
  Lentati, L.~T., et al.\ 2015, Science, 349, 1522 
 \bibitem[Shen(2018)]{2018arXiv180909359S} Shen, R.-F.\ 2018, ApJL, submtied, arXiv:1809.09359 
 \bibitem[Shen \& Gebhardt(2009)]{SaG09}Shen, J., \& Gebhardt, K.\ 2010, ApJ, 711, 484
 \bibitem[Smith Castelli et al.(2013)]{2013ApJ...772...68S} Smith Castelli,
  A.~V., Gonz{\'a}lez, N.~M., Faifer, F.~R., \& Forte, J.~C.\ 2013, \apj, 772,
  68 
 \bibitem[Soria et al.(2010)]{Soria:2010}Soria, R., Hau, G.K.T., Graham, A.W.,
  Kong, A.K.H., Kuin, N.P.M., Li, I.-H., Liu, J.-F. \& Wu, K.\ 20\ 10, MNRAS,
  405, 870
 \bibitem[Soria et al.(2017)]{2017MNRAS.469..886S} Soria, R., Musaeva, A., Wu,
   K., et al.\ 2017, \mnras, 469, 886 
 \bibitem[Sofue(1977)]{1977A&A....60..327S} Sofue, Y.\ 1977, \aap, 60, 327 
 \bibitem[Sofue(2000)]{2000ApJ...540..224S} Sofue, Y.\ 2000, \apj, 540, 224 
 \bibitem[Sofue \& Handa(1984)]{1984Natur.310..568S} Sofue, Y., \& Handa,
  T.\ 1984, \nat, 310, 568 
 \bibitem[Su et al.(2010)]{2010ApJ...724.1044S} Su, M., Slatyer, T.~R., \&
   Finkbeiner, D.~P.\ 2010, \apj, 724, 1044 
 \bibitem[Sutton et al.(2012)]{2012MNRAS.423.1154S} Sutton, A.~D., Roberts,
  T.~P., Walton, D.~J., Gladstone, J.~C., \& Scott, A.~E.\ 2012, \mnras, 423,
  1154
 \bibitem[T{\'a}pai(2015)]{2015mgm..conf..957T} T{\'a}pai, M.~K., Zolt{\'a}n
  Gergely, L{\'a}szl{\'o} {\'A}.\ 2015, Thirteenth Marcel Grossmann Meeting,
  edited by Rosquist Kjell et al., Published by World Scientific Publishing,
  957 (arXiv:1212.4973)
 \bibitem[Tenorio-Tagle \& Bodenheimer(1988)]{1988ARA&A..26..145T}
  Tenorio-Tagle, G., \& Bodenheimer, P.\ 1988, \araa, 26, 145
 \bibitem[Terashima \& Wilson(2003)]{2003ApJ...583..145T} Terashima Y., Wilson
   A.S.\ 2003, \apj, 583, 145
 \bibitem[Terlevich et al.(2001)]{2001MNRAS.326.1547T} Terlevich, A.~I.,
   Caldwell, N., \& Bower, R.~G.\ 2001, \mnras, 326, 1547 
 \bibitem[Toloba et al.(2014)]{2014ApJS..215...17T} Toloba, E., Guhathakurta,
  P., Peletier, R.~F., et al.\ 2014, \apjs, 215, 17 
 \bibitem[Tonry(1981)]{1981ApJ...251L...1T} Tonry, J.~L.\ 1981, \apjl, 251, L1 
 \bibitem[Tremonti et al.(2004)]Tremonti, C.A., et al.\ 2004, ApJ, 613, 898
 \bibitem[Tsuboi et al.(1985)]{1985PASJ...37..359T} Tsuboi, M., Inoue, M.,
  Handa, T., Tabara, H., \& Kato, T.\ 1985, \pasj, 37, 359 
 \bibitem[Tsuboi et al.(2017)]{2017ApJ...850L...5T} Tsuboi, M., Kitamura, Y.,
   Tsutsumi, T., et al.\ 2017, \apjl, 850, L5 
 \bibitem[Valencia-S.~et al.(2012)]{2012A&A...544A.129V} Valencia-S., M.,
  Zuther, J., Eckart, A., et al.\ 2012, \aap, 544, A129 
 \bibitem[Walsh et al.(2010)]{Walsh}Walsh, J.L., Barth, A.J., \& Sarzi, M.\ 2010, ApJ, 721, 762
 \bibitem[Wandel(1999)]{1999ApJ...519L..39W} Wandel, A.\ 1999,ApJ, 519, L39
 \bibitem[Webb et al.(2010)]{2010ApJ...712L.107W} Webb, N.~A., Barret, D.,
   Godet, O., et al.\ 2010, \apjl, 712, L107
 \bibitem[Webb et al.(2014)]{2014ApJ...780L...9W} Webb, N.~A., Godet, O.,
   Wiersema, K., et al.\ 2014, \apjl, 780, L
 \bibitem[Webb et al.(2017)]{2017A&A...602A.103W} Webb, N.~A., Gu{\'e}rou, A.,
   Ciambur, B., et al.\ 2017, \aap, 602, A103 
 \bibitem[Yee(1992)]{1992ASPC...31..417Y} Yee, H.~K.~C.\ 1992, in Relationships
  Between Active Galactic Nuclei and Starburst Galaxies, ed. A. V. Filippenko,
  ASP Conference Series (ASP: San Francisco), 31, 417
 \bibitem[Yusef-Zadeh et al.(2000)]{2000Sci...287...85Y} Yusef-Zadeh, F.,
   Melia, F., \& Wardle, M.\ 2000, Science, 287,  
 \bibitem[Zhang et al.(2009)]{2009ApJ...699..281Z} Zhang, W.~M., Soria, R.,
  Zhang, S.~N., Swartz, D.~A., \& Liu, J.~F.\ 2009, \apj, 699, 281 

\end{thebibliography}
\end{document}